\documentclass[11pt, a4paper]{article}

\usepackage[a4paper]{geometry}
\usepackage{mathtools}
\usepackage{amsmath}
\numberwithin{equation}{section}
\usepackage{amssymb}
\usepackage{amsthm}    
\usepackage{float}    
\usepackage{subcaption}
\usepackage{multirow}    
\usepackage{booktabs}    
\usepackage{array}    
\usepackage{tabularx}    
\usepackage{rotating}    
\usepackage{siunitx}
\usepackage{wrapfig}    
\usepackage{array}    
\usepackage{makecell}    
\usepackage{xcolor}    

\usepackage{tocloft}    
\usepackage{url}
\usepackage{hyperref}

\usepackage[boxruled]{algorithm2e}    

\usepackage{bbm}    
\usepackage{xfrac}    
\usepackage{siunitx}    

\usepackage[square, numbers, sort&compress]{natbib}
\usepackage{cleveref}    

\crefname{algocf}{alg.}{algs.}
\Crefname{algocf}{Algorithm}{Algorithms}
\crefname{mdef}{definition}{definitions}
\Crefname{mdef}{Definition}{Definitions}
\crefname{prop}{proposition}{propositions}
\Crefname{prop}{Proposition}{Propositions}
\crefname{lem}{lemma}{lemmas}
\Crefname{lem}{Lemma}{Lemmas}
\crefname{cor}{corollary}{corollaries}
\Crefname{cor}{Corollary}{corollaries}

\renewcommand{\cref}{\Cref}

\newcolumntype{Y}{>{\centering\arraybackslash}X}    

\newenvironment{itemize*}%
{\begin{itemize}%
		\setlength{\itemsep}{0pt}%
		\setlength{\parskip}{0pt}}%
	{\end{itemize}}
\newenvironment{enumerate*}%
{\begin{enumerate}%
		\setlength{\itemsep}{0pt}%
		\setlength{\parskip}{0pt}}%
	{\end{enumerate}}

\newcommand{\lword}[1]{\leavevmode\nobreak\hskip0pt plus \linewidth\penalty50\hskip0pt plus-\linewidth\nobreak{#1}}    

\newcommand{\TNOChallenge}{TNO OLYMPUS Field Development Optimisation Challenge}    
\newcommand{\TNOChallengeshort}{TNO OLYMPUS Challenge}    
\newcommand{\TNOChallengeWC}{TNO OLYMPUS Well Control Optimisation Challenge}    

\newcommand{\OLYMPUSWorkshop}{EAGE/TNO Workshop on OLYMPUS Field Development Optimization}    

\newcommand{\pardiff}{\delta}    
\newcommand{\pardiffi}{\pardiff_i}    

\newcommand{\pardiffvec}{\boldsymbol{\pardiff}}    

\newcommand{\dcp}{d_{jk,t_i}}    

\newcommand{\cpgeneral}{c}    
\newcommand{\cplgeneral}{\cpgeneral{}^l}    
\newcommand{\cpugeneral}{\cpgeneral{}^u}    

\newcommand{\cpl}[2]{\cpgeneral{}_{#1,#2}^l}    
\newcommand{\cppl}[2]{\cpl{P#1}{#2}}    

\newcommand{\cpu}[2]{\cpgeneral{}_{#1,#2}^u}    
\newcommand{\cppu}[2]{\cpu{P#1}{#2}}    

\newcommand{\extrapcogeneral}{b}    

\newcommand{\extrapco}[2]{\extrapcogeneral{}_{#1,#2}}    
\newcommand{\extrapcop}[2]{\extrapco{P#1}{#2}}    

\newcommand{\NPV}{\mathrm{NPV}}    
\newcommand{\NPVjd}{\NPV_j(\boldd)}    

\newcommand{\approxNPV}{\widetilde{\NPV}}    
\newcommand{\approxNPVjd}{\approxNPV_j(\boldd)}    

\newcommand{\ENPV}{\E[\NPV]}    
\newcommand{\ENPVd}{\ENPV(\boldd)}    

\newcommand{\meanNPV}{\overline{\NPV}}    
\newcommand{\meanNPVd}{\meanNPV(\boldd)}    

\DeclareMathOperator{\E}{\mathbb{E}}    
\DeclareMathOperator{\Var}{\mathrm{Var}}    
\DeclareMathOperator{\Cov}{\mathrm{Cov}}    

\newcommand{\boldx}{\mathbf{x}}    

\newcommand{\boldd}{\mathbf{d}}    
\newcommand{\bolddprime}{\boldd^\prime}    

\newcommand{\djkti}{d_{jk, t_i}}    
\newcommand{\djktiminusone}{d_{jk, t_i - 1}}    

\newcommand{\Dprime}{D^\prime}    

\newcommand{\Djk}{D^{(jk)}}    

\newcommand{\indicator}[1]{\mathbbm{1}_{ \left\{ #1 \right\} }}    

\newcommand{\indicatorequalboldd}{\indicator{ \boldd = \bolddprime }}    

\newcommand{\fd}{f(\boldd)}    

\newcommand{\fid}{f_i(\boldd)}    
\newcommand{\fidprime}{f_i(\bolddprime)}    

\newcommand{\fjd}{f_j(\boldd)}    

\newcommand{\approxf}{\widetilde{f}}    
\newcommand{\approxfd}{\approxf(\boldd)}    
\newcommand{\approxfid}{\approxf_i(\boldd)}    

\newcommand{\fbar}{\bar{f}}    
\newcommand{\fbard}{\fbar(\boldd)}    

\newcommand{\boldg}{\mathbf{g}}    

\newcommand{\loo}{leave-one-out}    
\newcommand{\Loo}{Leave-one-out}    

\newcommand{\adjRsq}{adjusted $R^2$}    
\newcommand{\AdjRsq}{Adjusted $R^2$}    

\newcommand{\Ntilde}{\tilde{N}}    
\newcommand{\mathEGES}{\mathrm{ES}}    
\newcommand{\epsilonEGES}{\varepsilon_{\mathEGES}}    

\newcommand{\sigmahatEGES}{\hat{\sigma}_\mathEGES}    

\newcommand{\betahatEGES}{\hat{\boldbeta}_\mathEGES}   

\newcommand{\XEGESd}{X_\mathEGES(\boldd)}    

\newcommand{\SigmabetaEGES}{\Sigma_{\beta, \mathEGES}}    

\newcommand{\Ud}{U(\boldd)}    

\newcommand{\alignnumber}{\addtocounter{equation}{1}\tag{\theequation}}    
\newcommand{\alignmultilineeq}{\\ &{} \phantom{=}\;}    

\DeclareMathOperator*{\argmax}{arg\,max}    
\DeclareMathOperator*{\argmin}{arg\,min}    

\newcommand{\Transpose}{\mathrm{T}}    

\newcommand{\boldzero}{\mathbf{0}}    

\newcommand{\mathBL}{\mathrm{BL}}    

\newcommand{\boldtheta}{\boldsymbol{\theta}}    

\newcommand{\boldbeta}{\boldsymbol{\beta}}    

\newcommand{\mubeta}{\boldsymbol{\mu}_{\beta}}    

\newcommand{\boldF}{\mathbf{F}}    
\newcommand{\boldFi}{\boldF_i}    
\newcommand{\boldFj}{\boldF_j}    

\newcommand{\EFfd}{\E_\boldF[\fd]}    
\newcommand{\VarFfd}{\Var_\boldF[\fd]}    

\newcommand{\EFifd}{\E_{\boldFi}[ \fid ]}    
\newcommand{\VarFifd}{\Var_{\boldFi}[ \fid ]}    
\newcommand{\CovFifdfdprime}{\Cov_{\boldFi}[ \fid, \fidprime ]}    

\newcommand{\Varbetalimit}{\Var[\boldbeta] \rightarrow \infty}    

\title{Bayesian Emulation of Grey-Box Multi-Model Ensembles Exploiting Known Interior Structure}
\author{Jonathan Owen\footnote{Corresponding author: jonathan.owen@sheffield.ac.uk} $^1$ \& Ian Vernon$^2$ \\
	\hspace{1cm} \\
	\small $^1$School of Mathematical and Physical Sciences, University of Sheffield, Sheffield, United Kingdom  \\
	\small $^2$Department of Mathematical Sciences, Durham University, Durham, United Kingdom
}
\date{\today}

\begin{document}

\maketitle

%
%

\begin{abstract}
	Computer models are widely used to study complex real world physical systems. However, there are major limitations to their direct use including: their complex structure; large numbers of inputs and outputs; and long evaluation times. Bayesian emulators are an effective means of addressing these challenges providing fast and efficient statistical approximation for computer model outputs. It is commonly assumed that computer models behave like a ``black-box'' function with no knowledge of the output prior to its evaluation. This ensures that emulators are generalisable but potentially limits their accuracy compared with exploiting such knowledge of constrained or structured output behaviour. We assume a ``grey-box'' computer model and develop a methodological toolkit for its analysis. This includes: multi-model ensemble subsampling to identifying a representative model subset to reduce computational expense; constructing a targeted Bayesian design for optimisation or decision support; a ``divide-and-conquer’’ approach to emulating sums of outputs; structured emulators exploiting known constrained and structured behaviour of constituent outputs through splitting the parameter space and imposing truncations; emulation of sums of time series outputs; and emulation of multi-model ensemble outputs. Combining these methods establishes a hierarchical emulation framework which achieves greater physical interpretability and more accurate emulator predictions. This research is motivated by and applied to the commercially important \TNOChallengeWC{} from the petroleum industry which we re-express as a decision support under uncertainty problem. We thus encourage users to examine their ``black-box'' simulators to achieve superior emulator accuracy.
\end{abstract}

\section{Introduction} \label{sec:Introduction}

Mathematical models of complex real world physical systems in the form of numerical codes known as computer models or simulators are prevalent across many scientific disciplines, industry, and government. They are used to: study the dynamics of physical systems \cite{2018:Vernon:Bayesian-uncertainty-analysis-systems-biology}; calibrate or history match to observation data \cite{2001:Craig:Bayesian-Forecasting-Using-Computer-Simulators,1996:Craig:Bayes-linear-strategies-matching-hydrocarbon-reservoir-history}; and to guide decision making processes \cite{2020:Owen:A-Bayesian-statistical-approach-to-decision-support-for-TNO-OLYMPUS-well-control-optimisation-under-uncertainty,2023:Kennedy:Multilevel-emulation-for-stochastic-computer-models-with-application-to-large-offshore-wind-farms}. However, computer models commonly exhibit a complex structure; possess large numbers of inputs and outputs, including spatial-temporal fields; and crucially have a high computational expense of evaluation. In order to address such challenges, a suite of Bayesian uncertainty analysis methodology has been developed for using computer models to perform inferences about real world systems. Of principal importance are Bayesian emulators, also known as surrogate models, which provide fast statistical approximations to (functions of) the computer model outputs for as yet unevaluated parameter settings, along with a corresponding statement of the associated uncertainty \cite{2001:Craig:Bayesian-Forecasting-Using-Computer-Simulators,2006:O'Hagan:Bayesian-analysis-of-computer-code-outputs-a-tutorial,2010:Vernon:Bayesian-Uncertainty-Analysis-Galaxy-Formation}. They are typically many orders of magnitude quicker to evaluate than the computer model. Emulators have been successfully employed across a wide range of applications including: climate science \cite{2006:Goldstein:Bayes-Linear-Calibrated-Prediction-for-Complex-Systems,2009:Rougier:Expert-Knowledge-and-Multivariate-Emulation-The-TIE-GCM,2013:Williamson:History-matching-for-exploring-and-reducing-climate-model-parameter-space-using-observations-and-a-large-perturbed-physics-ensemble,2019:Edwards:Revisiting-Antarctic-ice-loss-due-to-marine-ice-cliff-instability}; cosmology \cite{2010:Vernon:Bayesian-Uncertainty-Analysis-Galaxy-Formation,2014:Vernon:Bayesian-History-Matching-Galaxy-Formation,2009:Heitmann:The-Coyote-Universe-II-Cosmological-Models-and-Precision-Emulation-of-the-Nonlinear-Matter-Power-Spectrum,2011:Kaufman:Efficient-emulators-of-computer-experiments-using-compactly-supported-correlation-functions-with-an-application-to-cosmology}; epidemiology \cite{2015:Andrianakis:Bayesian-History-Matching-of-Complex-Infectious-Disease-Models-Using-Emulation-A-Tutorial-and-a-Case-Study-on-HIV-in-Uganda,2017:Andrianakis:History-matching-of-a-complex-epidemiological-model-of-HIV-transmission-by-using-variance-emulation,2022:Vernon:Bayesian-Emulation-and-History-Matching-of-JUNE}; and petroleum reservoir engineering \cite{1996:Craig:Bayes-linear-strategies-matching-hydrocarbon-reservoir-history,1997:Craig:Pressure-matching-Bayes-linear-strategies-large-computer-experiments,1998:Craig:Constructing-partial-prior-specifications,2001:Craig:Bayesian-Forecasting-Using-Computer-Simulators,1998:O'Hagan:Uncertainty-Analysis-and-other-Inference-Tools-for-Complex-Computer-Codes,2010:Cumming:Multiscale-Bayes-Linear-Uncertainty-Analysis-Oil-Reservoirs}.

Emulation is frequently based on the assumption that a computer model behaves like a ``black-box'' function: the output at a given parameter setting is unknown prior to model evaluation; as well as users' possessing no insight of the structure or links between individual physical processes. Whilst this assumption ensures that emulation methodology is generalisable, it potentially limits the emulator accuracy compared to when a user has an understanding of how certain outputs behave with respect to changes in the inputs. In this paper we assume a ``grey-box'' simulator which we define as possessing insight into the model behaviour prior to its evaluation, but without any specific knowledge of the underlying physics, model structure or equations governing the model. Physics informed approaches encompass: emulators for functions with known boundaries \cite{2019:Vernon:Known-Boundary-Emulation-of-Complex-Computer-Models,2023:Jackson:Efficient-Emulation-of-Computer-Models-Utilising-Multiple-Known-Boundaries-of-Differing-Dimensions}; emulators for functions possessing structured (partial) discontinuities in their input parameter space \cite{2023:Owen:Bayesian-Emulation-of-Complex-Computer-Models-with-Structured-Partial-Discontinuities,2024:Vernon:Bayesian-Emulation-for-Computer-Models-with-Multiple-Partial-Discontinuities}; and physics-informed neural networks which encode prior information of physical laws in non-linear partial differential equations which are used in the loss function when fitting a neural network \cite{2019:Raissi:Physics-informed-neural-networks-A-deep-learning-framework-for-solving-forward-and-inverse-problems-involving-nonlinear-partial-differential-equations}, however these are unsuitable for application to ``grey-box'' computer models of the described form.

In this paper we address the problem of constructing an accurate and efficient emulator for an output of interest obtained from a multi-model ensemble whilst exploiting known behaviour of individual ensemble members and components of the output. Combining these methods yields a novel hierarchical emulation framework and toolkit to incorporate specific features in the analysis of ``grey-box'' computer models resulting in a physically interpretable emulator with accurate predictions. The toolkit includes:
\begin{enumerate*}
	\item Subsampling from multi-model ensembles technique to identify a representative subset of models enabling more efficient use of available computational resources (\cref{sec:Subsampling-from-Multi-Model-Ensembles}),	
	
	\item Targeted Bayesian design of computer model simulations geared towards the optimisation or decision support objective (\cref{sec:Targeted-Bayesian-Design}),
	
	\item Divide-and-conquer approach to emulation where the output of interest is represented by a linear combination of constituent model outputs (\cref{sec:Divide-and-Conquer-Approach}),
	
	\item Structured emulation of ``grey-box'' computer models exploiting a known form of simulator behaviour to divide the parameter space by mode of behaviour and employing emulator truncation to enforce known physical effects (\cref{sec:Structured-Emulators-Exploiting-Known-Simulator-Behaviour}),
	
	\item Emulation of outputs formed as sums of time series outputs (\cref{sec:Emulating-Sums-of-Time-Series-Outputs}),
	
	\item Emulation of ensemble mean outputs (\cref{sec:Emulation-of-a-Multi-Model-Ensemble-Mean}).
\end{enumerate*}
All of these methods were required to address the challenges encountered in the motivating problem, although for the analysis of other computationally expensive computer models and multi-model ensembles it may be appropriate to employ a subset of these techniques.

This research is motivated by the highly complex and commercially significant \TNOChallengeWC{} \cite{2017:TNO:OLYMPUS-oil-Reservoir-Model,2018:ISAPP:Website} from the petroleum industry. The aim is to maximise the expected Net Present Value (NPV) objective function over the field lifetime with respect to well control decision parameters (target production and injection rates), whilst accounting for geological uncertainty represented through a multi-model ensemble of 50 realisations from an underlying stochastic geology model. We recast this as a decision support problem for which emulation and Bayesian uncertainty analysis techniques are essential due to the computational expense of the ensemble and high-dimensionality of the decision parameter space. An initial attempt at the \TNOChallengeWC{} from a Bayesian statistical perspective is presented in \cite{2020:Owen:A-Bayesian-statistical-approach-to-decision-support-for-TNO-OLYMPUS-well-control-optimisation-under-uncertainty} which was only moderately successful as it failed to exploit a number of challenging features in the multi-model ensemble mean NPV output. We therefore propose a series of methodological advances in this paper for which all were required in the \TNOChallengeWC{}.
Implementation of the efficient multi-model ensemble subsampling technique to identify a representative subset of models constitutes a novel application to petroleum reservoir engineering and greatly reduces the computational expense of the analysis. For each ensemble member the NPV is computed as the sum of discounted time series model outputs. Many of these exhibit known constrained or structured behaviour with respect to their corresponding well control parameters for which we formulate structured emulators. These are combined within our hierarchical emulator construction. We demonstrate a notable reduction in the emulator uncertainty compared with uninformed Bayes linear emulators. Whilst we establish our techniques in the context of decision support for well control optimisation under uncertainty, the overarching framework is flexible and adaptable to handle other structured forms of simulator outputs.

In \cref{sec:TNO-OLYMPUS-Well-Control-Optimisation-Challenge} we describe the motivating \TNOChallenge{}. \Cref{sec:Bayesian-Emulation} provides an overview of Bayesian emulation methodology and an application of Bayes linear emulators to the \TNOChallengeWC{} ensemble mean NPV objective function. For \cref{sec:Subsampling-from-Multi-Model-Ensembles,sec:Targeted-Bayesian-Design,sec:Divide-and-Conquer-Approach,sec:Structured-Emulators-Exploiting-Known-Simulator-Behaviour,sec:Emulating-Sums-of-Time-Series-Outputs,sec:Emulation-of-a-Multi-Model-Ensemble-Mean}, as in the above numbered list, we detail each methodological developments and its application to the \TNOChallengeWC{} in turn. In \cref{sec:Emulation-of-a-Multi-Model-Ensemble-Mean} a comparison is also performed of the developed approach to emulation with direct Bayes linear emulation of the ensemble mean NPV. A conclusion and future research directions are discussed in \cref{sec:Conclusion}.

\section{TNO OLYMPUS Well Control Optimisation Challenge} \label{sec:TNO-OLYMPUS-Well-Control-Optimisation-Challenge}

First we provide an overview of the \TNOChallengeWC{} in \cref{subsec:TNO-OLYMPUS-Well-Control-Optimisation-Challenge-Summary} before performing an exploratory analysis to highlight several important features of this challenge in \cref{subsec:OLYMPUS-Exploratory-Analysis} which motivate the subsequent methodological development.

\subsection{Summary} \label{subsec:TNO-OLYMPUS-Well-Control-Optimisation-Challenge-Summary}

A major and commercially important challenge in the petroleum industry is field development under uncertainty for a green oil field\footnote{A green oil field is a new subsurface region believed to contain oil or gas which has yet to be exploited meaning that no drilling, production or injection has been performed.}. The Netherlands Organisation for Applied Scientific Research (TNO), as part of Integrated Systems Approach for Petroleum Production (ISAPP) research programme, devised the \TNOChallenge{} \cite{2018:ISAPP:Website} (abbreviated to \TNOChallengeshort{}) to encourage research and technological advancements to address the problem of optimisation under uncertainty. There is a particular emphasis on the uncertainty induced by the unknown underlying geology. The \TNOChallengeshort{} has received much attention across academia and industry with results of the competition phase presented at the EAGE/TNO Workshop on OLYMPUS Field Development Optimization \cite{2018:EAGE-TNO-Workshop-on-OLYMPUS-Field-Development-Optimization}.

The \TNOChallengeshort{} is based around the fictitious oil reservoir named OLYMPUS (inspired by a virgin oil field in the North Sea) and specifically designed by TNO for the challenge. OLYMPUS is a medium complexity model of size 9km by 3km, with a depth of 50m split into 16 layers for modelling purposes. The design was conceived to imitate a real oil field possessing many of the features encountered in actual oil fields including: boundary and minor geological faults; two vertical zones separated by an impermeable shale layer (the top layer contains fluvial channel sands embedded in floodplain shale, whilst the bottom layer consists of alternating layers of coarse, medium and fine sands); as well as multiple types of facies (body of rock of specified characteristics) including channel sands, shale, and multiple types of sand. Geological uncertainty (unknown porosity, permeability, net-to-gross, and initial water saturation) is represented via a multi-model ensemble of $N = 50$ OLYMPUS realisations of a stochastic geology model. These are labelled as OLYMPUS 1 to 50. Full details of the model can be found in \cite{2017:TNO:OLYMPUS-oil-Reservoir-Model}.

The \TNOChallengeshort{} consists of three sub-challenges:
\begin{enumerate*}
	\item Well control optimisation,
	
	\item Field development optimisation,
	
	\item Joint optimisation of well placement and well control.
\end{enumerate*}
In this paper we focus on the first where the aim is to develop an optimal strategy with respect to maximising the expected Net Present Value (NPV) objective function over the 20 year field lifetime (starting January 1, 2016) with accumulation and discounting at 3 month intervals. The NPV for an individual OLYMPUS model is denoted $\NPVjd$ and is defined in \cref{eq:NPV-general-formula-in-decision-parameters} as a function of a vector of decision parameters, $\boldd$, consisting of target production and injection rates for producer and injector wells respectively.
\begin{align}
	\NPVjd &{} = \sum_{i = 1}^{N_t} \dfrac{R_j(\boldd, t_i)}{(1 + d)^{\frac{t_i}{\tau}}} \label{eq:NPV-general-formula-in-decision-parameters} \\
	\ENPVd &{} \approx \meanNPVd = \dfrac{1}{N} \sum_{j = 1}^{N} \NPVjd \label{eq:Expected-NPV-general-formula-in-decision-parameters}
\end{align}
The index $i$ refers to the time interval $\Delta t_i = t_i - t_{i - 1}$, total number of time intervals $N_t$, fixed discount factor $d = 0.08$, time interval for discounting $\tau = 365$ days, and $R_j(\boldd, t_i)$ as the difference of all revenue and expenditure during the interval $\Delta t_i$, and $j = 1, \ldots, N = 50$ indexes the particular OLYMPUS realisation of the stochastic geology model. The expected NPV is approximated by the ensemble mean NPV defined in \cref{eq:Expected-NPV-general-formula-in-decision-parameters}, as dictated by the \TNOChallengeshort{}, and hence forms our quantity of interest.

For the well control optimisation challenge a fixed well configuration is provided by TNO based on oil reservoir engineering principles with $R_j(\boldd, t_i)$ defined in \cref{eq:Well-control-optimisation-challenge-NPV-R(d_t_i)}, where $Q_{j,op}(\boldd, t_i)$, $Q_{j,wp}(\boldd, t_i)$, and $Q_{j,wi}(\boldd, t_i)$ are the Field Oil Production Total (FOPT), Field Water Production Total (FWPT), and Field Water Injection Total (FWIT) volumes in time interval $\Delta t_i$ under controls $\boldd$ respectively. 
\begin{align} \label{eq:Well-control-optimisation-challenge-NPV-R(d_t_i)}
	R_j(\boldd, t_i) &{} = Q_{j,op}(\boldd, t_i) \cdot r_{op} - Q_{j,wp}(\boldd, t_i) \cdot r_{wp} - Q_{j,wi}(\boldd, t_i) \cdot r_{wi}
\end{align}
The analogous quantities for an individual well are labelled as WOPT, WWPT, and WWIT respectively. TNO stipulate fixed oil revenue $r_{op} = 45$ \$ per bbl (where bbl are the units for a standard barrel of oil, approximately 159L), water production cost $r_{wp} = 6$ \$ per bbl, and water injection cost $r_{wi} = 2$ \$ per bbl.

For demonstrative purposes we focus on the control of a subset of the wells enclosed between two partial fault boundaries and in close proximity consisting of two producer wells: 2 \& 10, and two injector wells 2 \& 3, with eight control intervals starting on January 1, 2016, 2018, 2020, 2022, 2024, 2026, 2028 \& 2032; thus a total of $D = 32$ decision parameters. Throughout we represent specific individual decision parameters by $\djkti$, where $j \in \{ P, I \}$ refers to the well type ($P$ producer, $I$ injector), $k$ is the well number, and $t_i$ is the control interval start date. Note that each control interval consists of multiple 3 month discounting periods. Collectively these wells are referred to as the Controlled Wells Group (CWG) which provides a sub-problem of interacting wells on which to illustrate the presented methodology. All remaining wells within OLYMPUS use the fixed controls specified in the TNO reference strategy \cite{2018:ISAPP:Website}. The expected NPV objective function is computed from contributions of wells in the CWG only.

We believe that the \TNOChallengeshort{} setup does not faithfully represent the real world field development under uncertainty problem where ensembles of computer models are used to aid decision makers. Instead, there is an emphasis on developing efficient ensemble optimisation algorithms to identify a single optimal strategy. A full critique and discussion of these limitations is presented in \cite[Sec.~3.1]{2022:Owen:PhD-Thesis} and \cite{2020:Owen:A-Bayesian-statistical-approach-to-decision-support-for-TNO-OLYMPUS-well-control-optimisation-under-uncertainty}. We therefore re-formulate well control optimisation as a decision support problem. The aim of this paper is to develop accurate emulators for the expected NPV objective function by exploiting known simulator behaviour in order to efficiently perform decision support.

%
%

%
%
%

\subsection{OLYMPUS Exploratory Analysis} \label{subsec:OLYMPUS-Exploratory-Analysis}

\begin{figure}[!h]
	\centering
	\begin{subfigure}[t]{0.4955\linewidth}
		\includegraphics[page=1,width=\linewidth]{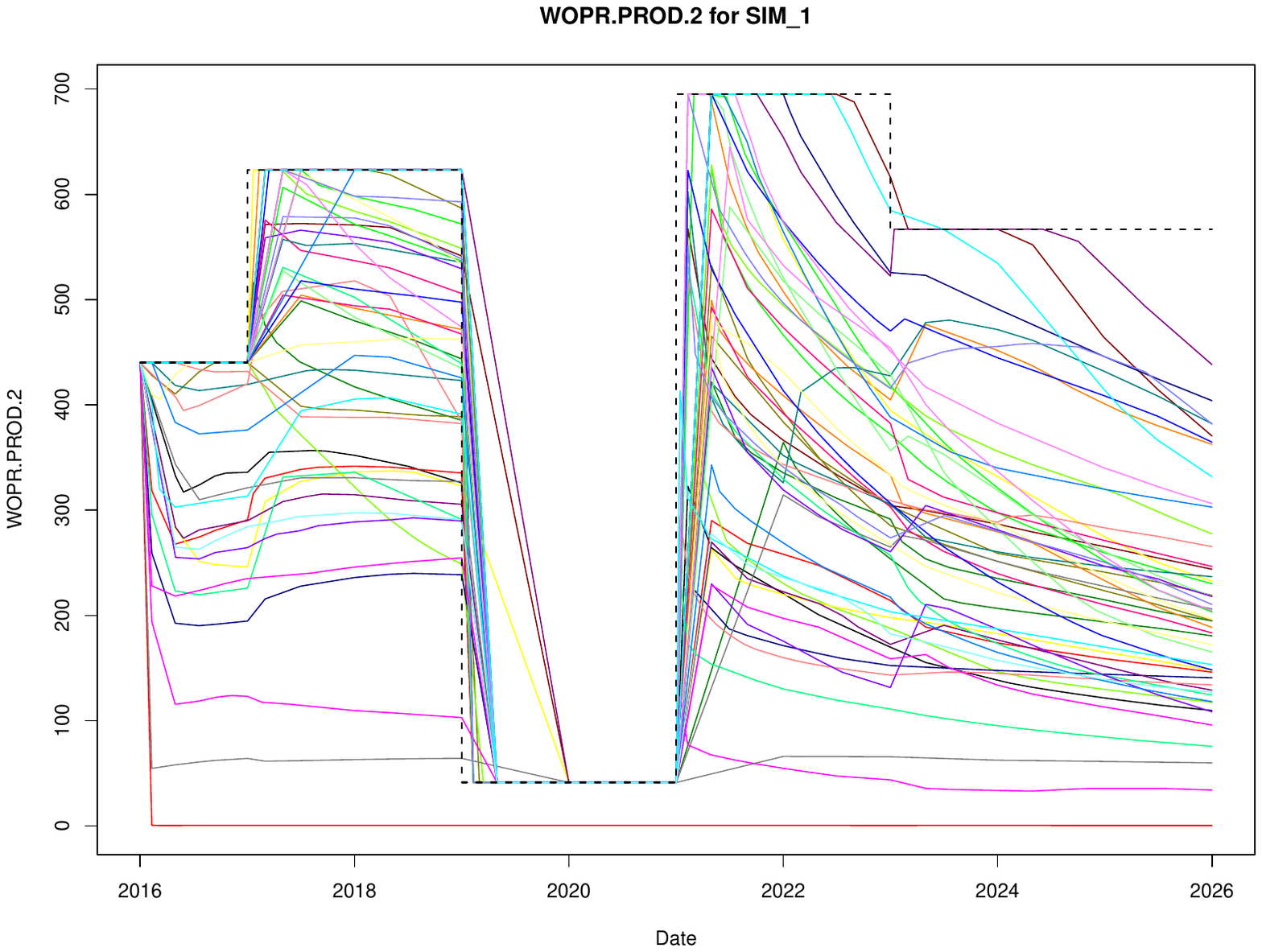}
		\caption{WOPR.PROD.2.}
		\label{subfig:OLYMPUS-Exploratory-Analysis-WOPRPROD2-vs-time-all-decisions-SIM-1}
	\end{subfigure}
	\hfill
	\begin{subfigure}[t]{0.4955\linewidth}
		\includegraphics[page=2,width=\linewidth]{Results/Graphics/Exploratory-Analysis/Decision-Output-vs-Time/Decision-Output-vs-Time-SIM_1}
		\caption{WOPR.PROD.10.}
		\label{subfig:OLYMPUS-Exploratory-Analysis-WOPRPROD10-vs-time-all-decisions-SIM-1}
	\end{subfigure}
	
	\begin{subfigure}[t]{0.4955\linewidth}
		\includegraphics[page=3,width=\linewidth]{Results/Graphics/Exploratory-Analysis/Decision-Output-vs-Time/Decision-Output-vs-Time-SIM_1}
		\caption{WWIR.INJ.2.}
		\label{subfig:OLYMPUS-Exploratory-Analysis-WWIRINJ2-vs-time-all-decisions-SIM-1}
	\end{subfigure}
	\hfill
	\begin{subfigure}[t]{0.4955\linewidth}
		\includegraphics[page=4,width=\linewidth]{Results/Graphics/Exploratory-Analysis/Decision-Output-vs-Time/Decision-Output-vs-Time-SIM_1}
		\caption{WWIR.INJ.3.}
		\label{subfig:OLYMPUS-Exploratory-Analysis-WWIRINJ3-vs-time-all-decisions-SIM-1}
	\end{subfigure}
	\caption{Comparison of the target and achieved control rate decision parameters for an OLYMPUS wave 0 exploratory simulation over the full multi-model ensemble. Plots show the OLYMPUS output actual control rates time series as coloured lines for each ensemble member for the Well Oil Production Rates (WOPR, top) or Well Water Injection Rates (WWIR, bottom) for the four wells within the CWG. The black dashed lines show the corresponding decision parameters of target production or injection rates respectively. The differences between the output traces and the inputs highlights the deviations between the target control and what is achieved due to physical constraints applied in the model.}
	\label{fig:OLYMPUS-Exploratory-Analysis-Decision-Output-Rates-vs-time-all-decisions-SIM-1}
\end{figure}

We first construct a maximin Latin hypercube design and run a wave 0 of $n = 20$ exploratory simulations using all $N = 50$ OLYMPUS models. An important feature is the adherence of OLYMPUS simulations to target production and injection rate decision parameters. For one vector of decision parameter settings \cref{fig:OLYMPUS-Exploratory-Analysis-Decision-Output-Rates-vs-time-all-decisions-SIM-1} compares the input target control rates (black dashed lines) with the corresponding outputted achieved rates over the 50 ensemble members (coloured traces). It is immediately evident in all plots that the input targets are not strictly adhered to for the full duration of the control intervals; a consequence of the underlying physics programmed into the OLYMPUS model, including constraints on BHP, resulting in such deviations between the actual and targeted control values. It is this behaviour which motivates our structured emulation approach described and demonstrated in \cref{sec:Structured-Emulators-Exploiting-Known-Simulator-Behaviour} .
Another interesting facet with potential ramifications for emulation and decision support is the vastly different relative absolute contributions of oil and water to the NPV objective function. An assessment is shown in \cref{fig:OLYMPUS-25-Exploratory-Analysis-NPV-Oil-Water-Contribution-all-decisions}, along with further discussion in \cref{subsec:Extended-Results-OLYMPUS-Exploratory-Analysis}.

%
%
%

\section{Bayesian Emulation} \label{sec:Bayesian-Emulation}

\subsection{Methodology} \label{subsec:Bayesian-Emulation-Methodology}

An emulator is a stochastic belief specification for a deterministic or stochastic function that provides a fast and efficient statistical approximation, yielding predictions for as yet unevaluated parameter settings, along with a corresponding statement of the associated uncertainty \cite{2001:Craig:Bayesian-Forecasting-Using-Computer-Simulators,2010:Vernon:Bayesian-Uncertainty-Analysis-Galaxy-Formation,1997:Craig:Pressure-matching-Bayes-linear-strategies-large-computer-experiments}. They are frequently employed for computationally expensive simulators across a range of scientific and industrial applications to perform tasks including: calibration \cite{2001:Kennedy:Bayesian-calibration-of-computer-models,2008:Higdon:Computer-Model-Calibration-Using-High-Dimensional-Output}; history matching \cite{1996:Craig:Bayes-linear-strategies-matching-hydrocarbon-reservoir-history,1997:Craig:Pressure-matching-Bayes-linear-strategies-large-computer-experiments,2010:Vernon:Bayesian-Uncertainty-Analysis-Galaxy-Formation,2019:Edwards:Revisiting-Antarctic-ice-loss-due-to-marine-ice-cliff-instability}; uncertainty quantifications \cite{1998:O'Hagan:Uncertainty-Analysis-and-other-Inference-Tools-for-Complex-Computer-Codes}; sensitivity analyses \cite{2004:Oakley-and-O'Hagan:Probabilistic-sensitivity-analysis-of-complex-models-a-Bayesian-approach,2023:Kennedy:Multilevel-emulation-for-stochastic-computer-models-with-application-to-large-offshore-wind-farms}; and decision support \cite{2009:Oakley:Decision-Theoretic-Sensitivity-Analysis-for-Complex-Computer-Models,2012:Williamson:Fast-linked-analyses-for-scenario-based-hierarchies}.

For a computer model $\fd$ the $i$\textsuperscript{th} univariate output is denoted by the function $\fid$, where $\boldd \in \Omega \subset \mathcal{R}^D$ is a vector of (decision) parameters in space $\Omega$. We employ Bayesian emulators of the general form in \cref{eq:Bayes-linear-emulator-structure-in-decision-parameters-univariate-output} \cite{1997:Craig:Pressure-matching-Bayes-linear-strategies-large-computer-experiments,2010:Vernon:Bayesian-Uncertainty-Analysis-Galaxy-Formation,2018:Vernon:Bayesian-uncertainty-analysis-systems-biology}:
\begin{align} \label{eq:Bayes-linear-emulator-structure-in-decision-parameters-univariate-output} 
	\fid &{} = \boldg_i(\boldd_{A_i})^\Transpose \boldbeta_i + u_i(\boldd_{A_i}) + w_i(\boldd) \nonumber \\
	&{} = \sum_{j = 1}^{p} \beta_{ij}g_{ij}(\boldd_{A_i}) + u_i(\boldd_{A_i}) + w_i(\boldd)
\end{align}
The subscript $A_i$ denotes a subset of active inputs which are the parameters deemed to be most influential for $\fid$, where $| A_i | = \Dprime \leq D$. Within the emulator the first term models the global function behaviour of $\fid$ where the $g_{ij}(\cdot)$ are deterministic functions of the active inputs with unknown scalar regression coefficients, $\beta_{ij}$ for $j = 1, \ldots, p$, where $p \in \mathbb{N}$. Collectively, these are denoted by the vector function $\boldg_i^\Transpose(\cdot) = \begin{pmatrix} g_{i1}(\cdot) & \cdots & g_{ip}(\cdot) \end{pmatrix}$, and the vector $\boldbeta_i^\Transpose = \begin{pmatrix} \beta_{i1} & \cdots & \beta_{ip} \end{pmatrix} \in \mathbb{R}^p$ respectively. The second term, $u_i(\cdot)$, models the local behaviour of $\fid$ and is a weakly stationary stochastic process with zero mean and a pre-specified covariance structure. A common choice is the squared exponential covariance function in \cref{eq:Squared-exponential-covariance-structure-long} \cite{2010:Vernon:Bayesian-Uncertainty-Analysis-Galaxy-Formation,2006:Rasmussen-Williams:Gaussian-Processes}, where $\sigma_{u_i}^2$ is a variance hyperparameter, and $\boldtheta_i = (\theta_{i1}, \ldots, \theta_{i\Dprime})$ is a $\Dprime$-vector of (distinct) correlation lengths.
\begin{align}
	\Cov[ u_i(\boldd_{A_i}), u_i(\bolddprime_{A_i}) ] = \sigma_{u_i}^2 \exp\left \{ - \sum_{k = 1}^{\Dprime} \left( \dfrac{d_{A_i, k} - d_{A_i, k}^\prime}{\theta_{ik}} \right)^2 \right \} \label{eq:Squared-exponential-covariance-structure-long}
\end{align}
The third term in \cref{eq:Bayes-linear-emulator-structure-in-decision-parameters-univariate-output}, $w_i(\boldx)$, is an uncorrelated, zero-mean nugget term with covariance:
\begin{align} \label{eq:Emulator-nugget-term-covariance-structure}
	\Cov[ w_i(\boldd), w_i(\bolddprime) ] = \sigma_{w_i}^2 \indicatorequalboldd
\end{align}
This is a white noise process which is included to account for the inactive variables \cite{2010:Cumming:Multiscale-Bayes-Linear-Uncertainty-Analysis-Oil-Reservoirs,2010:Vernon:Bayesian-Uncertainty-Analysis-Galaxy-Formation} and ensure numerical stability \cite{2001:Kennedy:Bayesian-calibration-of-computer-models}. Further arguments for the inclusion of a nugget term are presented in \cite{2012:Andrianakis-and-Challenor:The-effect-of-the-nugget-on-Gaussian-process-emulators-of-computer-models,2012:Gramacy-and-Lee:Cases-for-the-nugget-in-modeling-computer-experiments}. 

We follow a Bayes linear paradigm following the foundations of De Finetti \cite{1974:DeFinetti:Theory-of-Probability,1975:DeFinetti:Theory-of-Probability} using expectation as a primitive within a second-order belief specification. Moreover, we subscribe to subjective Bayesianism to provide a coherent framework to structure and combine expert prior beliefs with observed data to achieve posterior inferences \cite{2006:Goldstein:Subjective-Bayesian-Analysis-Principles-and-Practice}. Bayes linear methods have numerous advantages including: quick and simple elicitation of subjective prior beliefs; computational tractability; and robust inferences by removing the specification of full prior probability distributions, with Bayes linear emulators having been successfully implemented across numerous applications \cite{1997:Craig:Pressure-matching-Bayes-linear-strategies-large-computer-experiments,2010:Vernon:Bayesian-Uncertainty-Analysis-Galaxy-Formation,2018:Vernon:Bayesian-uncertainty-analysis-systems-biology,2006:Goldstein:Bayes-Linear-Calibrated-Prediction-for-Complex-Systems,2022:Vernon:Bayesian-Emulation-and-History-Matching-of-JUNE}. An in-depth discussion of Bayes linear statistics can be found in \cite{2007:Goldstein-and-Wooff:Bayes-Linear-Statistics-Theory-and-Methods}, with shorter summaries presented in \cite{2006:Goldstein:Bayes-Linear-Analysis,2011:Goldstein:External-Bayesian-analysis-for-computer-simulators}. Given a design $\mathcal{D} = \{ \boldd^{(1)}, \ldots, \boldd^{(n)} \}$ and computer model evaluations for output $i$, $\boldFi = \{ f_i(\boldd^{(1)}), \ldots, f_i(\boldd^{(n)}) \}$, the Bayes linear adjusted expectation, variance, and covariance are:
\begin{align}
	\EFifd &{}= \E[\fid] + \Cov[\fid, \boldFi] \Var[\boldFi]^{-1} (\boldFi - \E[\boldFi]) \label{eq:Bayes-linear-emulator-adjusted-expectation} \\
	\VarFifd &{}= \Var[\fid] - \Cov[\fid, \boldFi] \Var[\boldFi]^{-1} \Cov[\boldFi, \fid] \label{eq:Bayes-linear-emulator-adjusted-variance} \\
	\CovFifdfdprime &{}= \Cov[\fid, \fidprime] - \Cov[\fid, \boldFi] \Var[\boldFi]^{-1} \Cov[\boldFi, \fidprime] \label{eq:Bayes-linear-emulator-adjusted-covariance}
\end{align}
Full derivation of the Bayes linear emulator adjustment formulae is presented in \cite[Sec.~2.4.5]{2022:Owen:PhD-Thesis}. An alternative full Bayesian approach is Gaussian Process (GP) emulators, as discussed in \cite{2001:Kennedy:Bayesian-calibration-of-computer-models}. Under the Bayes linear formulation we opt for a hyperparameter plug-in approach where they are specified a priori, utilising expert elicitation, before validating using emulator diagnostic techniques \cite{2009:Bastos:Diagnostics-for-Gaussian-Process-Emulators}, as performed in \cite{1996:Craig:Bayes-linear-strategies-matching-hydrocarbon-reservoir-history,2010:Vernon:Bayesian-Uncertainty-Analysis-Galaxy-Formation,2018:Vernon:Bayesian-uncertainty-analysis-systems-biology}.

\subsection{Bayes Linear Emulation of the Expected NPV} \label{subsec:Bayes-Linear-Emulation-of-the-Expected-NPV}

Bayes linear emulation is directly applied to the expected NPV to explore the 32-dimensional wave 1 decision parameter space utilising the above design and linear model predictions for the ensemble mean NPV for training and validation. This serves as a comparison with our proposed hierarchical emulation approach in \cref{subsec:Emulator-Comparison}. Following the methodology summarised in \cref{sec:Bayesian-Emulation}, an emulator with a nugget term (see \cref{eq:Bayes-linear-emulator-structure-in-decision-parameters-univariate-output}) is employed with $\fd = \Ud = \ENPVd$. Investigations using linear modelling, stepwise selection with the AIC criterion, and with all parameters transformed onto $[-1, 1]$, as in \cite{2010:Vernon:Bayesian-Uncertainty-Analysis-Galaxy-Formation}, yields a subset of 12 active decision parameters, $\boldd_{A_\mathBL}$ (this includes all 8 target production rates for producer well 2, and target production rates for producer well 10 for the control intervals starting 1\textsuperscript{st} January 2016, 2018, 2020, and 2026), and a suggested second-order polynomial mean function form:
\begin{align} \label{eq:Expected-NPV-Bayes-linear-emulator-active-variable-selection-linear-model}
	\ENPV(\boldd_{A_\mathBL}) = \beta_{0} + \sum_{d_i \in A_\mathBL} \{ \beta_{i, 1} d_i + \beta_{i, 2} d_i^2 \} + \varepsilon
\end{align}
The residual uncertainty is captured through $\varepsilon$ which possesses an estimate residual standard error $\sigma_{lm}$. The unknown regression coefficients are assumed to have prior expectation $\mubeta = \boldzero$ and an infinite prior uncertainty, with emulator updates exploiting limiting results as $\Varbetalimit$ for which formulae are presented in \cite[Sec.~2.4.5]{2020:Owen:A-Bayesian-statistical-approach-to-decision-support-for-TNO-OLYMPUS-well-control-optimisation-under-uncertainty}.

For the residual process it is assumed that $\E[u(\boldd_{A_\mathBL})] = 0$ and $\E[w(\boldd)] = 0$ with a squared exponential covariance structure (\cref{eq:Squared-exponential-covariance-structure-long}) using a single common correlation length hyperparameter. Following the substitution approach for the hyperparameters: \lword{$\sigma_{u}^2 = (1 - \rho) \sigma_{lm}^2$} and $\sigma_{w}^2 = \rho \sigma_{lm}^2$ where $\rho = 0.05$; whilst the correlation length parameter is set to half of the parameter range, hence $\theta = 1$. These choices are validated via emulator diagnostics discussed below. Bayes linear emulator adjustment is performed using \cref{eq:Bayes-linear-emulator-adjusted-expectation,eq:Bayes-linear-emulator-adjusted-variance,eq:Bayes-linear-emulator-adjusted-covariance}.

\begin{figure}[!t]
	\centering
	\includegraphics[page=1,width=\linewidth]{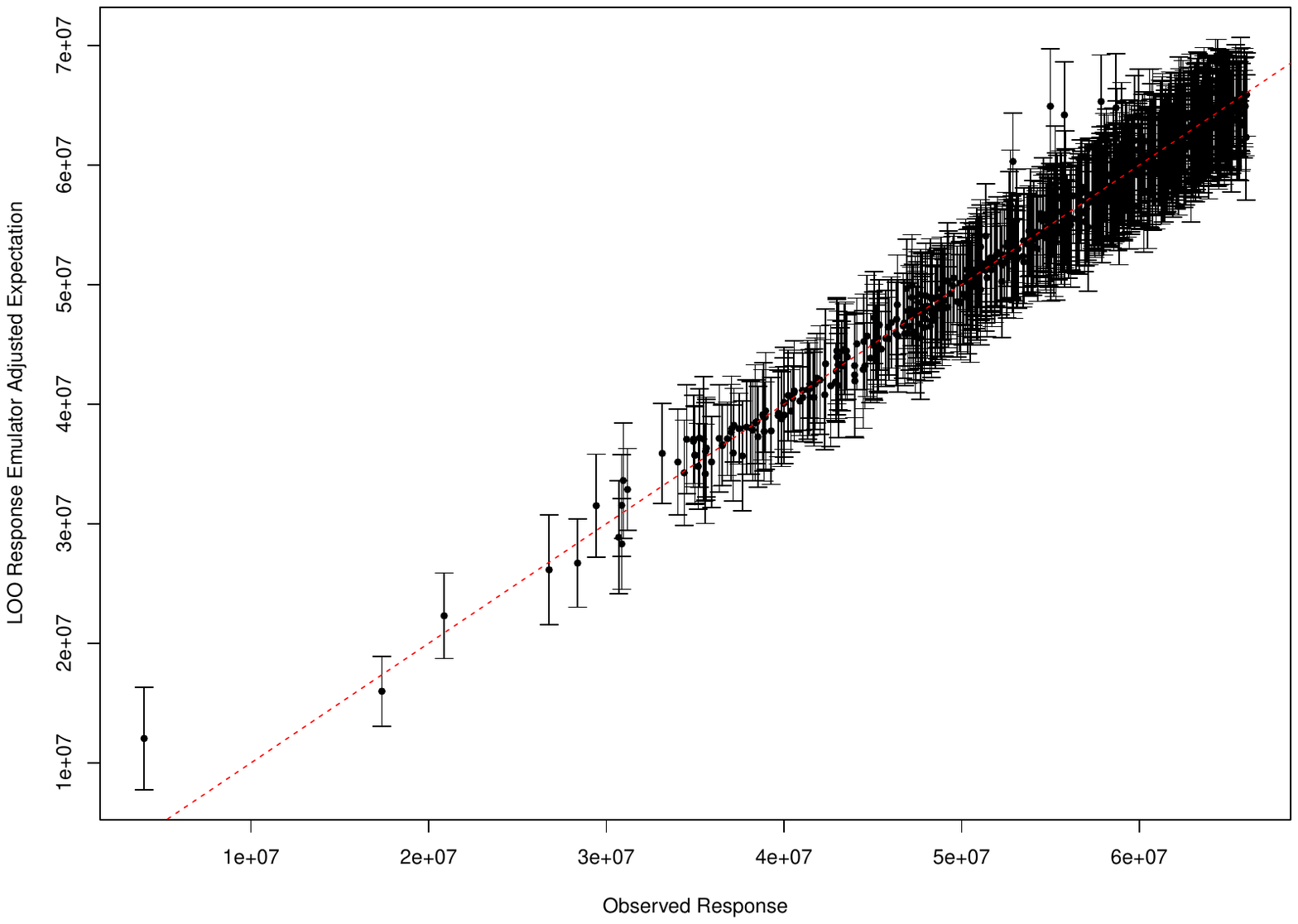}
	\caption{Bayes linear emulator for the expected NPV \loo{} diagnostics plot showing the emulator adjusted expectations with 95\% 3 adjusted standard deviation credible interval error bars versus the simulated expected NPV. The red dashed line denotes equality of the emulator prediction and the simulator output.}
	\label{fig:OLYMPUS-w1-BL-emulator-theta-0_5-LOO-diagnostics-plot-adj-exp-3-adj-sd-CI-vs-sim-NPV}
\end{figure}

\Loo{} diagnostics suggest that the emulator fits well across the decision space, as shown in \cref{fig:OLYMPUS-w1-BL-emulator-theta-0_5-LOO-diagnostics-plot-adj-exp-3-adj-sd-CI-vs-sim-NPV} of the adjusted expectation with an approximate 95\% credible interval of width 3 adjusted standard deviations (following Pukelsheim's 3-sigma rule \cite{1994:Pukelsheim:the-3-sigma-rule}) versus the expected NPV. 691 of the 702 (98.4\%) credible intervals contain the simulated expected NPV, as highlighted by the red dashed line representing equality. Moreover, if we instead employed a Gaussian process emulator, then the 95\% credible intervals contain 679 of the simulated expected NPVs; a 96.7\% coverage. It is noted that the few cases where these diagnostics are not satisfied tend to yield over-prediction. With a view to decision support this is less of a concern as these regions will not be incorrectly ruled out due to low expected NPVs, while iterative refinement enables more accurate emulation at later waves.

\section{Subsampling from Multi-Model Ensembles} \label{sec:Subsampling-from-Multi-Model-Ensembles}

Multi-model ensembles are frequently employed to characterise various forms of uncertainty. For example, multi-model ensembles are particularly prevalent in climate science such as in \cite{2007:Tebaldi:The-use-of-the-multi-model-ensemble-in-probabilistic-climate-projections,2021:Edwards:Projected-land-ice-contributions-to-twenty-first-century-sea-level-rise,2023:Harris:Multimodel-ensemble-analysis-with-neural-network-Gaussian-processes} where they are used to represent uncertainty pertaining to differing choices of aspects such as model structure, encoding of physical laws, discretisation scheme, and numerical solvers, in simulating parts of the earth system. In the petroleum industry these are prevalent for representing uncertainty induced by the unknown underlying field geology, reflecting the geologist's beliefs. In scenarios where these are sampled stochastically from an underlying geology model it may be appropriate to employ emulation methods for stochastic computer models such as stochastic Kriging \cite{2010:Ankenman:Stochastic-Kriging-for-Simulation-Metamodeling}, quantile Kriging \cite{2002:Rannou:Kriging-the-quantile:-application-to-a-simple-transmission-line-model,2014:Plumlee:Building-Accurate-Emulators-for-Stochastic-Simulations-via-Quantile-Kriging}, and heteroscedastic Gaussian Process (hetGP) emulation \cite{2018:Binois:Practical-Heteroscedastic-Gaussian-Process-Modeling-for-Large-Simulation-Experiments}. For a comprehensive review of stochastic model emulation methodology see \cite{2022:Baker-et-al:Stochastic-Simulators-An-Overview-with-Opportunities} and the references therein. In the \TNOChallengeshort{} the multi-model ensemble consists of 50 versions of the OLYMPUS model realised from an underlying stochastic geology model. However, this model was not released and so no further realisations were possible, hence the setup of our motivating application is unsuited to stochastic emulation approaches. 

Running all 50 models to obtain a representation of geological uncertainty requires a greater number of simulations placing a higher strain on computational resources. In many analyses the multi-model outputs are amalgamated such as through averaging, termed the ensemble mean. This is the case in the \TNOChallengeshort{} where the ensemble mean NPV is the focus. Whilst such quantities are easier to analyse and use, the averaging process reduces the benefits of starting with an ensemble by collapsing the uncertainty onto a single value. It is therefore desirable to establish a small collection of models to use as a surrogate, whilst acknowledging any reduction in information gained from the simulations through an appropriate quantification of the uncertainty, and thus develop a multi-model ensemble subsampling technique. Note that a superior choice encompassing uncertainties in the ensemble construction process would be to use an expected utility function over all possible ensembles. However in the \TNOChallengeshort{} it is stipulated that the expected NPV objective function is approximated by the ensemble mean NPV, hence for the purpose of this analysis we adhere to this choice, noting the critique of this choice in \cite[sec.~3.1]{2022:Owen:PhD-Thesis} and \cite{2020:Owen:A-Bayesian-statistical-approach-to-decision-support-for-TNO-OLYMPUS-well-control-optimisation-under-uncertainty}.

\subsection{Methodology} \label{subsec:Subsampling-from-Multi-Model-Ensembles-Methodology}

The process of identifying a representative subset requires a small exploratory design using all models in the ensemble; a wave 0 design, for example, constructed using a maximin Latin hypercube design \cite{1989:Sacks:Design-and-Analysis-of-Computer-Experiments,2003:Santner:The-Design-and-Analysis-of-Computer-Experiments}, in order to assess how well a given subset of models represents the ensemble mean for certain key outputs. First we propose an initial graphical investigation using plots of ensemble mean outputs versus that obtained using individual models. This provides insight into patterns where a strong linear correlation indicates that an individual model may be a good representative for the ensemble mean. Note that such plots are unable to capture the interaction between multiple models' outputs, thus missing where two or more models are jointly able to characterise the ensemble mean, often to a better extent than any one individual model. For large ensembles, this can be a useful screening process to identify a preliminary ensemble subset for further analysis.

Linear models provide a fast and effective tool for predicting the ensemble mean output, $\fbar(\cdot)$, for example, $\fbard = \meanNPVd$, from individual model outputs, $f^{(i_k)}(\cdot)$, $i_k \in \{ 1, \ldots, N \}$ distinct, as well as for quantifying the induced uncertainty. For an ensemble of size $N$ the aim is to select the best subset of $\Ntilde < N$ (to be determined) models. Since the ensemble mean is a linear combination of the individual models' output, an affine linear transformation of a subset of models is expected to yield good approximations. We propose the linear model in \cref{eq:EGES-ensemble-mean-resp-linear-model-on-individual-model-resp-N_EGES-models-unknown} where $\alpha_\mathEGES$ and $\beta_{k, \mathEGES}$ are unknown regression coefficients to be estimated, and $\varepsilon_{\mathEGES}(\boldd)$ is an uncorrelated error term. Here $\mathrm{ES}$ refers to ``Ensemble Subsampling''.
\begin{align} \label{eq:EGES-ensemble-mean-resp-linear-model-on-individual-model-resp-N_EGES-models-unknown}
	\fbard &{} = \alpha_\mathEGES + \sum_{k = 1}^{\Ntilde} \beta_{k, \mathEGES} f^{(i_k)}(\boldd) + \varepsilon_{\mathEGES}(\boldd)
\end{align}
Depending on the size of $N$, either an exhaustive or stepwise model selection may be performed to identify the most suitable choice of $\Ntilde$ and model subset, for example, using the Akaike Information Criterion (AIC) or Bayesian information Criterion (BIC). Note that at later stages within an analysis it is always possible to modify this choice if increased accuracy is required; a scenario that naturally occurs within iterative procedures such as history matching and decision support. In the context of petroleum reservoir field development optimisation we refer to this as Efficient Geological Ensemble Subsampling (EGES) \cite{2020:Owen:A-Bayesian-statistical-approach-to-decision-support-for-TNO-OLYMPUS-well-control-optimisation-under-uncertainty}. Using a subset of models does result in some information loss compared with simulating from the full ensemble. However, the described statistical approach provides a quantification of the additional induced uncertainty as a consequence of using fewer models. In setups where the primary focus is the ensemble mean output the treatment of ensemble variability has been collapsed to a single number and thus neglected. Moreover, running fewer models enables greater exploration of the parameter space and hence presents opportunities to address other sources of uncertainty such as parametric uncertainty. This technique is related to second-order multi-model ensemble exchangeability in \cite{2013:Rougier:Second-Order-Exchangeability-Analysis-for-Multimodel-Ensembles} where coexchangeability is used to establish a link between: the output of individual models; a common ``representative simulator'', which we interpret as the ensemble mean simulator; the output for the real world system as the true expected NPV with respect to all possible geological configurations; as well as any system observations.

%

\subsection{Subsampling from a Geological Multi-Model Ensemble Results} \label{subsec:Subsampling-from-Geological-Multi-Model-Ensemble-Results}

At the \OLYMPUSWorkshop{} \cite{2018:EAGE-TNO-Workshop-on-OLYMPUS-Field-Development-Optimization} a number of participant teams employed ensemble sub-setting techniques on the 50 Olympus models to aid computational tractability of their chosen optimisation procedure for obtaining a well control strategy. This included: selecting a single representative model in \cite{2018:Onwunalu:Stochastic-Oilfield-Optimization-For-Hedging-Against-Uncertain-Future-Development-Plans}, or based on geological modelling insight, specifically using the net hydrocarbon thickness map for upper and lower layers of the reservoir \cite{2018:Harb:OLYMPUS-Field-Development-Optimization-Challenge-American-University-Of-Beirut}; a risk averse approach to optimisation using the 4 worst performing ensemble members according to the Conditional Value at Risk (CVaR) criterion evaluated for the TNO defined base strategy \cite{2018:Schulze-Riegert:Standardized-Workflow-Design-For-Field-Development-Plan-Optimization-Under-Uncertainty}; and a stochastic rank-based realisation selection process used within a evolutionary strategy optimisation algorithm to produce a different subset at each step of the algorithm \cite{2018:Alrashdi:Well-Control-Field-Development-And-Joint-Optimization-Using-mu+lambda-Evolutionary-Strategy-Algorithm-And-A-Stochastic-Rank}. The described Efficient Geological Ensemble Subsampling technique provides a principled approach to selecting ensemble members which also makes efficient use of available computational resources, captures some of the ensemble variability, and not limited to potentially non-robust choices made using the base strategy. 

A preliminary graphical investigation is performed using plots of the ensemble mean versus individual models for a range of outputs of interest for the $n = 20$ wave 0 exploratory simulations described in \cref{subsec:OLYMPUS-Exploratory-Analysis}. Examples are provided in \cref{fig:OLYMPUS-EGES-ensemble-mean-vs-model-outputs} (see \cref{subsec:Extended-Results-Subsampling-from-Geological-Multi-Model-Ensembles}). It is unnecessary to sub-select models exhibiting close individual model output relations with the ensemble mean. Instead we screen for cases where the relationship is easy to model, for example, with a preference for linear associations with small output variation, identifying an initial set of 9 OLYMPUS models for further exploration via linear modelling. Further discussion is found in \cref{subsec:Extended-Results-Subsampling-from-Geological-Multi-Model-Ensembles}, and in \cite[Sec.~4.2]{2022:Owen:PhD-Thesis}.

In order to capture the interacting effects of the different OLYMPUS models, the subsampling technique utilising the linear model in \cref{eq:EGES-ensemble-mean-resp-linear-model-on-individual-model-resp-N_EGES-models-unknown} is implemented. First it is applied to the proposed OLYMPUS subset before extending to all models via both directions stepwise selection starting from the full model and using AIC as the model selection criterion. It is established that a subset of only $\Ntilde = 3$ models is sufficient for a large number of the investigated outputs, yielding high \AdjRsq{} values shown in \cref{fig:OLYMPUS-EGES-LM-all-resp-and-NPV-adj-R-Sq-3-models} in \cref{subsec:Extended-Results-Subsampling-from-Geological-Multi-Model-Ensembles}. The optimal collection for the ensemble mean NPV is OLYMPUS 25, 33, \& 45. The fitted linear model provides an efficient means of prediction and uncertainty quantification by using only 3 ensemble members yielding substantial computational savings. This is a novel application of such multi-model ensemble subsampling techniques in petroleum reservoir engineering.

For computational reasons only 20 runs were available for this pilot study. Discussions with petroleum reservoir engineers suggested that these 20 runs would be sufficient for the purpose of identifying a representative subset since the complexity of this part of the model was not expected to be too high. It is therefore possible to reliably select 3 of the 50 OLYMPUS models based off this collection of simulations, as well as fit linear models to ensemble mean quantities of interest and use these over unexplored regions of the parameter space. If more simulations over the full ensemble were available then it is possible to use diagnostics to verify the robustness of this subset choice. As highlighted above, other participants in the \TNOChallengeWC{} employed model sub-selection, but via less statistically principled approaches.


%

\section{Targeted Bayesian Design of Simulations} \label{sec:Targeted-Bayesian-Design}

\subsection{Methodology} \label{subsec:Targeted-Bayesian-Design-Methodology}

We propose a targeted Bayesian design algorithm to efficiently sample from the decision parameter space by incorporating prior knowledge of both the parameter range and their time ordered consecutive differences. This is a generalisation of the design approach presented in \cite{2020:Owen:A-Bayesian-statistical-approach-to-decision-support-for-TNO-OLYMPUS-well-control-optimisation-under-uncertainty} and \cite[Sec.~4.3]{2022:Owen:PhD-Thesis} where it is tailored towards the well control optimisation problem. Without loss of generality let $\boldd \in \mathbb{R}^D$ be a time ordered vector of decision parameters, thus they are not mutually independent. A difference constraint stipulates that $\lvert d_i - d_{i - 1} \rvert \leq \Delta_i, \ i = 2, \ldots, D$. The decision parameter space is no longer a hypercube, thus motivating the need for a Bayesian design informed by this prior information.

Our targeted Bayesian design algorithm involves a re-parameterisation and sampling the sum of the parameters and their time consecutive differences. First assume each $d_i \in [0, 1]$. Define $t = \sum_{i = 1}^{D} d_i \in [0, D]$ to be the sum of the parameters and $\pardiffi = \frac{d_i - d_{i - 1}}{\sqrt{2}}$ for $i = 2, \dots, D$, to be the scaled differences where the scaling by $\frac{1}{\sqrt{2}}$ is required due to the rotation of the parameter space in this alternative parametrisation. The new parameters are mutually orthogonal with $t \in [0, D]$ and $\lvert \pardiffi \rvert \leq \Delta_i^\prime = \frac{\Delta_i}{\sqrt{2}}$. This re-parametrisation is represented by the linear transformation $(t, \pardiffvec)^\Transpose = L \boldd$ in \cref{eq:Sampling-Parameter-Sum-Differences-Matrix-Equation} where $L$ is the transformation matrix. Sampling in the re-parametrised space automatically satisfies the difference constraints with a sample for $\boldd$ obtained via $\boldd = L^{-1} (t, \pardiffvec)^\Transpose$. The range constraints, $d_i \in [0, 1]$, must also then be verified.
\begin{align} \label{eq:Sampling-Parameter-Sum-Differences-Matrix-Equation}
	\begin{pmatrix}
			t \\
			\pardiff_2 \\
			\pardiff_3 \\
			\pardiff_4 \\
			\vdots \\
			\pardiff_D
	\end{pmatrix}
	=
	\begin{pmatrix}
			1 & 1 & 1 & 1 & \cdots & 1 & 1\\
			\sfrac{-1}{\sqrt{2}} & \sfrac{1}{\sqrt{2}} & 0 & 0 & \cdots & 0 & 0\\
			0 & \sfrac{-1}{\sqrt{2}} & \sfrac{1}{\sqrt{2}} & 0 & \cdots & 0 & 0\\
			0 & 0 & \sfrac{-1}{\sqrt{2}} & \sfrac{1}{\sqrt{2}} & \cdots & 0 & 0\\
			\vdots & \vdots & \vdots & \vdots & \ddots & \vdots & \vdots \\
			0 & 0 & 0 & 0 & \cdots & \sfrac{-1}{\sqrt{2}} & \sfrac{1}{\sqrt{2}}
		\end{pmatrix}
	\begin{pmatrix}
			d_1 \\
			d_2 \\
			d_3 \\
			d_4 \\
			\vdots \\
			d_D
		\end{pmatrix}
\end{align}
In order to ensure good exploration along the $t$-direction, which is thought to be important by petroleum reservoir engineers for the application, we propose preserving an initial sample of the parameter sums, and then uniformly resample $\pardiffvec$ until both constraint types are satisfied. There is freedom to choose the sampling distribution of $t$ with probability density function, $f_T(t)$, dependent on the analysis. We use a truncated normal distribution with other examples being: uniform; mixture of uniforms; and transformed beta distributions. Orthogonal projection of the samples onto the line $d_1 = d_2 = \cdots = d_D$ will approximately follow the specified distribution.

The process of generating a sample of size $n$ for an independent subgroup of parameters is described in the rejection style \cref{alg:Sampling-of-Parameter-Sum-and-Differences-preserving-Parameter-Sum-Sampling-Distribution-adjust-Difference-Sampling-Range} yielding matrix $B$ in which each column is a sampled vector of decision parameters. If $t$ is close to 0 or $D$ the rejection step can be computationally time consuming with the efficiency improved by sampling the differences conditional on $t$. This algorithm may be applied separately to each independent subgroup of decision parameters for improved efficiency before combing to obtain a design over the full decision parameter space. Design optimisation may be performed using standard design selection criteria, for example, minimax or maximin design \cite{2003:Santner:The-Design-and-Analysis-of-Computer-Experiments}, both for the designs by subgroup and the full design by combining using random permutations. Further discussion of the targeted Bayesian design method is presented in \cite[Sec.~4.3]{2022:Owen:PhD-Thesis}.

\begin{algorithm}[!t]
	\caption{Sampling of parameter sums and differences preserving the initial sum of parameters sample.}
	\label{alg:Sampling-of-Parameter-Sum-and-Differences-preserving-Parameter-Sum-Sampling-Distribution-adjust-Difference-Sampling-Range}    
	\SetAlgoLined    
	\KwResult{Matrix $B$ of columns of sampled parameter vectors}    
	Let $\mathbf{t}$ be a vector of $n$ samples of $t \sim f_T(t)$\;
	Let $B$ be an empty matrix of $D$ rows\;
	Define $dimension(\cdot)$ to be a function to obtain the length of a vector\;
	\While{$dimension(\mathbf{t}) > 0$}{
		\ForEach{t in $\mathbf{t}$}{
			Set $\epsilon_i = \min\{t, D - t, \Delta_i^\prime\}$\;
			Generate $\pardiffi \mid t \sim \mathcal{U}[- \epsilon_i, \epsilon_i]$, for $i = 2, \dots, D$\;
		}
		Row bind $\mathbf{t}, \pardiffvec_2, \dots, \pardiffvec_{D}$ to form matrix $B_{r, \mathrm{prop}}$\;
		Compute $B_{\mathrm{prop}} = L^{-1} B_{r, \mathrm{prop}}$\;
		\ForEach{Column in $B_{\mathrm{prop}}$}{
			\eIf{Range \textit{conditions} of parameter vector are satisfied}{
				Join \textit{Column} to $B$\;
				Remove corresponding $t$ from $\mathbf{t}$\;
			}{
				Discard \textit{Column}\;
			}
		}
	}
\end{algorithm}

\subsection{Targeted Bayesian Design of Simulations Results} \label{subsec:Targeted-Bayesian-Design-Results}

We employ the targeted Bayesian design methodology from \cref{subsec:Targeted-Bayesian-Design-Methodology} to perform decision support through targeted sampling based on prior beliefs of experienced petroleum reservoir engineers regarding the location of optimal decision parameter settings, as well as imposing practical and physical constraints. Firstly, TNO stipulate operational range constraints on the control parameters to be $[ 0 , 900 ]$m\textsuperscript{3}/day and $[ 0, 1600 ]$m\textsuperscript{3}/day for production and injection rates respectively, leading to a 32-dimensional hypercube parameter space. Oil reservoir engineers deem large temporal variation in controls to be unphysical and poor practice, thus suggesting a difference constraint between time consecutive controls. Here we use the notation $\djkti$ for the decision parameters where the index $i$ is replaced by the indices tuple $(jk, t_i)$, where $j \in \{ P, I \}$ refers to the well type ($P$ producer, $I$ injector), $k$ is the well number, and $t_i$ is the control interval start date. The 32 decision inputs naturally split into four independent subgroups by well with difference constraints $\lvert \djkti - \djktiminusone \rvert \leq \Delta_i, \ i = 2, \ldots, \Djk$, where $\Djk = 8$ is the number of control intervals for well $jk$. A conservative choice is that the maximum permitted change over a two year time interval is $\Delta_i = \frac{1}{3}$ of the operational range for the well type. Consequently the decision parameter space is no longer a hypercube with a volume of 3.45\% of the initial hypercube due to the range constraints only.

The targeted Bayesian design algorithm is implemented to generate a $n = 700$ point design. First, for each of the four subgroups of eight decision parameters the normalised parameter sums are sampled from a truncated normal distribution in order to facilitate the exploration of more extreme values of the total sums of the eight normalised parameters than would be the case using a standard uniform or Latin hypercube design. This is perceived to be important based on reservoir engineering insight. Next, the differences are sampled according to the specified value of $\Delta_i$ before imposing the operational range constraints (after transforming the normalised parameters to their physical values). Each parameter subgroup and the overall design are approximately optimised with respect to the minimax design selection criterion by comparing candidate designs to a large $20,000$ point uniform random sample (over the constrained parameter space) \cite{2003:Santner:The-Design-and-Analysis-of-Computer-Experiments}. Moreover, the optimised design is augmented to include two further decision parameter vectors with all parameters set to either their minimum or maximum values since it is of interest to observe the model behaviour at these extremes.

\begin{figure}[!t]
	\centering
	\includegraphics[width=\linewidth]{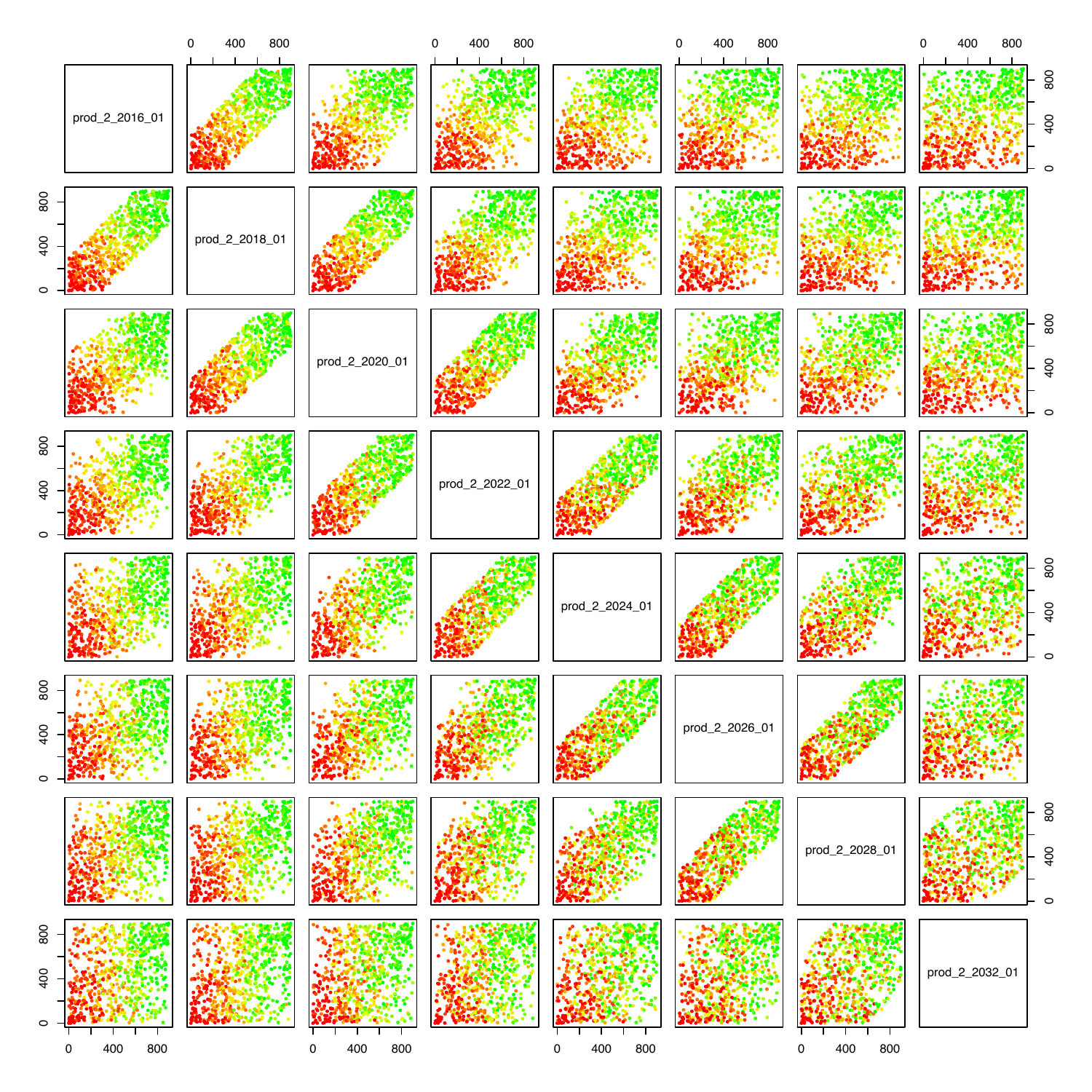}
	\caption{Wave 1 702 point design OLYMPUS producer well 2 decision parameters pairs plots. Points are coloured by the multi-model ensemble subsampling linear model predicted NPV where green, yellow and red correspond to high, moderate and low NPVs respectively.}
	\label{fig:OLYMPUS-w1-sampleSumDiff-pres-par-sum-truncnorm-minimax-design-prod2-Pairs-Plots}
\end{figure}

The final wave 1 design for eight producer well 2 parameters is illustrated in \cref{fig:OLYMPUS-w1-sampleSumDiff-pres-par-sum-truncnorm-minimax-design-prod2-Pairs-Plots}. The plots next to the diagonal highlights the difference constraints as points are clustered between two clearly defined diagonal parallel bounds. Since the final two control intervals are of length 4 years a greater change of up to $\frac{2}{3}$ of the parameter operational range is permitted, hence the wider bands. In addition, the difference constraints affect decision parameters at larger time separations where there are fewer points away from the diagonal, although is less pronounced for greater time gaps. This design is evaluated for the identified subset of 3 OLYMPUS models with the linear model used to predict the ensemble mean NPV for which points are coloured green, yellow and red for high, moderate and low NPVs respectively in \cref{fig:OLYMPUS-w1-sampleSumDiff-pres-par-sum-truncnorm-minimax-design-prod2-Pairs-Plots}.
Note that the presented emulation methodology also works with more traditional space filling designs. Employment of a targeted Bayesian design is to enhance the overall decision support aim, and to incorporate expert knowledge regarding the reservoir behaviour and practical decision implementation.

\section{Divide-and-Conquer Approach} \label{sec:Divide-and-Conquer-Approach}

\subsection{Methodology} \label{subsec:Divide-and-Conquer-Approach-Methodology}

Outputs of interest often consist of linear combinations of computer model outputs, as is the case in the \TNOChallengeWC{} where the focus is the ensemble mean NPV objective function. One approach is to directly emulate this output using methodology such as that described in \cref{subsec:Bayesian-Emulation-Methodology}. However, this ignores the potential gains which can be achieved by decomposing these sums into their constituent simulation outputs and instead emulating each of these before recombining. We term this the ``divide-and-conquer'' approach. In full generality assume an output $\fd$ can be expressed as:
\begin{align} \label{eq:divide-and-conquer-simulator-output}
	\fd = \sum_{i = 1}^{q} a_i \fid
\end{align}
where each $\fid$, $i = 1, \ldots, q$, is a constituent simulation output. The emulator update equations for the expectation and variance are:
\begin{align}
	\EFfd &{}= \sum_{i = 1}^{q} a_i \EFifd \label{eq:divide-and-conquer-emulator-adjusted-expectation} \\
	\VarFfd &{}= \sum_{i = 1}^{q} a_i^2 \VarFifd \label{eq:divide-and-conquer-emulator-adjusted-variance}
\end{align}
where $\boldFi = \{ f_i(\boldd^{(1)}), \ldots, f_i(\boldd^{(n)}) \}$, and $\boldF = \{ \boldFi \}_{i = 1}^q$. Note that \cref{eq:divide-and-conquer-emulator-adjusted-variance} it is assumed that independent emulators are fitted for each $\fid$, although this naturally extends to where multivariate emulators are employed by introducing the relevant covariance terms.

\subsection{Results} \label{subsec:Divide-and-Conquer-Approach-Results}

The divide-and-conquer approach is applied in our analysis of the \TNOChallengeWC{}. First we emulate the NPV for an individual OLYMPUS model, denoted here by $\fd$, and defined in \cref{eq:NPV-general-formula-in-decision-parameters,eq:Well-control-optimisation-challenge-NPV-R(d_t_i)}, and omitting the OLYMPUS model index $j$ for clarity of notation. This is a linear combination of the well contributions with weights determined by the associated cost parameters and the discount factor. A natural decomposition is thus:
\begin{align} \label{eq:Well-control-optimisation-challenge-NPV-R(d_t_i)-by-well-decomposition}
	\fd = \sum_{i = 1}^{8} a_i \left\{ r_{op} \left( \sum_{ k \in \{2, 10\} } f_{Pk, t_i}^{op}(\boldd) \right) - r_{wp} \left( \sum_{ k \in \{2, 3\} } f_{Ik, t_i}^{wp}(\boldd) \right) - r_{wi} \left( \sum_{ k \in \{2, 3\} } f_{Ik, t_i}^{wi}(\boldd) \right) \right\}
\end{align}
where $f_{Pk, t_i}^{op}(\boldd)$, $f_{Ik, t_i}^{wp}(\boldd)$, and $f_{Ik, t_i}^{wi}(\boldd)$ are the Well Oil Production Total (WOPT), Well Water Production Total (WWPT), and Well Water Injection Total (WWIT) within the 8 control intervals ending at time $t_i$ respectively. The index $P$ and $I$ refer to producer and injector wells respectively, and $k$ is the well number over the set of wells used in this analysis. The coefficients $a_i$ are average discounting factors computed as described in \cref{subsec:Emulating-Sums-of-Time-Series-Outputs-Results}. It is these 48 constituents which are to be emulated.

In principle we could employ this approach for all ensemble members which comes at the cost of requiring simulations from all models. Within this application it is unnecessary due to the geological multi-model ensemble subsampling performed in \cref{subsec:Subsampling-from-Geological-Multi-Model-Ensemble-Results}, hence this process is only performed for the NPV for OLYMPUS 25, 33, \& 48. The importance of the divide-and-conquer approach will become evident in \ref{subsec:Structured-Emulators-Exploiting-Known-Simulator-Behaviour-Results} where we exploit known behavioural structure in the constituent outputs to obtain more accurate emulators for the WOPT and WWIT outputs compared with a ``black-box'' approach.


\section{Structured Emulators Exploiting Known Simulator Behaviour} \label{sec:Structured-Emulators-Exploiting-Known-Simulator-Behaviour}

\subsection{Methodology} \label{subsec:Structured-Emulators-Exploiting-Known-Simulator-Behaviour-Methodology}

\begin{figure}[!th]
	\centering
	\includegraphics[page=1,width=\textwidth]{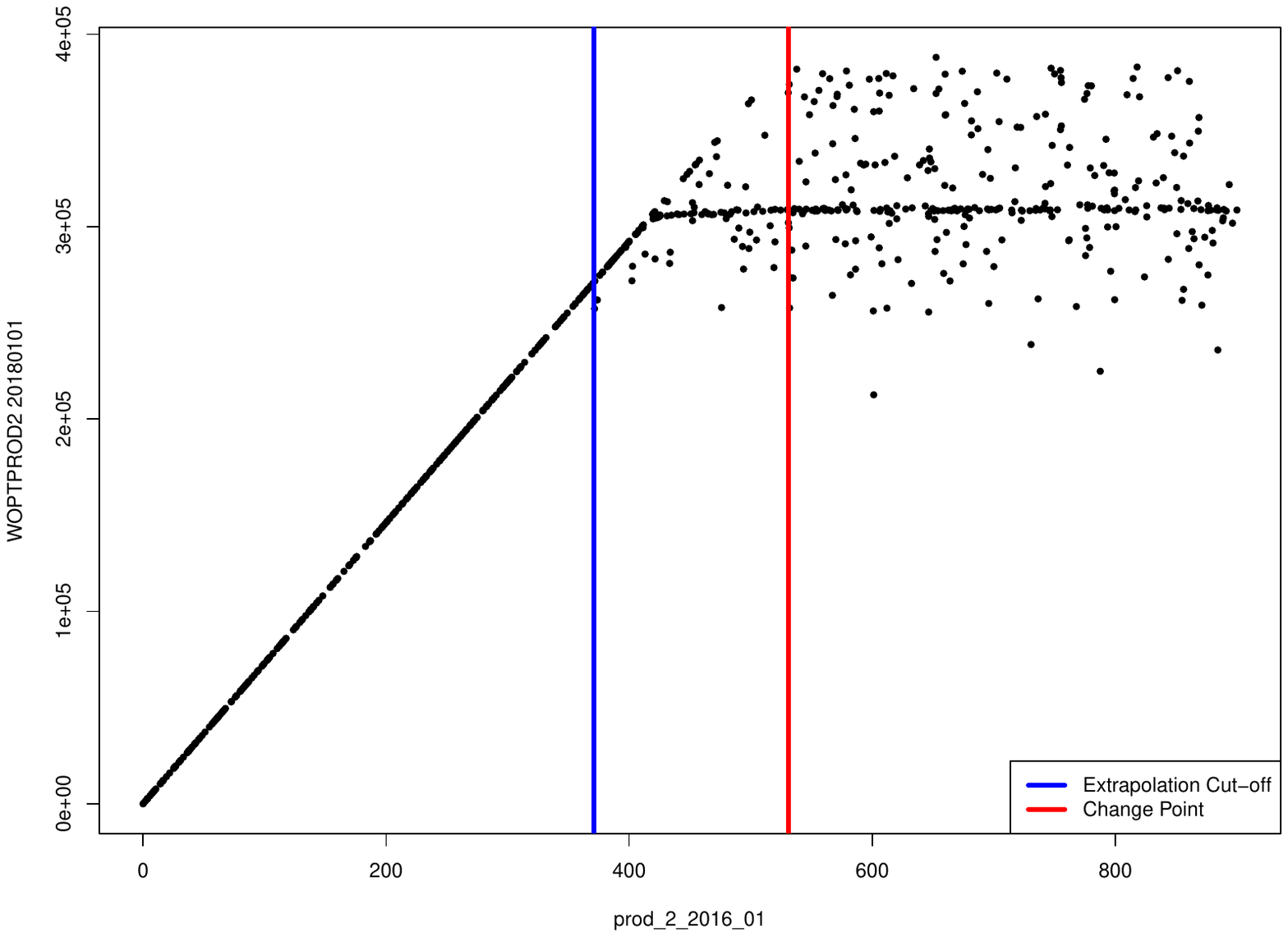}
	\caption[OLYMPUS 25 WOPTPROD2\_20180101 versus prod\_2\_2016\_01]{OLYMPUS 25 WOPT output for producer well 2 during the first two years (ending 01/01/2018) versus the corresponding decision input parameter, the target production rate prod\_2\_2016\_01. For small values of prod\_2\_2016\_01 the target is achieved resulting in the perfect linear behaviour up to a change point beyond which the WOPT plateaus as a maximum threshold on the production rate is achieved. The vertical blue and red lines denote the extrapolation cut-off and change point upper bounds respectively.}
	\label{fig:OLYMPUS-25-WOPTPROD2_20180101_diff-vs-prod_2_2016_01-behaviour}
\end{figure}

The emulation methodological development presented here is motivated by the partially known behavioural form of the WOPT and WWIT outputs within control intervals with respect to their corresponding decision input parameters, the target production or injection rate for the control interval respectively, as shown in \cref{fig:OLYMPUS-25-WOPTPROD2_20180101_diff-vs-prod_2_2016_01-behaviour}. Within \cref{eq:Well-control-optimisation-challenge-NPV-R(d_t_i)} for OLYMPUS model $j$ the WOPT and WWIT are the by well constituents of the respective field totals $Q_{j,op}(\boldd, t_i)$ and $Q_{j,wi}(\boldd, t_i)$.
For small values of the decision parameter the output is known as a linear function of this input up to a small tolerance. Beyond a certain value of this parameter there is a departure from this linear behaviour before reaching a plateau with output fluctuations depending on variation in the other parameter values.

These distinct function behavioural regimes within different regions of the parameter space should be exploited to achieve more accurate emulators than applying the general Gaussian Process (GP) or Bayes linear emulators. Another option is to employ partition based emulation approaches such as treed Gaussian processes \cite{2008:Gramacy-and-Lee:Bayesian-Treed-Gaussian-Process-Models-with-an-Application-to-Computer-Modeling} which split the parameter space parallel and perpendicular to the input axes and fits independent GPs within each region. However whilst treed GPs are a flexible model they do not exploit the physical behaviour that is known to exist in the model. In addition, a further limitation to their accuracy is the number of design points within each partition region which may constitute a small volume of the overall parameter space, particularly important in the often narrow intermediate region for the developed methodology. These considerations have implications during subsequent (decision) analyses utilising the emulator.

Here we present novel methodology which is able to capture both the observed output structure in \cref{fig:OLYMPUS-25-WOPTPROD2_20180101_diff-vs-prod_2_2016_01-behaviour} and an output upper bound which involves splitting the parameter space and output modes of behaviour into three regions: slop, intermediate, and plateau. In this section we denote the computer model output by $\fd$, with decision inputs $\boldd$, and assume without loss of generality that the behaviour manifests with respect to decision parameter $d_1$. In \cref{subsec:Structured-Emulators-Exploiting-Known-Simulator-Behaviour-Results} we link this notation to the \TNOChallengeshort{} application.

\subsubsection{Change Points} \label{subsubsec:Structured-Emulation-Change-Points}


In the slope region $d_1$ determines the output according to a known functional relationship up to a small tolerance $\delta \geq 0$ whilst also imposing an upper bound on $\fd$. Here we focus on a linear map with respect to $d_1$ with known intercept, $\alpha$, and gradient, $\beta$, noting that the methodology also extends to other known relationships. Within the plateau region this known behaviour is not the case with uncertainty induced by $\boldd \setminus d_1$. In principle this leads to these two regions only separated by a change point denoted $\cpgeneral{}$. However, the value of $d_1$ of the transition from slope to plateau is unknown, depends on all other decision parameters, and given only a finite number of simulations is impossible to exactly determine. Consequently, in practical application, there are three distinct regions of behaviour: the slope and plateau separated by an additional uncertain region believed to contain the unknown change point. This is labelled as the intermediate region where there is a mixture of adherence to this known (linear) relationship and model output exhibiting uncertainty around the plateau. Given a design $\mathcal{D}$ and simulator output $\boldF = \{ f(\boldd^{(1)}), \ldots, f(\boldd^{(n)}) \}$, one option it to estimate the mean change point location.

A conservative change point upper bound estimate, $\cpugeneral{}$, is defined in \cref{eq:Change-point-upper-bound}, where $f_{ \mathrm{max}} = \max_{\boldd \in \mathcal{D}} \fd$, and $\delta_{u} \geq 0$ is a tolerance included for numerical stability and to ensure that an upper bound is obtained.
\begin{align} \label{eq:Change-point-upper-bound}
	\cpugeneral{} = \min_{d_1} \{ d_1 \mid \alpha + d_1 \cdot \beta \geq f_{\mathrm{max}} + \delta_{u} \}
\end{align}
This is the smallest value of $d_1$ such that if simulator output achieved the upper bound, $\alpha + d_1 \cdot \beta$, then this exceeds the largest simulated value (plus a tolerance) over $\mathcal{D}$.

An estimate for the change point lower bound is defined in \cref{eq:Change-point-lower-bound}, where $f_{\mathrm{diff}}(\boldd) = \alpha + d_1 \cdot \beta - \fd$ is the difference between the theoretical maximum and the simulated output, %
and $\delta_{l} \geq 0$ is another numerical stability tolerance.
\begin{align*} \label{eq:Change-point-lower-bound}
	\cplgeneral{} &{} = \dfrac{1}{2} \left( \argmin_{d_1 \mid \boldd \in \mathcal{D}} \{ \fd < \alpha + d_1 \cdot \beta - \delta_{l} \} \right. \alignmultilineeq \left. + \argmax_{d_1 \mid \boldd \in \mathcal{D}} \left\{ d_1 < \argmin_{d_1 \mid \boldd \in \mathcal{D}} \{ \fd < \alpha + d_1 \cdot \beta - \delta_{l} \} \right\} \right) \alignnumber
\end{align*}
\Cref{fig:OLYMPUS-25-WOPTPROD2_20180101_diff-vs-prod_2_2016_01-for-cp-lwr-zoom} provides an illustration of the change point lower bound for the output WOPTPROD2\_20180101 with respect to the target rate prod\_2\_2016\_01.

\subsubsection{Extrapolation Cut-Offs} \label{subsubsec:Structured-Emulation-Extrapolation-Cut-Offs}

The two distinct modes of behaviour suggests an emulator be fitted piecewise using a combination of the more accurate knowledge in the slope region, and the less well understood behaviour in the plateau region. Noting the change point location uncertainty, for the plateau region we propose fitting an emulator based only on data which is almost certainly on the plateau using the change point upper bound estimate. For $f(\cdot)$ this is design points with $d_1 \geq \cpugeneral{}$. In order to connect the slope and plateau regions we must extrapolate the plateau emulator. It is necessary to introduce an extrapolation cut-off, denoted $\extrapcogeneral{}$, beyond which the emulator should not be extrapolated. This is due to limited plateau training data issues. For simulator output $\fd$ this is defined with respect to the same decision parameter, $d_1$. The decision space is thus split into three distinct regions:
\begin{enumerate}
	\item \textbf{Slope Region:} where $d_1 < \extrapcogeneral{}$;
	
	\item \textbf{Intermediate Region:} where $\extrapcogeneral{} \leq d_1 < \cpugeneral{}$, for which there is uncertainty as to whether simulator output falls on the slope or in the plateau;
	
	\item \textbf{Plateau Region:} where $d_1 \geq \cpugeneral{}$.
\end{enumerate}
There is an estimation trade-off between overly cautious small values which fails to alleviate the above issue, and large values risking points being wrongly classified as on the slope. A suitable and sufficiently conservative approach is to use the change point lower bound, so $\extrapcogeneral = \cplgeneral{}$ (see \cref{eq:Change-point-lower-bound}).

\subsubsection{Structured Emulation with Upper Truncation} \label{subsubsec:Structured-Emulation-Methodology}

Prior information for $\fd$ stipulates that it cannot exceed a maximum (up to a tolerance) determined by $d_1$ and the coefficients $\alpha$ \& $\beta$. This upper bound is $\alpha + d_1 \cdot \beta$. Alongside the above parameter space dichotomy, this constraint is imposed through an upper truncation with structured emulators respecting partially known behaviour of these simulator outputs. First a preliminary Bayes linear emulator for $\fd$ is fitted using a sub-design, $\mathcal{D}^\prime = \{ \boldd \mid \boldd \in \mathcal{D},\ d_1 > \cpugeneral{} \}$, with corresponding simulator output $\boldF^\prime = \{ \fd \mid \boldd \in \mathcal{D}^\prime \}$. This represents the behaviour in the plateau region. By construction, all parameter settings in $\mathcal{D}^\prime$ do not adhere to the target rate and hence are in the plateau, thus providing reliable information on which to construct this part of the emulator. This preliminary emulator is only evaluated for $\boldd$ satisfying $d_1 \geq \extrapcogeneral{}$ and is compared with the upper bound in a classification step determining the structured emulator form:
\begin{enumerate*}
	\item \textbf{Slope Region:} If $d_1 < \extrapcogeneral{}$ \textbf{or} for the preliminary emulator  \lword{$\E_{\boldF^\prime}[\fd] - 3 \sqrt{\Var_{\boldF^\prime}[\fd]} > \alpha + d_1 \cdot \beta$}, then collapse the emulator such that for the structured emulator $\E_{\boldF^\prime}[\fd] = \alpha + d_1 \cdot \beta$ with fixed maximum absolute errors of size $\delta$.
	
	\item \textbf{Intermediate Region:} If the preliminary emulator satisfies \lword{$\E_{\boldF^\prime}[\fd] - 3 \sqrt{\Var_{\boldF^\prime}[\fd]} \leq \alpha + d_1 \cdot \beta < \E_{\boldF^\prime}[\fd] + 3 \sqrt{\Var_{\boldF^\prime}[\fd]}$}, a truncated Gaussian process (truncated GP) emulator is used with mean and variance determined by \cref{eq:Truncated-normal-distribution-mean,eq:Truncated-normal-distribution-variance} respectively \cite{1994:Johnson:Continuous-Univariate-Distributions-Vol-1}.
	
	\item \textbf{Plateau Region:} In all other cases where $\E_{\boldF^\prime}[\fd] + 3 \sqrt{\Var_{\boldF^\prime}[\fd]} \leq \alpha + d_1 \cdot \beta$, the preliminary emulator output is used.
\end{enumerate*}
Alternative width credible intervals may be used depending on the level of conservativeness desired within an analysis with justification based on the Vysochanskij-Petunin inequality \cite{1980:Vysochanskij:3-sigma-rule-for-unimodal-distributions}.

\subsubsection{Structured Emulation with Two-Sided Truncation} \label{subsubsec:Structured-Emulation-with-Two-Sided-Truncation-Methodology}

An alternative variant of structured emulation uses a two-sided truncation where a lower truncation is also imposed. This may be either a constant $\gamma$, such as to enforce that an output is non-negative via $\gamma = 0$, or a function of the parameters $\gamma(\boldd)$. For clarity of notation we denote this by $\gamma$. Both upper and lower constraints are utilised alongside the preliminary Bayes linear emulator in a modified classification step:
\begin{enumerate*}
	\item \textbf{Slope Region:} If $d_1 < \extrapcogeneral{}$ \textbf{or} for the preliminary emulator \lword{$\E_{\boldF^\prime}[\fd] - 3 \sqrt{\Var_{\boldF^\prime}[\fd]} > \alpha + d_1 \cdot \beta$}, then collapse the emulator such that for the structured emulator $\E_{\boldF^\prime}[\fd] = \alpha + d_1 \cdot \beta$ with fixed maximum absolute errors of size $\delta$.
	
	\item \textbf{Intermediate Region:} As for the upper truncation version, if the preliminary emulator satisfies $\E_{\boldF^\prime}[\fd] - 3 \sqrt{\Var_{\boldF^\prime}[\fd]} \leq \alpha + d_1 \cdot \beta < \E_{\boldF^\prime}[\fd] + 3 \sqrt{\Var_{\boldF^\prime}[\fd]}$, \textbf{or} if the additional criterion of $\E_{\boldF^\prime}[\fd] - 3 \sqrt{\Var_{\boldF^\prime}[\fd]} <  \gamma$ \textbf{whilst} $\E_{\boldF^\prime}[\fd] + 3 \sqrt{\Var_{\boldF^\prime}[\fd]} \leq \alpha + d_1 \cdot \beta$, a truncated GP emulator is evaluated. The mean and variance are determined by \cref{eq:Truncated-normal-distribution-mean,eq:Truncated-normal-distribution-variance} respectively.
	
	\item \textbf{Plateau Region:} In all other cases where $\E_{\boldF^\prime}[\fd] - 3 \sqrt{\Var_{\boldF^\prime}[\fd]} \geq \gamma$ \textbf{and} $\E_{\boldF^\prime}[\fd] + 3 \sqrt{\Var_{\boldF^\prime}[\fd]} \leq \alpha + d_1 \cdot \beta$, use the preliminary emulator.
\end{enumerate*}

The structured emulation methodology utilises a truncated Gaussian process (truncated GP) emulator for which the mean and variance are determined by \cref{eq:Truncated-normal-distribution-mean,eq:Truncated-normal-distribution-variance} respectively \cite{1994:Johnson:Continuous-Univariate-Distributions-Vol-1}, where $\phi(\cdot)$ and $\Phi(\cdot)$ represent the probability density and cumulative distribution functions respectively of a standard normal distribution. These are computed assuming a preliminary Gaussian process emulator with posterior mean and variance, abbreviated to $\mu$ and $\sigma^2$ respectively, equal to the computed adjusted expectation and variance, and truncation bounds $p = \gamma$ and $q = \alpha + d_1 \cdot \beta$, with $\nu = \frac{p - \mu}{\sigma}$, and $\omega = \frac{q - \mu}{\sigma}$. This form of emulation is used in the intermediate uncertain region around the change point's true location.
\begin{align}
	\E_{\boldF^\prime}[\fd \mid p < \fd < q] &{} = \mu + \sigma \dfrac{\phi(\nu) - \phi(\omega)}{\Phi(\omega) - \Phi(\nu)} \label{eq:Truncated-normal-distribution-mean} \\
	\Var_{\boldF^\prime}[\fd \mid p < \fd < q] &{} = \sigma^2 \left[ 1 + \dfrac{\nu \phi(\nu) - \omega \phi(\omega)}{\Phi(\omega) - \Phi(\nu)} - \left( \dfrac{\phi(\nu) - \phi(\omega)}{\Phi(\omega) - \Phi(\nu)} \right)^2 \right] \label{eq:Truncated-normal-distribution-variance}
\end{align}

Compared to using standard GP, Bayes linear, or treed GP emulators, the structured approach yields improved accuracy by encapsulating the known output structure and constraints, along with an increase in speed and efficiency; a consequence of using fewer design points in the fitting. Moreover, both the change point upper bound and extrapolation cut-off estimation processes are computationally very cheap, whilst the use of a truncated GP helps reduce the reliance on accurate estimation of the extrapolation cut-off. Another benefit to this approach is that in principle few points are required within the intermediate region between the slope and plateau since the emulator is only trained on those which exceed the change point upper bound, although they do form a useful emulator diagnostic check. Further commentary on the accuracy and speed of structured versus Bayes linear emulators can be found in the application to NPV constituents exhibiting such constrained behaviour in \cref{subsec:Structured-Emulators-Exploiting-Known-Simulator-Behaviour-Results}, as well as in the comparison of employing Bayes linear emulators versus structured emulators within the combined hierarchical emulation framework for the ensemble mean NPV in \cref{subsec:Emulator-Comparison}.

\subsection{Results} \label{subsec:Structured-Emulators-Exploiting-Known-Simulator-Behaviour-Results}


For each OLYMPUS model the NPV is determined by the oil production, water injection, and water production. Following \cref{subsec:Divide-and-Conquer-Approach-Results} we decompose the NPV calculation into WOPT, WWIT, and WWPT, by both model and control interval, as in \cref{eq:Well-control-optimisation-challenge-NPV-R(d_t_i)-by-well-decomposition}. The WOPT and WWIT over a control interval are observed to follow the structured behaviour where the quantity is equal to the corresponding target rate decision parameter multiplied by the length of the time interval up to an unknown change point beyond which there is a plateau in the behaviour. In addition, the value of this decision parameter also imposes a maximum achievable output over this time interval. This is illustrated for the OLYMPUS 25 WOPT for producer well 2 over the first two year control interval (01/01/2016 to 01/01/2018) in \cref{fig:OLYMPUS-25-WOPTPROD2_20180101_diff-vs-prod_2_2016_01-behaviour} which is used as a running example.

The structured emulation with upper truncation methodology is employed separately for each of the WOPT and WWIT within control interval constituents for wells in the CWG. For outputs $f_{Pk, t_i}^{op}(\boldd)$ and $f_{Ik, t_i}^{wi}(\boldd)$ the corresponding decision parameter is the target production or injection rate for this interval and is denoted $\dcp$, where $j \in \{ P, I \}$, $k$ is the well number, and $t_i$ is the control interval end year, which equates to $d_1$ in \cref{subsec:Structured-Emulators-Exploiting-Known-Simulator-Behaviour-Methodology}. These target control rates should be adhered to for the entire duration of the interval, $\Delta t_i$. However, this is not always possible due to Bottom Hole Pressure (BHP) constraints which results in a departure from the target and the observed different modes of behaviour across the parameter space. The upper truncation is therefore obtained by specifying $\alpha = 0$ and $\beta = \Delta t_i$ throughout \cref{subsec:Structured-Emulators-Exploiting-Known-Simulator-Behaviour-Results}.

\begin{figure}[!t]
	\centering
	\includegraphics[page=1,width=\textwidth]{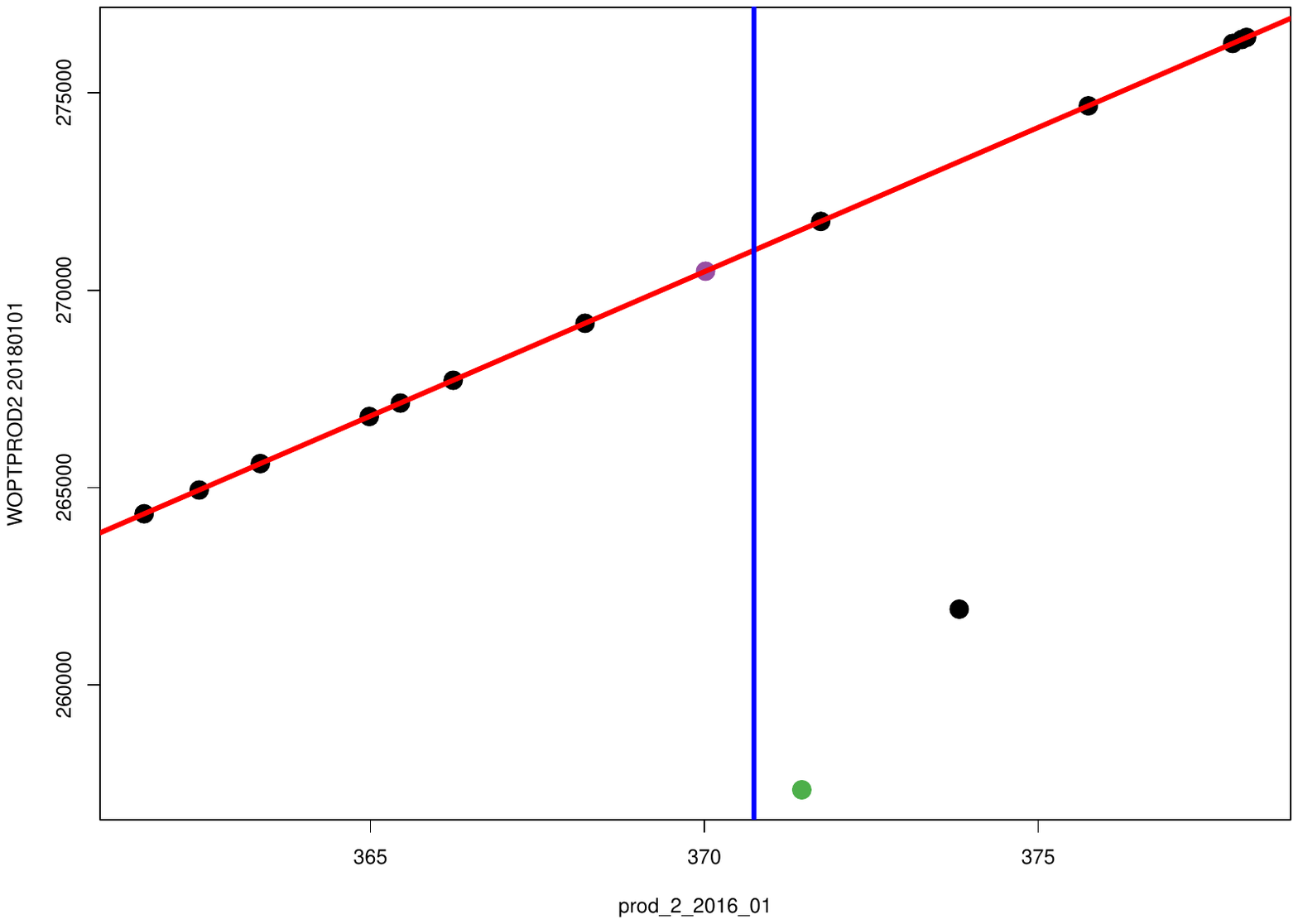}
	\caption[OLYMPUS 25 WOPTPROD2\_20180101 versus prod\_2\_2016\_01 focused around $\cppl{2}{2016}$]{OLYMPUS 25 WOPT for producer well 2 during the first two years (ending 01/01/2018) versus the corresponding target production rate, prod\_2\_2016\_01. The focus is the change point lower bound, $\cppl{2}{2016}$, computed using \cref{eq:Change-point-lower-bound}, employed as the extrapolation cut-off, and denoted by the vertical blue line. The red line depicts the slope upper bound computed as $\text{prod\_2\_2016\_01} \cdot \Delta t_{201801}$ and attained when the target production rate is adhered to for the full control interval. It is shown that $\cppl{2}{2016}$ is the midpoint of the first point not on the slope coloured green, and preceding point to the left of the vertical blue line which is on the slope coloured magenta.}
	\label{fig:OLYMPUS-25-WOPTPROD2_20180101_diff-vs-prod_2_2016_01-for-cp-lwr-zoom}
\end{figure}

Conservative change points upper bounds, $\cpu{jk}{t_i}$, and extrapolation cut-offs, $\extrapco{jk}{t_i}$, are estimated using \cref{eq:Change-point-upper-bound,eq:Change-point-lower-bound} respectively over the wave 1 simulations, with tolerances $\delta_{u} = \delta_{l} = 10$ chosen to ensure numerical stability. These are shown for WOPTPROD2\_20180101 versus the target rate prod\_2\_2016\_01 at $\extrapcop{2}{2016} = \cppl{2}{2016}$ and $\cppu{2}{2016}$ in \cref{fig:OLYMPUS-25-WOPTPROD2_20180101_diff-vs-prod_2_2016_01-behaviour} by the vertical blue and red lines respectively. In addition, the estimation process of the change point lower bound (or extrapolation cut-off) for this output is depicted in \cref{fig:OLYMPUS-25-WOPTPROD2_20180101_diff-vs-prod_2_2016_01-for-cp-lwr-zoom}  where the red line represents the slope and upper bound if the target is adhered to for the entire control interval. The vertical blue line denotes $\cpl{P2}{2016}$ as the midpoint between the first simulation decision parameter setting not on the slope; hence with $f_{\mathrm{diff}}(\boldd) > \delta_{l}$ (green point; first term in \cref{eq:Change-point-lower-bound}), and the decision parameter setting with the largest value of $d_{P2, 201601}$ which is less than this first departure point previously obtained (magenta point; second term in \cref{eq:Change-point-lower-bound}).
Further results are displayed in \Cref{fig:OLYMPUS-w1-WOPT-WWIT-cp-and-extrap-co-intervals} (in \cref{subsec:Extended-Results-Hierarchical-Emulation-of-the-Expected-NPV}) showing the regions in which the ``true'' change points are believed to be situated for all WOPT and WWIT for each of the three wave 1 OLYMPUS models.

For each NPV constituent the next stage is to fit a preliminary Bayes linear emulator where the deterministic functions of the active decision parameters are of the form:
\begin{align} \label{eq:NPV-constituent-preliminary-Bayes-linear-emulator-active-variable-selection-linear-model}
	m(\boldd_{A_{jk, t_i}}) = \boldg(\boldd_{A_{jk, t_i}})^\Transpose \boldbeta = \beta_{0} + \sum_{d_i \in A_{jk, t_i}} \{ \beta_{i, 1} d_i + \beta_{i, 2} d_i^2 \}
\end{align}
It is assumed that the active decision parameters comprise all decisions which take place in the past of the output for reasons of temporal consistency. For our running example this is $A_{jk, t_i} = \{${prod\_2\_2016\_01}, {prod\_10\_2016\_01}, {inj\_2\_2016\_01}, {inj\_3\_2016\_01}$\}$. This is a logical choice since future decisions are physically unable to impact on an output up to the current time, however any past decisions may potentially have an effect. The remainder of each emulator's prior specification is analogous to \cref{subsec:Bayes-Linear-Emulation-of-the-Expected-NPV}, but with the distinction that only those simulation points in $\mathcal{D}^\prime = \{ \boldd \mid \boldd \in \mathcal{D},\ \dcp > \cpu{jk}{t_i} \}$ with output $\boldF^\prime = \{ \fd \mid \boldd \in \mathcal{D}^\prime \}$ are used in the fitting.

\begin{figure}
	\centering
	\begin{subfigure}{\linewidth}
		\centering
		\includegraphics[width=0.75\linewidth]{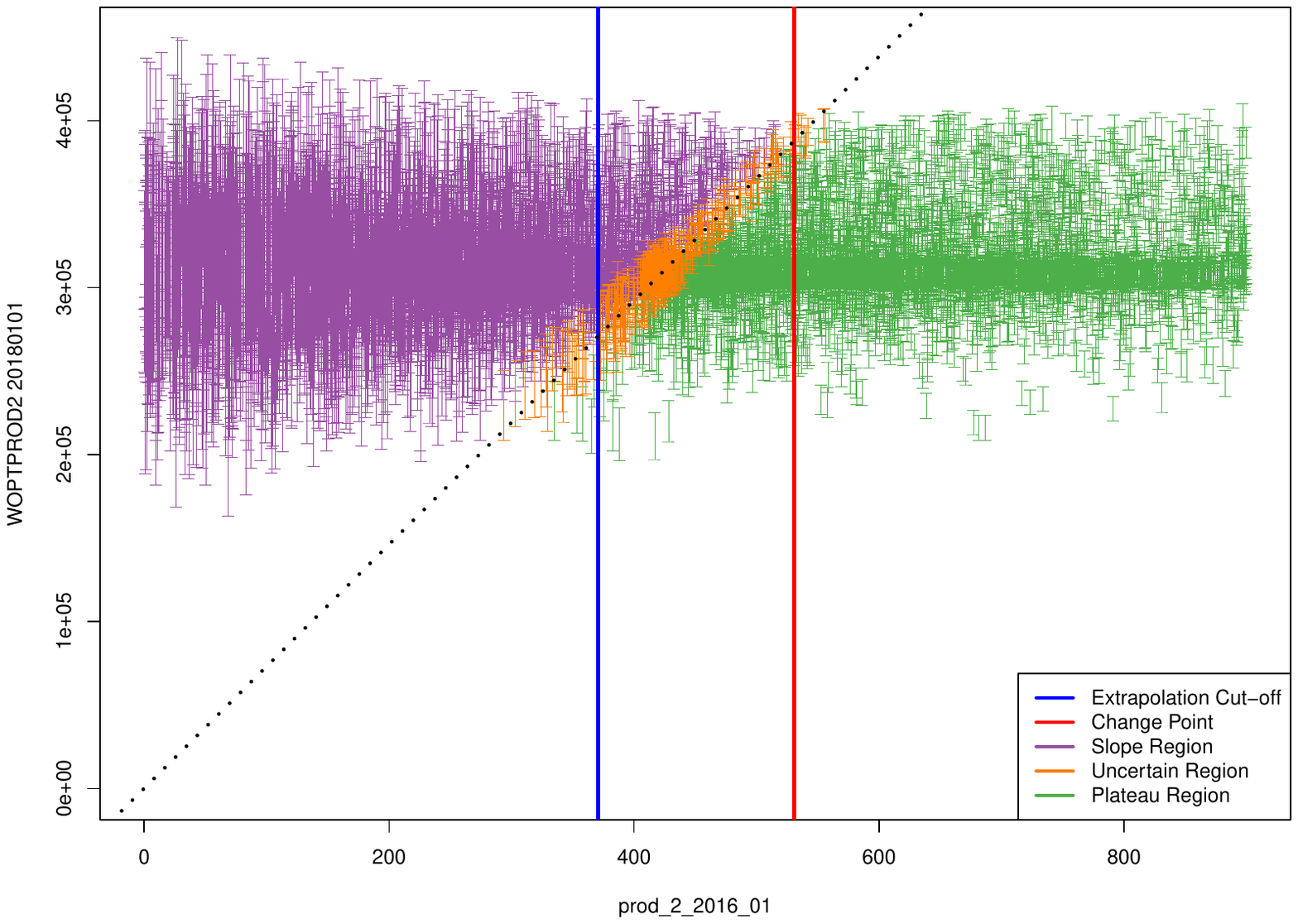}
		\vspace{-10pt}
		\caption{Preliminary Bayes linear emulator predictive CI versus prod\_2\_2016\_01.}
		\label{subfig:OLYMPUS-25-w1-WOPTPROD2_20180101_diff-vs-prod_2_2016_01-prelim-BL-emul-CI}
	\end{subfigure}
	
	\vspace{-18pt}
	
	\begin{subfigure}{\linewidth}
		\centering
		\includegraphics[width=0.75\linewidth]{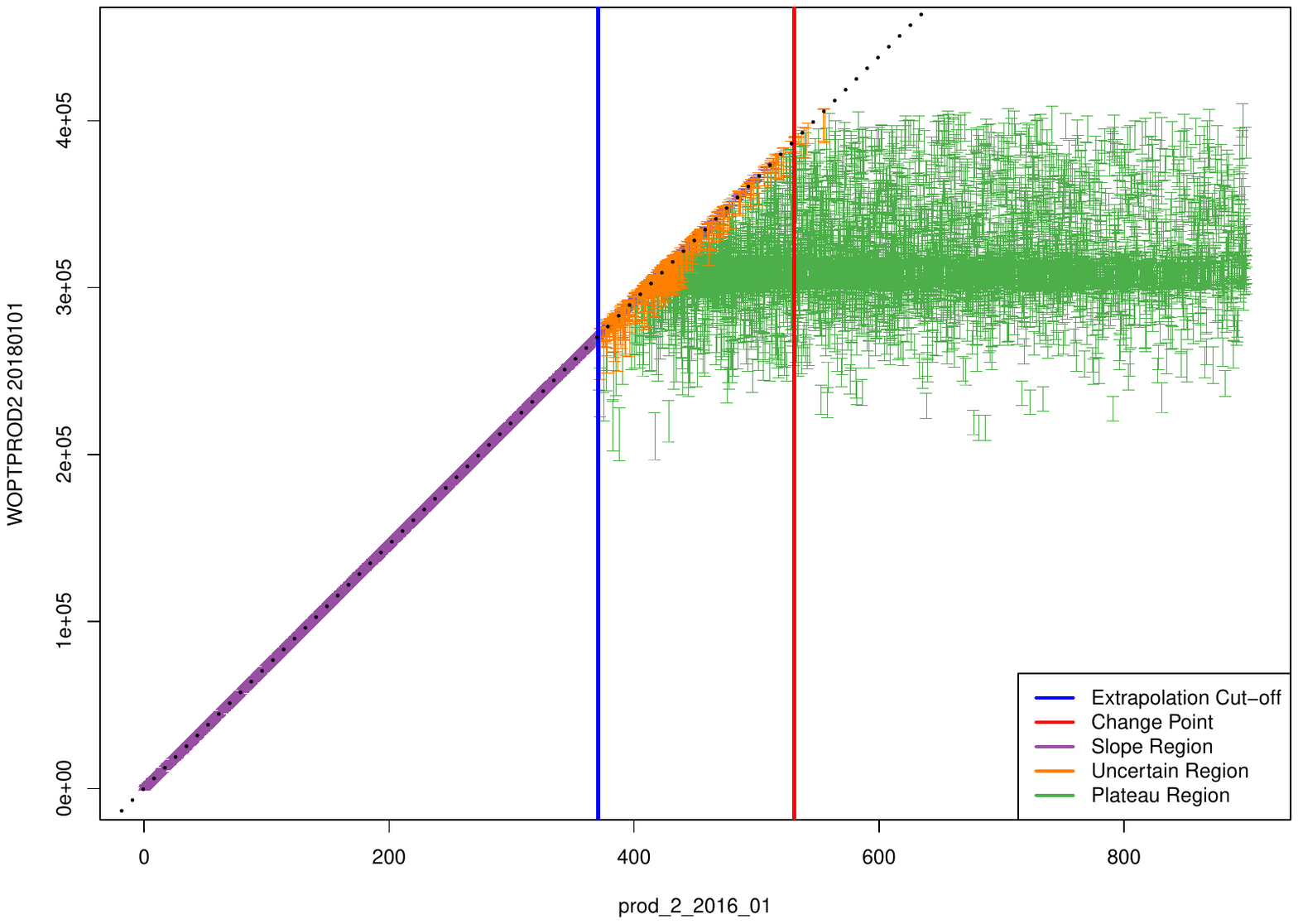}
		\vspace{-10pt}
		\caption{Structured emulator with upper truncation predictive CI versus prod\_2\_2016\_01.}
		\label{subfig:OLYMPUS-25-w1-WOPTPROD2_20180101_diff-vs-prod_2_2016_01-struc-emul-CI}
	\end{subfigure}
	\vspace{-5pt}
	\caption{OLYMPUS 25 wave 1 WOPT for producer well 2 during the first two years (ending 01/01/2018) versus the corresponding target production rate, prod\_2\_2016\_01. The top plot shows the wave 1 preliminary Bayes linear emulator predictive 3-sigma credible interval (CI) fitted using only the simulations with $d_{P2,2016} \geq \cppu{2}{2016}$. This is used within the structured emulation algorithm imposing the upper truncation due to the slope (black dotted line) with the CI shown in the bottom plot. The vertical blue and red lines are situated at $\extrapcop{2}{2016}$ and $\cppu{2}{2016}$ respectively. The purple, orange and green CI correspond to points in the slope, uncertain, and plateau regions respectively.}
	\label{fig:OLYMPUS-25-w1-WOPTPROD2_20180101_diff-vs-prod_2_2016_01-prelim-BL-emul-and-struc-emul-CI}
\end{figure}

For our example of the OLYMPUS 25 WOPT for producer well 2 in the first control interval (ending 01/01/2018) the preliminary emulator predictive 3-sigma credible intervals are illustrated in \cref{subfig:OLYMPUS-25-w1-WOPTPROD2_20180101_diff-vs-prod_2_2016_01-prelim-BL-emul-CI} versus the corresponding decision parameter, {prod\_2\_2016\_01}. This is compared to the theoretical maximum output depicted by the black dotted line determined by the effective target rate for the control interval. The vertical blue and red lines are situated at $\extrapcop{2}{2016}$ and $\cppu{2}{2016}$ respectively. The structured emulation methodology using an upper truncation yields the results shown in \cref{subfig:OLYMPUS-25-w1-WOPTPROD2_20180101_diff-vs-prod_2_2016_01-struc-emul-CI}. Within the slope region, shown in purple, the preliminary emulator credible intervals exceed the constraint and are thus collapsed onto the slope yielding very narrow intervals representing strong beliefs that these decisions will be adhered to for the full control interval. Included are all decision parameter vectors with $d_{P2,2016} < \extrapcop{2}{2016}$, as well as cases where $d_{P2,2016} \geq \extrapcop{2}{2016}$ in which the preliminary emulator credible interval lower bound exceeds the slope. The uncertain intermediate region, shown in orange, is handled by a truncated GP reflecting the uncertainty in whether the model output is on the slope or relatively close when a target rate is achieved for the majority of a control interval. All of these points lie close to the black dotted slope line with much narrower credible intervals than the preliminary Bayes linear emulator. For the plateau region, shown in green, the preliminary emulator credible interval is well below this slope. It is therefore unnecessary to impose a truncation due to the very small probability that an emulator realisation actually exceeds this physical constraint, hence these intervals are unchanged between the two plots.

\begin{figure}[!t]
	\centering
	\begin{subfigure}[t]{0.4955\linewidth}
		\centering
		\includegraphics[page=1,width=\linewidth]{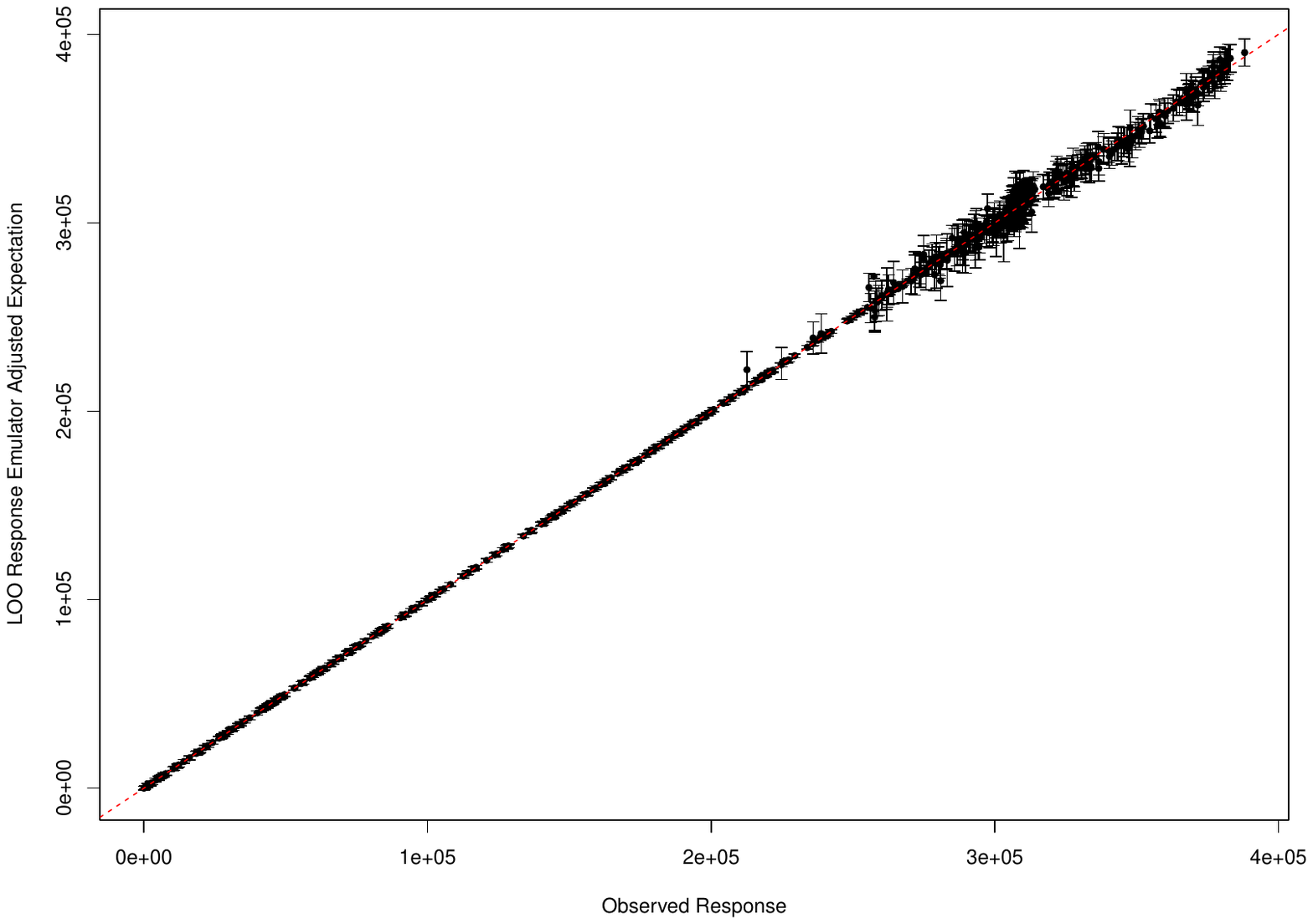}
		\caption{$\text{WOPTPROD2\_2018\_01 emulator CI}$ versus simulated output.}
		\label{subfig:OLYMPUS-25-w1-emul-WOPTPROD2_20180101-v6-k3-extrap-co-cp_lwr-LOO-diagnostics-adj-CI-vs-sim}
	\end{subfigure}
	\hfill
	\begin{subfigure}[t]{0.4955\linewidth}
		\centering
		\includegraphics[page=2,width=\linewidth]{Results/Graphics/Hierarchical-Emulators-Exploiting-Known-Simulator-Behaviour/OLYMPUS-25-w1-dec-par-GP-emulator-WOPTPROD2_20180101_diff-theta-0_5-V6-k3-extrap-co-cp_lwr-LOO-diagnostics-plots-std-all-dec-par}
		\caption{$\text{WOPTPROD2\_2018\_01 emulator CI}$ versus prod\_2\_2016\_01.}
		\label{subfig:OLYMPUS-25-w1-emul-WOPTPROD2_20180101-v6-k3-extrap-co-cp_lwr-LOO-diagnostics-adj-CI-vs-prod_2_2016_01}
	\end{subfigure}
	
	\begin{subfigure}[t]{0.4955\linewidth}
		\centering
		\includegraphics[page=1,width=\linewidth]{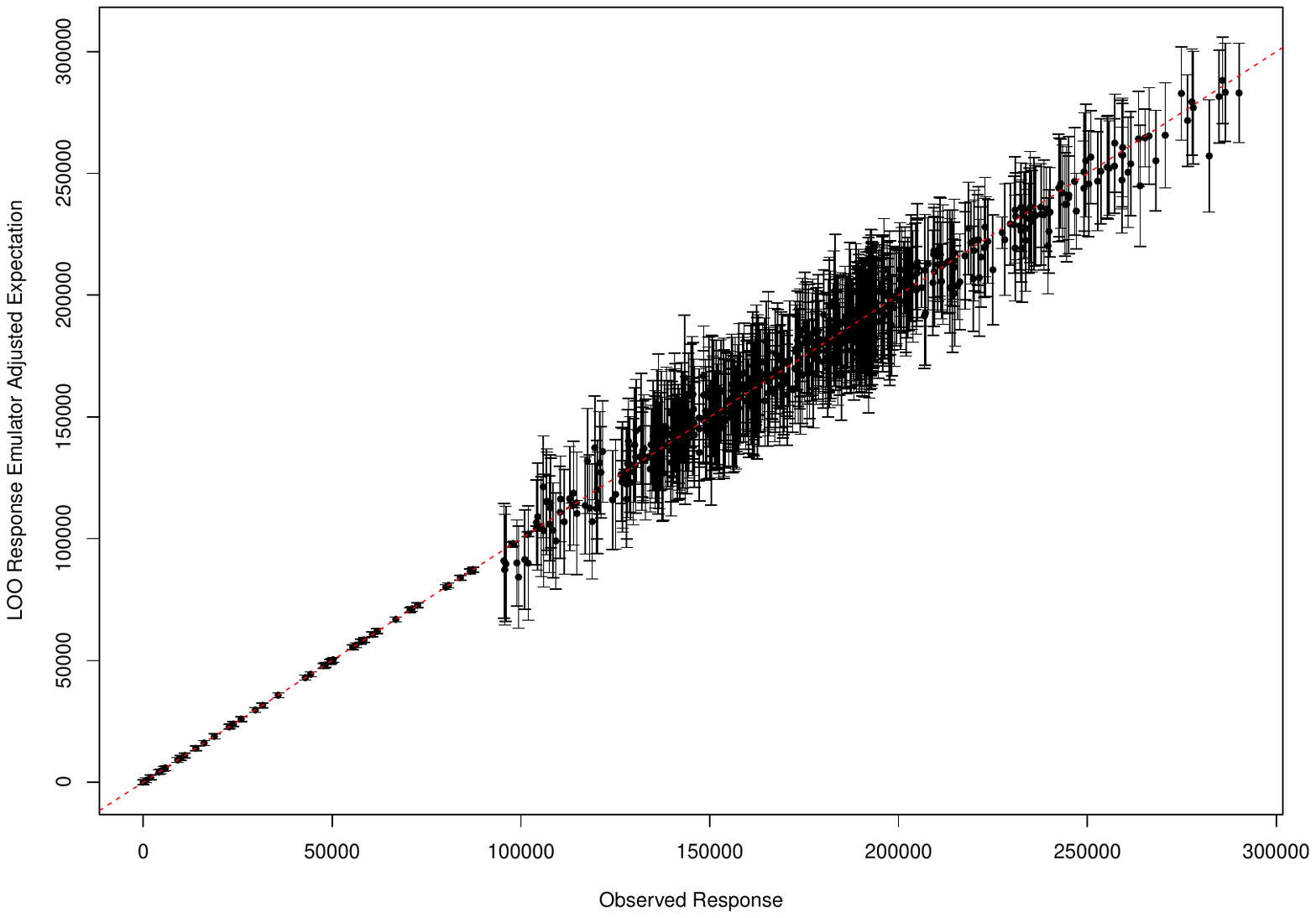}
		\caption{$\text{WWITINJ2\_2022\_01 emulator CI}$ versus simulated output.}
		\label{subfig:OLYMPUS-25-w1-emul-WWITINJ2_20220101-v6-k3-extrap-co-cp_lwr-LOO-diagnostics-adj-CI-vs-sim}
	\end{subfigure}
	\hfill
	\begin{subfigure}[t]{0.4955\linewidth}
		\centering
		\includegraphics[page=2,width=\linewidth]{Results/Graphics/Hierarchical-Emulators-Exploiting-Known-Simulator-Behaviour/OLYMPUS-25-w1-dec-par-GP-emulator-WWITINJ2_20220101_diff-theta-0_5-V6-k3-extrap-co-cp_lwr-LOO-diagnostics-plots-std-all-dec-par}
		\caption{$\text{WWITINJ2\_2022\_01 emulator CI}$ versus inj\_2\_2020\_01.}
		\label{subfig:OLYMPUS-25-w1-emul-WWITINJ2_20220101-v6-k3-extrap-co-cp_lwr-LOO-diagnostics-adj-CI-vs-inj_2_2020_01}
	\end{subfigure}
	\caption{Structured emulation \loo{} diagnostic plots for OLYMPUS 25 WOPTPROD2\_2018\_01 (top) and WWITINJ2\_2022\_01 (bottom). Left: Adjusted expectation with 95\% credible intervals (CI) of width 3 adjusted standard deviations versus the simulated value. The red dashed line denotes equality of the emulator and simulator. Right: Credible interval versus the output's corresponding target production and injection rate respectively. Red points denote the simulated output.}
	\label{fig:OLYMPUS-25-w1-emul-WOPTPROD2_20180101-and-WOPTPROD2_20280101-V6-k3-extrap-co-cp_lwr-LOO-diagnostics}
\end{figure}

\Loo{} structured emulator diagnostics demonstrate satisfactory results with examples shown in \cref{fig:OLYMPUS-25-w1-emul-WOPTPROD2_20180101-and-WOPTPROD2_20280101-V6-k3-extrap-co-cp_lwr-LOO-diagnostics} for OLYMPUS 25 NPV constituents WOPT and WWIT in the control intervals ending 01/01/2018 and 01/01/2022 respectively. The first is our running example.
\Cref{subfig:OLYMPUS-25-w1-emul-WOPTPROD2_20180101-v6-k3-extrap-co-cp_lwr-LOO-diagnostics-adj-CI-vs-sim,subfig:OLYMPUS-25-w1-emul-WWITINJ2_20220101-v6-k3-extrap-co-cp_lwr-LOO-diagnostics-adj-CI-vs-sim} show the emulator adjusted expectation with 95\% (3 adjusted standard deviations) credible intervals versus the simulated output. In each case the emulator is exceptionally accurate for smaller NPV constituent values corresponding to where the target rate is adhered to for the entire control interval. For larger simulated outputs believed to be on plateau, the credible interval is wider, whilst the use of a truncated GP emulator for intermediate values demonstrates a reduction in the uncertainty in these locations. The classification step emulator type is best observed in \cref{subfig:OLYMPUS-25-w1-emul-WOPTPROD2_20180101-v6-k3-extrap-co-cp_lwr-LOO-diagnostics-adj-CI-vs-prod_2_2016_01,subfig:OLYMPUS-25-w1-emul-WWITINJ2_20220101-v6-k3-extrap-co-cp_lwr-LOO-diagnostics-adj-CI-vs-inj_2_2020_01} of the credible intervals versus each NPV constituents' corresponding decision parameter. The structured emulation approach is applied to each of the 3 OLYMPUS models identified in \cref{subsec:Subsampling-from-Geological-Multi-Model-Ensemble-Results} for all of the NPV constituents. It is found that the majority of the 95\% credible intervals contain the simulated value with the maximum percentage of failures over each output type reported in \cref{tab:Structured-Emulation-LOO-Diagnostics-Maximum-Percentage-Failure-Rate-Results-all-OLYMPUS-Models-and-NPV-Constituents}. Moreover, no issues are detected in other \loo{} diagnostic analyses.
\begin{table}
	\centering
	\begin{tabular}{| l | c | c |}
		\hline
		& WOPT & WWIT \\
		\hline
		OLYMPUS 25 & $\leq 4.0\%$ & $\leq3.4\%$ \\
		\hline
		OLYMPUS 33 & $\leq 2.0\%$\textsuperscript{1} & $\leq2.7\%$ \\
		\hline
		OLYMPUS 45 & $\leq 1.2\%$\textsuperscript{2} & $\leq3.0\%$\textsuperscript{3} \\
		\hline
	\end{tabular}
	\caption{Summary of the maximum percentage of structured emulator with upper truncation 95\% credible intervals which do not contain the simulated values in \loo{} diagnostics for the WOPT and WWIT over the 8 control intervals for each of the 3 OLYMPUS models. The exceptions are: (1) for OLYMPUS 33 WOPT emulation of the output in one control interval yields a failure rate of 7.5\%; (2) for OLYMPUS 45 WOPT emulation in two control intervals yields a failure rate of 6.1\% \& 5.3\%; and (3) for OLYMPUS 45 WWIT emulation in one control interval yields a failure rate of 6.3\%.}
	\label{tab:Structured-Emulation-LOO-Diagnostics-Maximum-Percentage-Failure-Rate-Results-all-OLYMPUS-Models-and-NPV-Constituents}
\end{table}
This demonstrates how incorporating known structures within the emulator enables very accurate emulators for the WOPT and WWIT NPV constituents based on a relatively small number of simulations whilst also capturing the change in behaviour.

\begin{figure}[!t]
	\centering
	\includegraphics[width=0.75\linewidth]{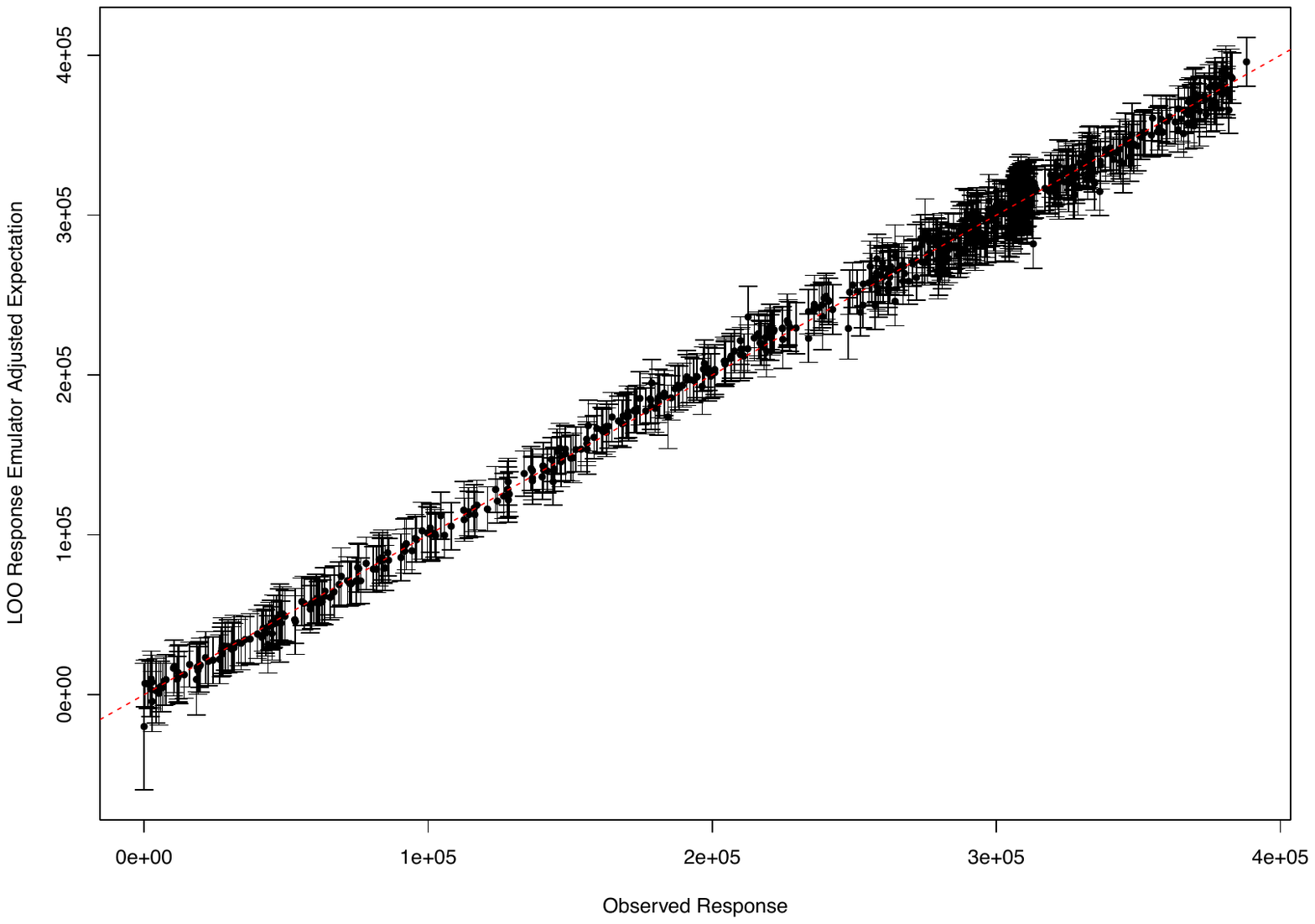}
	\vspace{-10pt}
	\caption{Bayes linear emulation \loo{} diagnostic plot for OLYMPUS 25 WOPTPROD2\_2018\_01 showing the adjusted expectation with 95\% credible intervals of width 3 adjusted standard deviations versus the simulated output. The red dashed line denotes equality of the emulator and simulator. For comparison, the structured emulation exploiting known simulator behaviour \loo{} diagnostics plot for the same output is shown in \cref{subfig:OLYMPUS-25-w1-emul-WOPTPROD2_20180101-v6-k3-extrap-co-cp_lwr-LOO-diagnostics-adj-CI-vs-sim}.}
	\label{fig:OLYMPUS-25-w1-WOPTPROD2_20180101_diff-BL-emulator-v0.1-LOO-diagnostics-adj-CI-vs-sim}
\end{figure}

A comparison with Bayes linear emulation of the WOPT or WWIT NPV constituents, in each case fitted using all simulations, unlike the preliminary Bayes linear emulator in the structured emulation approach which is fitted using the green points in \cref{fig:OLYMPUS-25-w1-WOPTPROD2_20180101_diff-vs-prod_2_2016_01-prelim-BL-emul-and-struc-emul-CI} only, highlights the superior performance. \Loo{} diagnostics are shown for the Bayes linear emulator of OLYMPUS 25 WOPTPROD2\_2018\_01, our running example, in \cref{fig:OLYMPUS-25-w1-WOPTPROD2_20180101_diff-BL-emulator-v0.1-LOO-diagnostics-adj-CI-vs-sim} depicting the emulator adjusted expectation with 95\% (3 adjusted standard deviations) credible intervals versus the simulated output. The corresponding \loo{} diagnostics plot using the structured emulation approach exploiting known simulator behaviour is shown in \cref{subfig:OLYMPUS-25-w1-emul-WOPTPROD2_20180101-v6-k3-extrap-co-cp_lwr-LOO-diagnostics-adj-CI-vs-sim}.
Over the plateau region (green in \ref{subfig:OLYMPUS-25-w1-WOPTPROD2_20180101_diff-vs-prod_2_2016_01-struc-emul-CI}) the accuracy and credible interval width for the two emulators is comparable. However, within the slope (purple) and intermediate (orange) regions the Bayes linear emulator credible interval width is much wider. In the slope region there is at least a two orders of magnitude difference, a consequence of not imposing the known physical constructs. Moreover, within these two regions of parameter space the emulator adjusted expectation often exceeds the maximum upper bound imposed by the target rate decision parameter governing this period, whilst the majority of the credible interval upper bounds also exceed this limit. This implies that unphysical emulator predictions are permitted which may go unchecked. The structured emulation approach protects against this facet.

Structured behaviour is not observed for WWPT within a control interval since there is no corresponding target rate; its behaviour is a consequence of attempting to achieve a given target production rate subject to BHP constraints with water present within the oil field. We separately employ Bayes linear emulators for each of the WWPT constituents following the same approach as above, but fitted using all simulations in $\mathcal{D}$. For each OLYMPUS model the collection of 48 emulators for the WOPT, WWPT, and WWIT for each of the 8 control intervals and for wells in the CWG are combined following the ``divide-and-conquer'' approach in \cref{subsec:Divide-and-Conquer-Approach-Results}.

\section{Emulating Sums of Time Series Outputs} \label{sec:Emulating-Sums-of-Time-Series-Outputs}

\subsection{Methodology} \label{subsec:Emulating-Sums-of-Time-Series-Outputs-Methodology}

The ``divide-and-conquer'' approach in \cref{sec:Divide-and-Conquer-Approach} permits the exploitation of known behaviour such as illustrated in \cref{sec:Structured-Emulators-Exploiting-Known-Simulator-Behaviour}. We develop accurate and efficient emulation methodology where the quantity of interest is the sum of time series outputs addressing the challenges arising from the discretisation of continuous time outputs. In \cref{subsubsec:Emulation-of-an-Approximation-to-the-Sum-of-Time-Series-Outputs} we first emulate an approximation to the sum of time series outputs computed over longer time periods, thus reducing the number of emulators required, focusing on the merger of discounting intervals, before linking to the exact quantity of interest in \cref{subsubsec:Linking-the-Exact-and-Approximate-Sums-of-Time-Series-Outputs}.

\subsubsection{Emulation of an Average Discounting Approximation to the Sum of Time Series Outputs} \label{subsubsec:Emulation-of-an-Approximation-to-the-Sum-of-Time-Series-Outputs}

Let $\fd$ be the sum of time series outputs:
\begin{align} \label{eq:Exact-fd-as-sum-of-time-series-outputs}
	\fd = \sum_{i = 1}^{N_t} \frac{1}{ (1 + d)^{ \frac{t_i}{\tau} } } \fid
\end{align}
where $i$ indexes the time point with $t_i < t_{i+1}$, $N_t$ is the total number of discounting intervals, $d$ is the discount factor, and $\tau$ the discounting period. This is analogous to in \cref{eq:divide-and-conquer-simulator-output} following the ``divide-and-conquer'' approach with $q \equiv N_t$ and $a_i = (1 + d)^{- \frac{t_i}{\tau} }$.

In situations where $N_t$ is very large it may be impractical to accurately emulate and validate for all $\fid$. An average discounting approximation to the exact quantity of interest, $\fd$, is denoted by $\approxfd$ which is computed as a sum of outputs formed by amalgamating multiple time consecutive discounting intervals labelled by $\approxfid$. A formula for $\approxfd$ is given in \cref{eq:Average-discounting-approximate-fd-formula}, where $\Ntilde_t < N_t$ is the numbered of combined time intervals, and $\lambda_i$ is a weighted average discounting factor for the $i$\textsuperscript{th} interval defined in \cref{eq:Approximate-fd-Average-discounting-coefficient} for which $k$ indexes the discounting intervals contained within the longer control interval, $N_{t_i}$ is the total number of such discounting intervals, and $t_{i, 0} = t_{i - 1}$.
\begin{align}
	\approxfd &{} = \sum_{i = 1}^{\Ntilde_t} \lambda_i \approxfid \label{eq:Average-discounting-approximate-fd-formula} \\
	\lambda_i &{} = \dfrac{1}{t_i - t_{i - 1}}\sum_{k = 1}^{N_{t_i}} \dfrac{t_{i, k} - t_{i, k - 1}}{(1 + d)^{\frac{t_{i, k}}{\tau}}} \label{eq:Approximate-fd-Average-discounting-coefficient}
\end{align}
Note that using an averaged discount factor yields a more accurate approximation compared with applying the discounting at the end of each time interval. Assuming the $\approxfid$ are uncorrelated and using a collection of univariate emulators the adjusted expectation and variance formulae for $\approxfd$ are obtained following \cref{eq:divide-and-conquer-emulator-adjusted-expectation,eq:divide-and-conquer-emulator-adjusted-variance} respectively.

\subsubsection{Linking the Exact and Approximate Sums of Time Series Outputs} \label{subsubsec:Linking-the-Exact-and-Approximate-Sums-of-Time-Series-Outputs}

The next step is to link $\approxfd$ with $\fd$. Due to the similar form of \cref{eq:Exact-fd-as-sum-of-time-series-outputs,eq:Average-discounting-approximate-fd-formula} there exists a strong linear relationship between the approximate and exact $\fd$ for which a simple linear regression in \cref{eq:Exact-fd-on-Approximate-fd-LM-form} provides a meaningful statistical link whilst capturing the additional induced uncertainties.
\begin{align} \label{eq:Exact-fd-on-Approximate-fd-LM-form}
	\fd = \beta_{0, \approxf} + \beta_{1, \approxf} \approxfd	+ \varepsilon_{\approxf}
\end{align}
The adjusted expectation and variance are then computed using \cref{eq:Exact-fd-on-Approximate-fd-emulation-via-LM-adjusted-expectation,eq:Exact-fd-on-Approximate-fd-emulation-via-LM-adjusted-variance} respectively.
\begin{align*}
	\E_{\boldF}[\fd] &{} = \hat{\beta}_{0, \approxf} + \hat{\beta}_{1, \approxf} \E_{\boldF}\left[ \approxfd \right] \alignnumber \label{eq:Exact-fd-on-Approximate-fd-emulation-via-LM-adjusted-expectation} \\
	\Var_{\boldF}[\fd] &{} = \Var\left[ \hat{\beta}_{0, \approxf} \right] + 2 \Cov\left[ \hat{\beta}_{0, \approxf}, \hat{\beta}_{1, \approxf} \right] \E_{\boldF}\left[ \approxfd \right] \alignmultilineeq
	+ \left\{ \hat{\beta}_{1, \approxf}^2 + \Var\left[ \hat{\beta}_{1, \approxf} \right] \right\} \Var_{\boldF}\left[ \approxfd \right] \alignmultilineeq
	+ \Var\left[ \hat{\beta}_{1, \approxf} \right] \left( \E_{\boldF}\left[ \approxfd \right] \right)^2 + \sigma_{\approxf}^2 \alignnumber \label{eq:Exact-fd-on-Approximate-fd-emulation-via-LM-adjusted-variance}
\end{align*}
Estimates of the regression coefficients, $\hat{\beta}_{0, \approxf}$ and $\hat{\beta}_{1, \approxf}$, along with their variances and covariance, are obtained using the wave 0 exploratory simulations data, whilst $\varepsilon_{\approxf}$ is treated as independent with residual standard error $\sigma_{\approxf}$. The collection of all simulation data, $\boldF$, is as defined in \cref{subsec:Divide-and-Conquer-Approach-Methodology}.

\subsection{Results} \label{subsec:Emulating-Sums-of-Time-Series-Outputs-Results}


The NPV objective function in the \TNOChallengeWC{} (see  \cref{eq:NPV-general-formula-in-decision-parameters,eq:Expected-NPV-general-formula-in-decision-parameters}) is of the form of \cref{eq:Exact-fd-as-sum-of-time-series-outputs}. For each OLYMPUS ensemble member $\fd = \NPVjd$ and $\fid$ are the NPV constituents. In this application the 8 decisions for each well are enacted over periods constructed by amalgamating consecutive 3-month discounting intervals. The by model average discounting approximate NPV, $\approxNPVjd$ is obtained from \cref{eq:Well-control-optimisation-challenge-NPV-R(d_t_i)-by-well-decomposition}, noting that each of $f_{Pk, t_i}^{op}(\boldd)$, $f_{Ik, t_i}^{wp}(\boldd)$, and $f_{Ik, t_i}^{wi}(\boldd)$ are calculated over periods longer than the discounting intervals, hence $\fd$ does correspond to the approximation $\approxfd$, and with $a_i = \lambda_i$ from \cref{eq:Approximate-fd-Average-discounting-coefficient}.

\begin{figure}[!t]
	\centering
	\begin{subfigure}[t]{0.4955\linewidth}
		\centering
		\includegraphics[page=1,width=\linewidth]{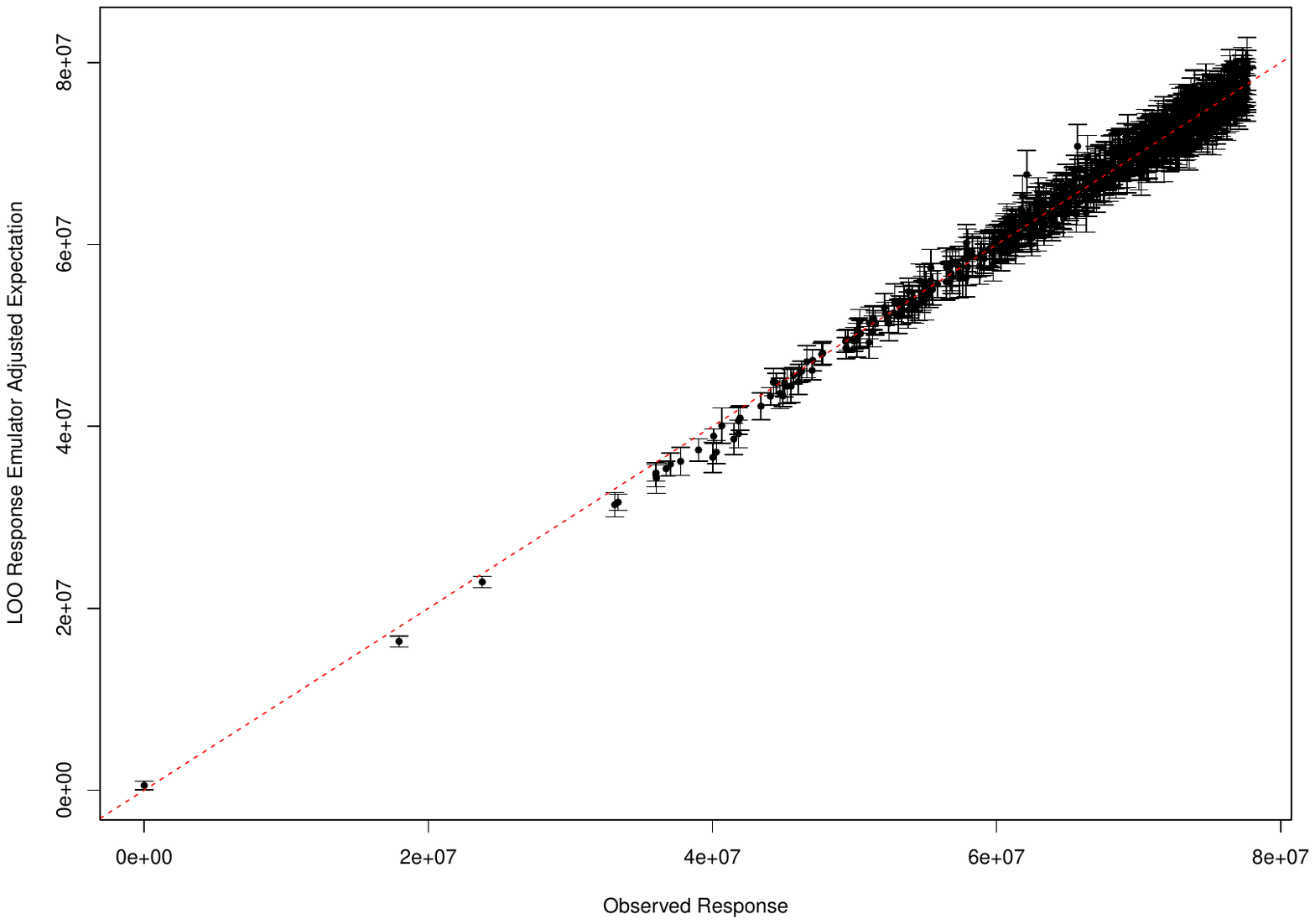}
		\caption{Approx NPV CI versus simulated values.}
		\label{subfig:OLYMPUS-25-w1-emul-avg-discount-approx-NPV-via-NPV-summation-LOO-diagnostics-v6-k3-adj-CI-vs-sim}
	\end{subfigure}
	\hfill
	\begin{subfigure}[t]{0.4955\linewidth}
		\centering
		\includegraphics[page=1,width=\linewidth]{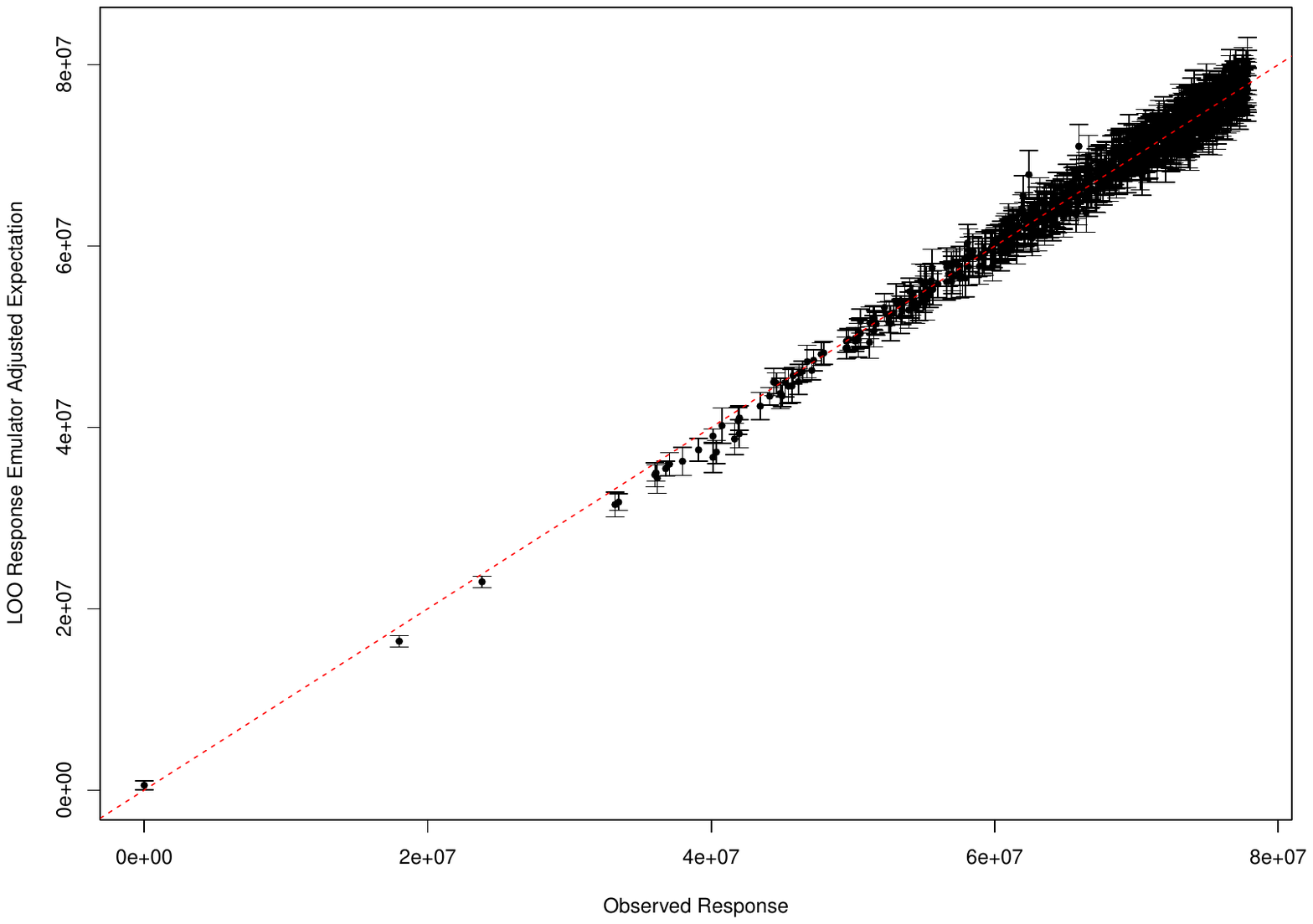}
		\caption{Exact NPV CI versus simulated values.}
		\label{subfig:OLYMPUS-25-w1-emul-NPV-via-lm-on-avg-discount-approx-NPV-sum-LOO-diagnostics-v6-k3-adj-CI-vs-sim}
	\end{subfigure}
	\caption{Emulator \loo{} diagnostic plots for OLYMPUS 25 average discounting approximate NPV (left) and exact NPV (right, obtained via a linear model of the form in \cref{eq:Exact-fd-on-Approximate-fd-LM-form} on the emulated approximate NPV) showing the adjusted expectation with 3 adjusted standard deviation Credible Intervals (CI) versus simulated values. The red dashed line denotes when the emulator and simulator coincide.}
	\label{fig:OLYMPUS-25-w1-emul-avg-discount-approx-NPV-and-NPV-via-lm-on-avg-discount-approx-NPV-sum-LOO-diagnostics-v6-k3}
\end{figure}

Emulation of $\approxNPVjd$ is performed following the method described in \cref{subsec:Emulating-Sums-of-Time-Series-Outputs-Methodology} summing structured emulators for the WOPT and WWIT contributors (details in \cref{subsec:Structured-Emulators-Exploiting-Known-Simulator-Behaviour-Results} and Bayes linear emulators for the WWPT constituents. \Loo{} diagnostics plots for the OLYMPUS 25 approximate NPV is shown in \cref{subfig:OLYMPUS-25-w1-emul-avg-discount-approx-NPV-via-NPV-summation-LOO-diagnostics-v6-k3-adj-CI-vs-sim}. There exists a strong linear relation between the emulator adjusted expectation and the simulated approximate NPV with the majority of points situated close to the red dashed equality line. It is observed that the uncertainty generally increases with the value of the approximate NPV. Petroleum reservoir engineering provides insight: higher target production and injection rates are generally necessary to achieve the largest NPVs. For the WOPT and WWIT structured emulators this occurs above the extrapolation cut-off and thus each constituent emulator exhibits a larger uncertainty. Furthermore, when many of the NPV constituents fall in their slope regions the structured emulator returns a small uncertainty determined by the tolerance. These linearly combine to produce a small uncertainty for the approximate NPV.

The exact and average discounting approximate NPV for each OLYMPUS model are linked using the simple linear regression framework in \cref{eq:Exact-fd-on-Approximate-fd-LM-form} where the coefficients are estimated using the wave 0 simulation data. This accounts for the discrepancy induced by coalescing the discounting intervals. \Loo{} emulator diagnostics for the OLYMPUS 25 NPV are shown in \cref{subfig:OLYMPUS-25-w1-emul-NPV-via-lm-on-avg-discount-approx-NPV-sum-LOO-diagnostics-v6-k3-adj-CI-vs-sim}. The results are very similar to those for the approximate NPV with our commentary and interpretation mirroring the above. The percentage of 95\% credible intervals containing the simulated value (computed using simulation output in the respective average discounting approximate or exact NPV formula) for each of the 3 OLYMPUS models is reported in \cref{tab:Emulation-LOO-Diagnostics-Percentage-Failure-Rate-Results-all-OLYMPUS-Models-Approximate-and-Exact-NPV}.
\begin{table}
	\centering
	\begin{tabular}{| l | c | c |}
		\hline
		& Approximate NPV & Exact NPV \\
		\hline
		OLYMPUS 25 & 6.7\% & 6.7\% \\
		\hline
		OLYMPUS 33 & 4.4\% & 4.3\% \\
		\hline
		OLYMPUS 45 & 3.7\% & 3.8\% \\
		\hline
	\end{tabular}
	\caption{Summary of the percentage of emulator 95\% credible intervals which do not contain the simulated values in \loo{} diagnostics for the average discounting approximation to the NPV and the exact NPV for each of the 3 OLYMPUS models.}
	\label{tab:Emulation-LOO-Diagnostics-Percentage-Failure-Rate-Results-all-OLYMPUS-Models-Approximate-and-Exact-NPV}
\end{table}

\section{Emulation of a Multi-Model Ensemble Mean} \label{sec:Emulation-of-a-Multi-Model-Ensemble-Mean}

\subsection{Methodology} \label{subsec:Emulation-of-a-Multi-Model-Ensemble-Mean-Methodology}

The objective is to emulate the ensemble mean output, $\fbard$, by combining emulators for the individual models' outputs, $\fjd$. A reasonable assumption is that the ensemble members are independent given the complexity of their differing constructions.

\subsubsection{When Simulations are Available for all Ensemble Members} \label{subsubsec:Emulation-of-the-Ensemble-Mean-NPV-when-Simulations-are-Available-for-all-Ensemble-Members}

When simulations are relatively quick to evaluate; large amounts of computing resources are available, or there is a desire to minimise the uncertainty (such as in the \TNOChallengeshort{} due to the underlying geology, which is particularly relevant when the ensemble mean NPV is assumed equal to the expected NPV), it may be possible to simulate from the entire ensemble. The ensemble mean output is computed as either the arithmetic or a weighted mean (with weights obtained from a prior probability distribution over the models) of the individual model outputs. This presents a natural method to emulate $\fbard$ with the adjusted expectation and variance defined in \cref{eq:Ensemble-Mean-fd-all-members-emulator-adjusted-expectation,eq:Ensemble-Mean-fd-all-members-emulator-adjusted-variance}, where $\boldF = \{ \boldFj \}_{j=1}^N$ denotes all necessary simulation data with $\boldFj$ being the outputs for ensemble member $j$, and weights $\omega_j$, with $\omega_j = \frac{1}{N}$ for the arithmetic mean.
\begin{align}
	\E_{\boldF}\left[ \fbard \right] &{} = \sum_{j = 1}^{N} \omega_j \E_{\boldFj}[\fjd] \label{eq:Ensemble-Mean-fd-all-members-emulator-adjusted-expectation} \\
	\Var_{\boldF}\left[ \fbard \right] &{} =  \sum_{j = 1}^{N} \omega_j^2 \Var_{\boldFj}[\fjd] \label{eq:Ensemble-Mean-fd-all-members-emulator-adjusted-variance}	
\end{align}
The variance formula may be adapted when the output for different ensemble members are believed to be correlated by introducing the relevant covariance terms in \cref{eq:Ensemble-Mean-fd-all-members-emulator-adjusted-variance}.

\subsubsection{Using an Ensemble Subsampling Linear Model} \label{subsubsec:Emulation-of-the-Ensemble-Mean-NPV-Using-the-Ensemble-Subampling-LM} 

A more realistic and practical scenario is that simulations are only performed for a subset of the ensemble, such as, but not limited to, those selected using the techniques described in \cref{subsec:Subsampling-from-Multi-Model-Ensembles-Methodology}. A linear model of the form shown in \cref{eq:EGES-ensemble-mean-resp-linear-model-on-individual-model-resp-N_EGES-models-unknown} is used to emulate $\fbard$ with the emulated output for each of the sub-selected models as inputs. These are $\{ f_{j_1}(\boldd), \ldots, f_{j_{\Ntilde}}(\boldd) \}$, for which $\Ntilde < N$, with $j_1, \ldots, j_{\Ntilde} \in \{ 1, \ldots, N \}$, and $j_k \neq j_l$ for $k \neq l$. The estimated coefficients are denoted by $\hat{\alpha}_\mathEGES$ and $\hat{\beta}_{k, \mathEGES}$. It is assumed that the individual emulator outputs and the regression coefficients are uncorrelated, which is justifiable if two distinct simulation data sets are used to construct the linear model and fit the emulators. Under this formulation the adjusted expectation is shown in \cref{eq:Ensemble-Mean-fd-Emulation-via-EGES-LM-adjusted-expectation}.
\begin{align*}
	\E_\boldF\left[ \fbard \right] &{} = \E_{\boldF}\left[ \hat{\alpha}_\mathEGES + \sum_{k = 1}^{\Ntilde} \hat{\beta}_{k, \mathEGES} f_{j_k}(\boldd) + \epsilonEGES(\boldd) \right] \\
	&{} = \hat{\alpha}_\mathEGES + \sum_{k = 1}^{\Ntilde} \hat{\beta}_{k, \mathEGES} \E_{\boldF_{j_k}}[f_{j_k}(\boldd)] \alignnumber \label{eq:Ensemble-Mean-fd-Emulation-via-EGES-LM-adjusted-expectation}
\end{align*}
Define $\betahatEGES = (\hat{\alpha}_\mathEGES, \hat{\beta}_{1, \mathEGES}, \ldots, \hat{\beta}_{\Ntilde, \mathEGES})^\Transpose \in \mathbb{R}^{\Ntilde + 1}$ with $\SigmabetaEGES = \Var\left[ \betahatEGES \right]$, and \lword{$\XEGESd = (1, f_{j_1}(\boldd), \ldots, f_{j_{\Ntilde}}(\boldd))^\Transpose \in \mathbb{R}^{\Ntilde + 1}$} with uncorrelated components, so $\Var_{\boldF}\left[ \XEGESd \right]$ is diagonal. The adjusted variance is presented in \cref{eq:Ensemble-Mean-fd-Emulation-via-EGES-LM-adjusted-variance}, where $\sigmahatEGES$ is the estimated residual standard error for $\epsilonEGES(\boldd)$.
\begin{align*}
	\Var_{\boldF}\left[ \fbard \right] &{}= \Var\left[ \hat{\alpha}_\mathEGES \right] \alignmultilineeq + \sum_{k = 1}^{\Ntilde} \Var\left[ \hat{\beta}_{k, \mathEGES} \right] \left( \Var_{\boldF_{j_k}}\left[ f_{j_k}(\boldd) \right] + \E_{\boldF_{j_k}}\left[ f_{j_k}(\boldd) \right]^2 \right) \alignmultilineeq + 2 \sum_{k = 1}^{\Ntilde} \Cov\left[ \hat{\alpha}_\mathEGES, \hat{\beta}_{k, \mathEGES} \right] \E_{\boldF_{j_k}}\left[ f_{j_k}(\boldd) \right] \alignmultilineeq + \sum_{\substack{k \neq l \\ k, l = 1, \ldots, \Ntilde}} \left( \Cov\left[ \hat{\beta}_{k, \mathEGES}, \hat{\beta}_{l, \mathEGES} \right] \E_{\boldF_{j_k}}\left[ f_{j_k}(\boldd) \right] \E_{\boldF_{j_l}}[f_{j_l}(\boldd)] \right) \alignmultilineeq + \sum_{k = 1}^{\Ntilde} \hat{\beta}_k^2 \Var_{\boldF_{j_k}}[f_{j_k}(\boldd)] + \sigmahatEGES^2 \alignnumber \label{eq:Ensemble-Mean-fd-Emulation-via-EGES-LM-adjusted-variance}
\end{align*}


\subsection{Results} \label{subsec:Emulation-of-a-Multi-Model-Ensemble-Mean-Results}


The process of building structured emulators for each of the NPV constituents in \cref{subsec:Structured-Emulators-Exploiting-Known-Simulator-Behaviour-Results}, their combination via the NPV formula to obtain the average discounting approximate NPV, and subsequent linking to the exact NPV in \cref{subsec:Emulating-Sums-of-Time-Series-Outputs-Results}, is repeated for each of the three sub-selected OLYMPUS models. For the \TNOChallengeWC{} and our decision support setup the ensemble mean NPV, $\fbard = \meanNPVd$, is the quantity of interest as the objective and utility function respectively. This is emulated using the ensemble subsampling linear model devised in \cref{subsec:Subsampling-from-Geological-Multi-Model-Ensemble-Results} to combine the emulators for the OLYMPUS 25, 33, \& 45 NPVs, following the approach in \cref{subsec:Emulation-of-a-Multi-Model-Ensemble-Mean-Methodology}.

\begin{figure}[!t]
	\centering
	\begin{subfigure}[t]{0.4955\linewidth}
		\centering
		\includegraphics[page=1,width=\linewidth]{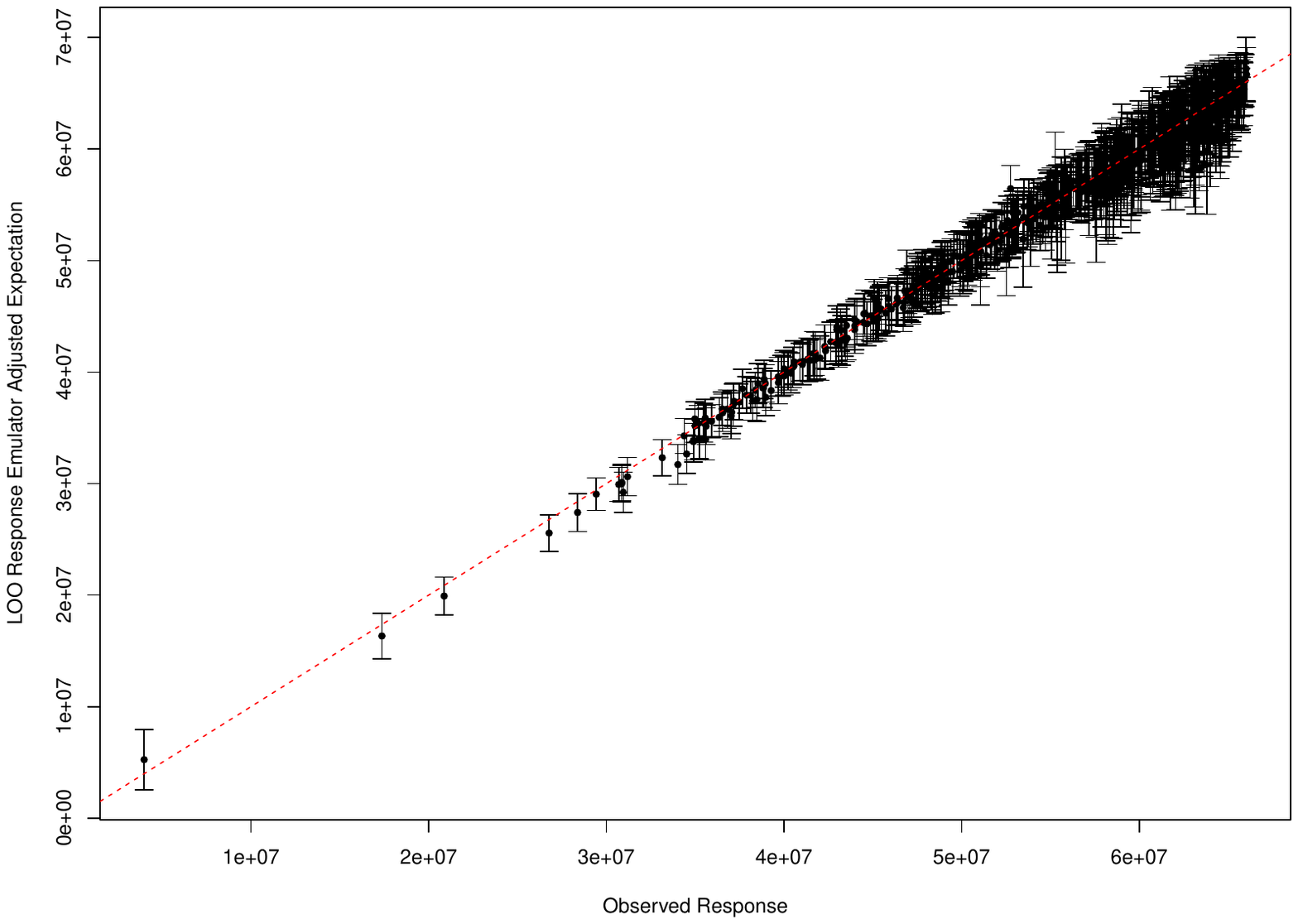}
		\caption{Hierarchical emulator CI versus subsampling predicted ensemble mean NPV. The red dashed line denotes emulator and simulator equality.}
		\label{subfig:OLYMPUS-w1-hier-emul-Ensemble-Mean-NPV-via-EGES-LM-diagnostics-adj-CI-vs-sim}
	\end{subfigure}
	\hfill
	\begin{subfigure}[t]{0.4955\linewidth}
		\centering
		\includegraphics[page=3,width=\linewidth]{Results/Graphics/Hierarchical-Emulators-Exploiting-Known-Simulator-Behaviour/OLYMPUS-w1-dec-par-hierarchical-emulator-ensemble-mean-CWG-NPV-LOO-diagnostics-plots-V6-k3-extrap-co-cp_lwr-std-variable-prior}
		\caption{Hierarchical emulator standardised residuals versus the simulated ensemble mean NPV.}
		\label{subfig:OLYMPUS-w1-hier-emul-Ensemble-Mean-NPV-via-EGES-LM-diagnostics-std-resid-vs-sim}
	\end{subfigure}
	\caption{OLYMPUS wave 1 hierarchical emulation diagnostics plots for the predicted ensemble mean NPV via the ensemble subsampling linear model combining the emulation output for the exact NPV of the three sub-selected OLYMPUS models.}
	\label{fig:OLYMPUS-w1-hier-emul-Ensemble-Mean-NPV-via-EGES-LM-diagnostics}
\end{figure}

It is not possible to perform \loo{} diagnostics for the true ensemble mean NPV because simulations have only been performed for the identified subset of OLYMPUS models. The wave 0 simulations were run for all 50 OLYMPUS models under a setup using a shorter field lifetime due to available computational resources, hence these cannot be used in emulator diagnostics. Note that the additional uncertainty pertaining to the ensemble subsampling linear model is accounted for within the hierarchical emulator construction. Instead we compare the hierarchical emulator with the predicted ensemble mean NPV in  \cref{fig:OLYMPUS-w1-hier-emul-Ensemble-Mean-NPV-via-EGES-LM-diagnostics} where \cref{subfig:OLYMPUS-w1-hier-emul-Ensemble-Mean-NPV-via-EGES-LM-diagnostics-adj-CI-vs-sim} demonstrates accurate predictions. Moreover, the increase in the uncertainty compared to individually emulating a single OLYMPUS model NPV, such as for OLYMPUS 25 NPV in \cref{subfig:OLYMPUS-25-w1-emul-NPV-via-lm-on-avg-discount-approx-NPV-sum-LOO-diagnostics-v6-k3-adj-CI-vs-sim}, is modest; thus the process of subsampling from the ensemble before reconstructing the ensemble mean NPV contributes relatively little additional uncertainty versus the structured emulation of the NPV constituents for each model.  \Cref{subfig:OLYMPUS-w1-hier-emul-Ensemble-Mean-NPV-via-EGES-LM-diagnostics-std-resid-vs-sim} shows no distinguishable pattern in the pseudo standardised residuals, whilst the majority are of magnitude less than three.

\subsection{Emulator Comparison} \label{subsec:Emulator-Comparison}

\begin{figure}
	\centering
	\begin{subfigure}[t]{0.4955\linewidth}
		\centering
		\includegraphics[width=\linewidth]{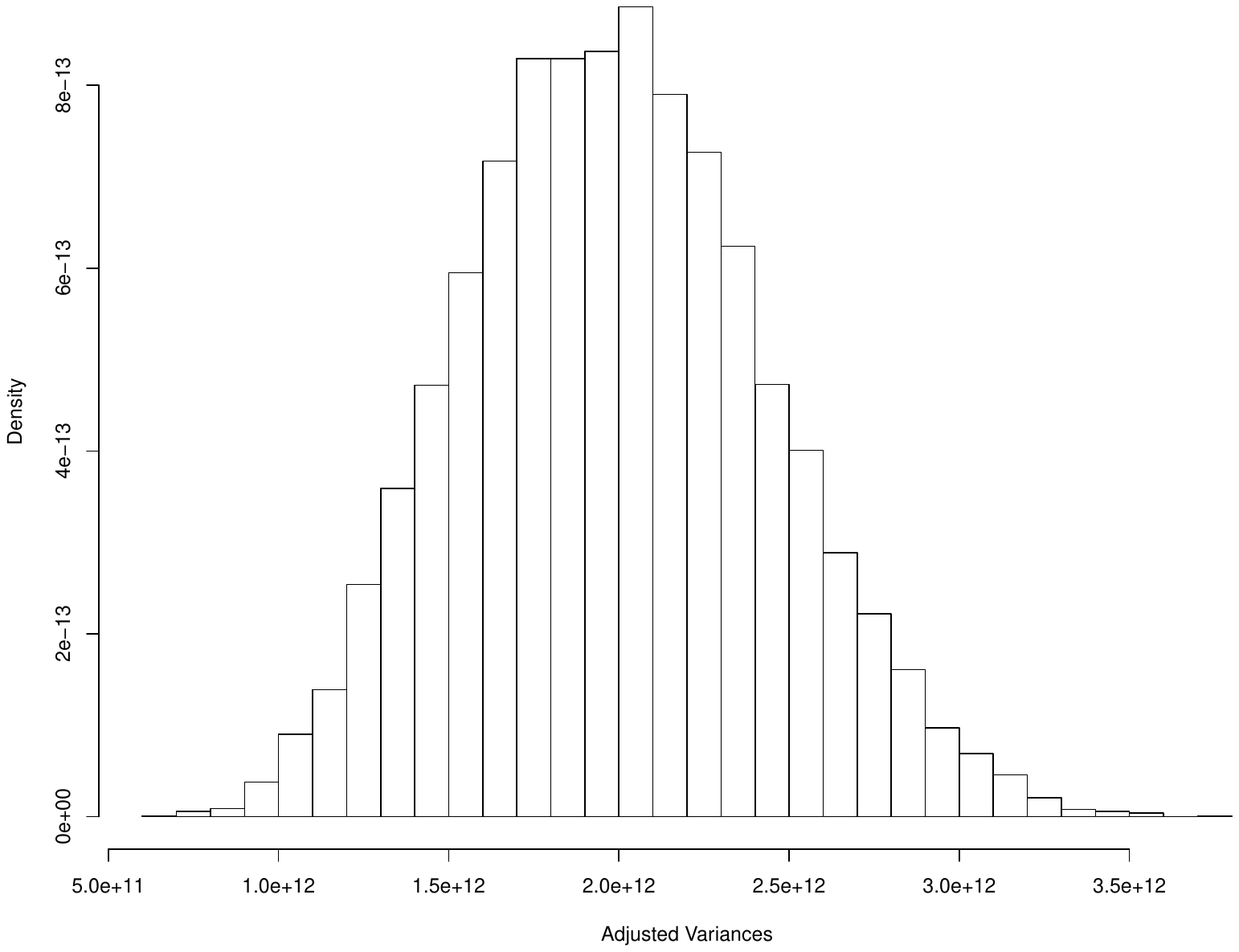}
		\caption{Bayes linear emulator.}
		\label{subfig:OLYMPUS-w1-BL-emulator-ensemble-mean-NPV-theta-0_5-hist-adj-var}
	\end{subfigure}
	\hfill
	\begin{subfigure}[t]{0.4955\linewidth}
		\centering
		\includegraphics[width=\linewidth]{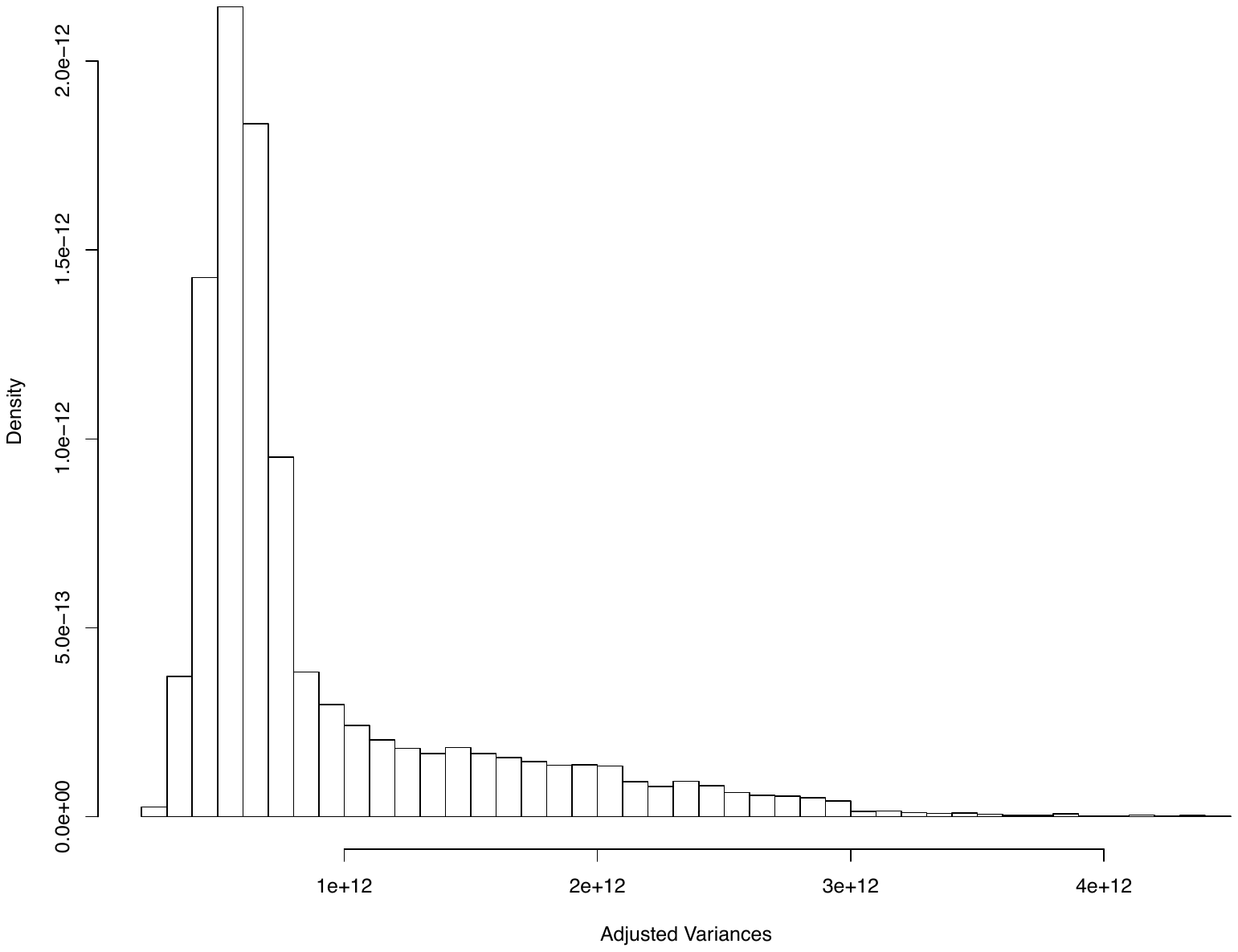}
		\caption{Hierarchical emulator.}
		\label{subfig:OLYMPUS-w1-hier-emulator-ensemble-mean-NPV-hist-adj-var}
	\end{subfigure}
	\caption{Histograms comparing the Bayes linear and hierarchical emulators adjusted variances for the ensemble mean NPV. Note that the seemingly large variances are inline with the simulated ensemble mean NPV which is of the order $\num{5.0e+07}$ \$ to $\num{6e+07}$ \$.}
	\label{fig:OLYMPUS-w1-comparison-BL-and-hier-emulator-ensemble-mean-NPV-hist-adj-var}
\end{figure}

Two approaches were implemented for emulating the ensemble mean NPV: a Bayes linear emulator in \cref{subsec:Bayes-Linear-Emulation-of-the-Expected-NPV}; and a hierarchical emulator which exploits known constrained behaviour for certain simulator outputs built up over \cref{subsec:Subsampling-from-Geological-Multi-Model-Ensemble-Results,subsec:Divide-and-Conquer-Approach-Results,subsec:Structured-Emulators-Exploiting-Known-Simulator-Behaviour-Results,subsec:Emulating-Sums-of-Time-Series-Outputs-Results,subsec:Emulation-of-a-Multi-Model-Ensemble-Mean-Results}. Firstly, comparing each emulator's adjusted variances evaluated for the same large collection of decision parameter vectors in \cref{fig:OLYMPUS-w1-comparison-BL-and-hier-emulator-ensemble-mean-NPV-hist-adj-var} demonstrates how the hierarchical emulator achieves a discernible reduction in the uncertainty versus the Bayes linear emulator. This feature is also evident when comparing the \loo{} diagnostics plots in \cref{fig:OLYMPUS-w1-BL-emulator-theta-0_5-LOO-diagnostics-plot-adj-exp-3-adj-sd-CI-vs-sim-NPV,subfig:OLYMPUS-w1-hier-emul-Ensemble-Mean-NPV-via-EGES-LM-diagnostics-adj-CI-vs-sim} where there is a prevalent reduction in the credible interval widths. A direct comparison of the adjusted variances for each decision parameter vector highlights an average reduction in the adjusted variance of more than a half. Note that there exist a small number of cases where there is a moderate increase in the uncertainty, although this is outweighed by the gains achieved across the majority of sampled locations within the decision parameter space.

A crucial motivation for employing emulators as a surrogate to computer models is their speed of evaluation in order to enable further analyses such as decision support. Bayes linear emulation is known to be a very fast and efficient means of constructing emulators. In this application we achieve a substantial reduction in computation time with over 2000 emulator evaluations for new decision parameter settings per second using a single core of a standard desktop computer or laptop. This is juxtaposed with approximately 30 minutes per OLYMPUS model simulation, or 25 hours when using the entire ensemble. The combination of ensemble subsampling and Bayes linear emulation equates to an efficiency gain of the order of $10^8$.

The full hierarchical emulation process applied to the ensemble mean NPV requires for each OLYMPUS model the fitting of 48 separate emulators: 32 of the structured type; and 16 Bayes linear emulators, a total of 144 emulators over the three sub-selected OLYMPUS models. Next, these are combined to obtain emulators for the approximate and exact NPVs for each OLYMPUS model, before emulating the ensemble mean NPV, and then linking to the expected NPV. The computational performance is more modest achieving emulator evaluations at approximately 4 new decision parameter vectors per second using a single core, and is thus slower than Bayes linear emulation. However, in comparison with direct simulation from the OLYMPUS ensemble there is a considerable efficiency improvement of the order of $10^4$. This is sufficient for comprehensively exploring the decision parameter space. Moreover, the additional computational expense of hierarchical emulation can be justified by the reduction in emulator uncertainty. This is highly beneficial to performing an iterative decision support analysis where reducing emulator uncertainty is imperative for efficiently eliminating non-implausible regions of the decision parameter space, thus avoiding extra waves of extremely expensive simulations at locations that would have been ruled out by more accurate emulators. The additional computational cost is therefore offset versus the need for extra simulations. Such arguments are also relevant to analyses using single-stage (or one-shot) designs where fewer expensive computer model evaluations are required to achieve similar emulator accuracy across the parameter space. Both forms of emulators are easily parallelisable, thus permitting further efficiency gains.

\section{Conclusion} \label{sec:Conclusion}

We have presented a methodological toolkit for the analysis of multi-model ensembles of ``grey-box'' computer models. This include: an efficient technique for obtaining a small representative subset of models by subsampling from a multi-model ensemble; targeted Bayesian design methodology incorporating relevant prior information to the objective of providing decision support under uncertainty; a ``divide-and-conquer'' approach to emulation of sums of outputs where it is preferable to emulate the constituents, for example due to knowledge of their underlying behaviour; structured emulation of outputs to exploit constrained and structured behaviour through the partitioning of the parameter space and use of truncated emulators; the efficient combination of multiple emulators for time series outputs through an average discounting approximation; and emulation of an ensemble mean output. Combining these methods yields a novel hierarchical emulator achieving more accurate predictions for quantities of interest, whilst each technique may also be employed separately depending on the problem specific features exhibited by the computer model.

This is motivated by and applied to the \TNOChallengeWC{} from the petroleum industry where the aim is to maximise the expected NPV, approximated by the ensemble mean NPV, as a function of well control decision parameters. We reconstrue this as a decision support problem where the utility function consists of a discounted sum of oil production, water injection, and water production, both by well and control interval. The first two simulator outputs exhibit partially known behaviour, constrained by choices of inputs and physical limits with respect to their corresponding target production and injection rate decision parameters respectively, with this feature encompassed within our structured emulator formulation. The application demonstrates superior accuracy versus Bayes linear emulators, whilst the slower speed of evaluation is mitigated by the need for fewer (waves of) simulations from the expensive computer model ensemble. Both factors are important for the overall aim of providing robust decision support under uncertainty. Moreover, we introduce multi-model ensemble subsampling techniques to efficiently identify a representative subset of models which collectively best characterises the ensemble mean output of interest, in this application, the ensemble mean NPV, whilst also providing a method for their prediction. This constitutes a novel application to the petroleum industry where multi-model ensembles are commonly used to represent geological uncertainty, greatly reducing the computational cost of our decision analysis.

The next step is to employ the presented hierarchical emulation methodology within iterative decision support, applied to the \TNOChallengeWC{}, which incorporates a comprehensive and realistic uncertainty quantification to statistically link inferences for the computer model (OLYMPUS) with the corresponding real world physical system. See \cite[Sec.~4.6 \& 4.7]{2022:Owen:PhD-Thesis} for details. In addition, further methodological development should focus on enhancing the overall hierarchical emulation framework. This may be achieved by revising the structured emulators change point estimation methods, classification and truncation, as well as via the refinement of the uncertainty propagation in \cref{subsec:Emulating-Sums-of-Time-Series-Outputs-Methodology,subsec:Emulation-of-a-Multi-Model-Ensemble-Mean-Methodology}. Another direction is multivariate structured emulation of the NPV constituents to assess their correlation and thus more accurately quantify the approximate NPV by ensemble member emulator variance in \cref{subsec:Structured-Emulators-Exploiting-Known-Simulator-Behaviour-Results,subsec:Emulating-Sums-of-Time-Series-Outputs-Results}. Such further methodological developments must also be efficient so as not add to the computational burden.

The methodological toolkit and their combination to form a hierarchical emulator presented in this paper, whilst motivated by and tailored to the petroleum well control optimisation problem, is sufficiently flexible and adaptable to handle other (partially) known forms of computer model outputs and functions thereof. Opening ``black-box'' simulators and exploring functions of their output to investigate their behaviours is evidently beneficial, as is using domain expert prior knowledge and small carefully designed collections of simulations. Another example of emulating ``grey-box'' models is in known boundary emulation \cite{2019:Vernon:Known-Boundary-Emulation-of-Complex-Computer-Models,2023:Jackson:Efficient-Emulation-of-Computer-Models-Utilising-Multiple-Known-Boundaries-of-Differing-Dimensions}. The additional prior information can then be used to guide the choice from existing emulation methods or to design novel forms which exploit known behavioural facets in order to achieve superior accuracy and enhance the usefulness of emulators for real world applications.

%
%



\bibliography{PhD-Paper-1-NPV-Emulation-Bibliography.bib}

\appendix

\section{TNO OLYMPUS Well Control Optimisation Challenge -- Extended Results} \label{sec:TNO-OLYMPUS-Well-Control-Optimisation-Challenge-Extended-Results}

In this appendix we extend our discussion of results of the application to the \TNOChallengeWC{} (see \cref{sec:TNO-OLYMPUS-Well-Control-Optimisation-Challenge} for an overview) using the methodology proposed in this paper.

\subsection{OLYMPUS Exploratory Analysis -- Additional Plots and Discussion} \label{subsec:Extended-Results-OLYMPUS-Exploratory-Analysis}

\begin{figure}
	\centering
	\begin{subfigure}{\linewidth}
		\centering
		\includegraphics[page=1,width=0.76\linewidth,viewport= 0 18 765 575, clip]{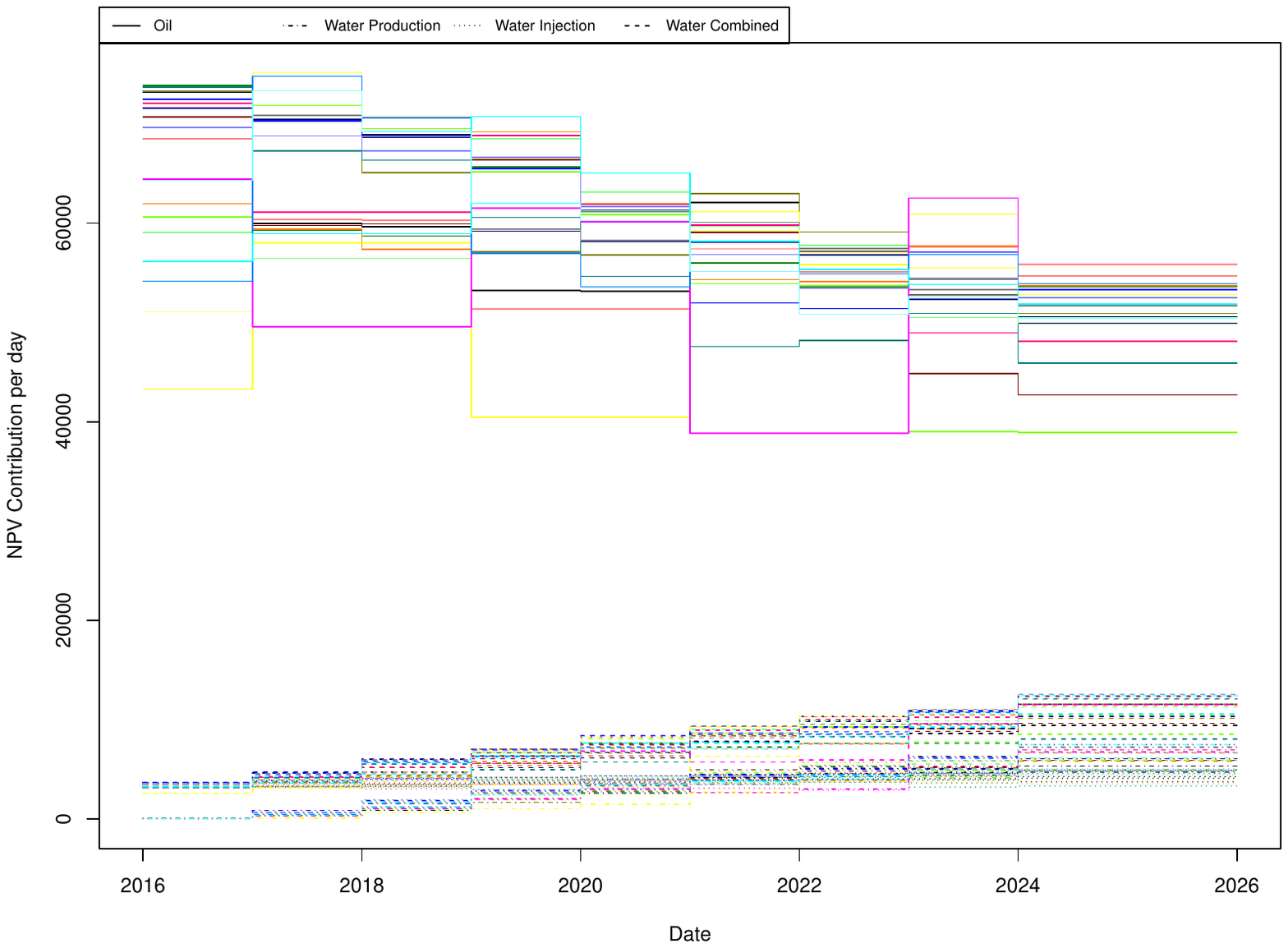}
		\vspace{-5pt}
		\caption{Raw scale.}
		\label{subfig:OLYMPUS-25-Exploratory-Analysis-NPV-Oil-Water-Contribution-all-decisions-raw-scale}
	\end{subfigure}
	
	
	\begin{subfigure}{\linewidth}
		\centering
		\includegraphics[page=2,width=0.76\linewidth,viewport= 0 18 765 585, clip]{Results/Graphics/Exploratory-Analysis/NPV-Oil-Water-Contribution/OLYMPUS_25-all-Decisions-NPV-Oil-Water-Contribution-plots}
		\vspace{-5pt}
		\caption{Logarithmic scale.}
		\label{subfig:OLYMPUS-25-Exploratory-Analysis-NPV-Oil-Water-Contribution-all-decisions-log-scale}
	\end{subfigure}
	\vspace{-8pt}
	\caption[OLYMPUS 25 approximate absolute annual contributions to the NPV for the exploratory simulations]{OLYMPUS 25 approximate absolute contribution to the NPV per year for each of the exploratory simulations shown as coloured lines. The NPV is decomposed into the oil production (solid line), absolute water production (dot-dashed line) and injection (dotted line), and the total water contribution (dashed line), with each scaled by the respective fixed NPV cost parameter. These are $\lvert Q_{j,op}(\boldd, t_i) \cdot r_{op} \rvert$, $\lvert Q_{j,wp}(\boldd, t_i) \cdot r_{wp} \rvert$, $\lvert Q_{j,wi}(\boldd, t_i) \cdot r_{wi} \rvert$ and $\lvert Q_{j,wp}(\boldd, t_i) \cdot r_{wp} + Q_{j,wi}(\boldd, t_i) \cdot r_{wi} \rvert$ in \cref{eq:Well-control-optimisation-challenge-NPV-R(d_t_i)} respectively. The top and bottom plots are on the raw and logarithmic scale respectively.}
	\label{fig:OLYMPUS-25-Exploratory-Analysis-NPV-Oil-Water-Contribution-all-decisions}
\end{figure}

Our exploratory analysis identifies large differences in the absolute contributions of oil and water, both production and injection, to the NPV objective function. This feature has potential ramifications for emulation and decision support. An assessment of the absolute contributions approximated within one year intervals for the OLYMPUS 25 NPV is shown in \cref{fig:OLYMPUS-25-Exploratory-Analysis-NPV-Oil-Water-Contribution-all-decisions} using each of the 20 exploratory analysis decision parameter vectors represented by different colours. In \cref{eq:Well-control-optimisation-challenge-NPV-R(d_t_i)} the oil contribution (solid lines), $\lvert Q_{j,op}(\boldd, t_i) \cdot r_{op} \rvert$, is dominant versus both the absolute water production contribution (dot-dashed lines), $\lvert Q_{j,wp}(\boldd, t_i) \cdot r_{wp} \rvert$, and water injection contribution (dotted lines), $\lvert Q_{j,wi}(\boldd, t_i) \cdot r_{wi} \rvert$, as well as their sum (dashed lines), $\lvert Q_{j,wp}(\boldd, t_i) \cdot r_{wp} + Q_{j,wi}(\boldd, t_i) \cdot r_{wi} \rvert$. For earlier time intervals the magnitude of the oil contribution to the NPV is typically of the order of 100 times the combined water contribution which decays towards 10 times larger for later time intervals. Plotting on the logarithmic scale in \cref{subfig:OLYMPUS-25-Exploratory-Analysis-NPV-Oil-Water-Contribution-all-decisions-log-scale} facilitates an easier comparison of the water contributions. It is observed that water injection contributes a much larger amount to the NPV, particularly for earlier time intervals. This is to be expected since production wells are drilled within regions containing a high oil concentration, hence at initial times there should be very little water production. At later times the contribution becomes more alike as an increased quantity of water is produced in order to maintain oil production, whilst also noting the higher fixed cost per barrel of water produced versus injected. Similar observations are made for other OLYMPUS ensemble members.

\subsection{Subsampling from Geological Multi-Model Ensembles -- Additional Plots} \label{subsec:Extended-Results-Subsampling-from-Geological-Multi-Model-Ensembles}

\begin{figure}[!t]
	\centering
	\begin{subfigure}[t]{0.4955\linewidth}
			\centering
			\includegraphics[page=25,width=\linewidth]{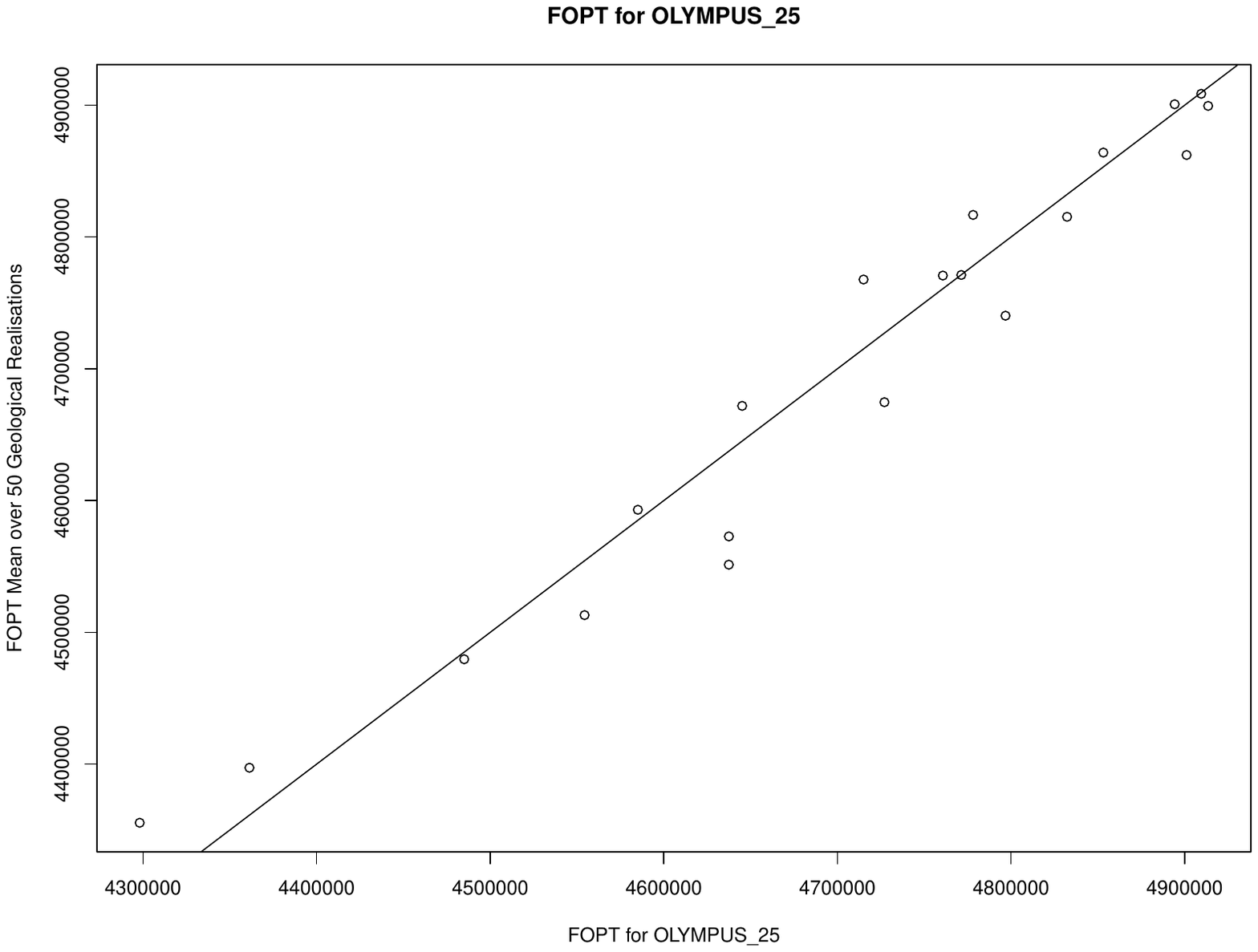}
			\caption{OLYMPUS 25 FOPT.}
			\label{subfig:OLYMPUS-EGES-FOPT-ensemble-mean-vs-model-OLYMPUS-25}
		\end{subfigure}
	\hfill
	\begin{subfigure}[t]{0.4955\linewidth}
			\centering
			\includegraphics[page=45,width=\linewidth]{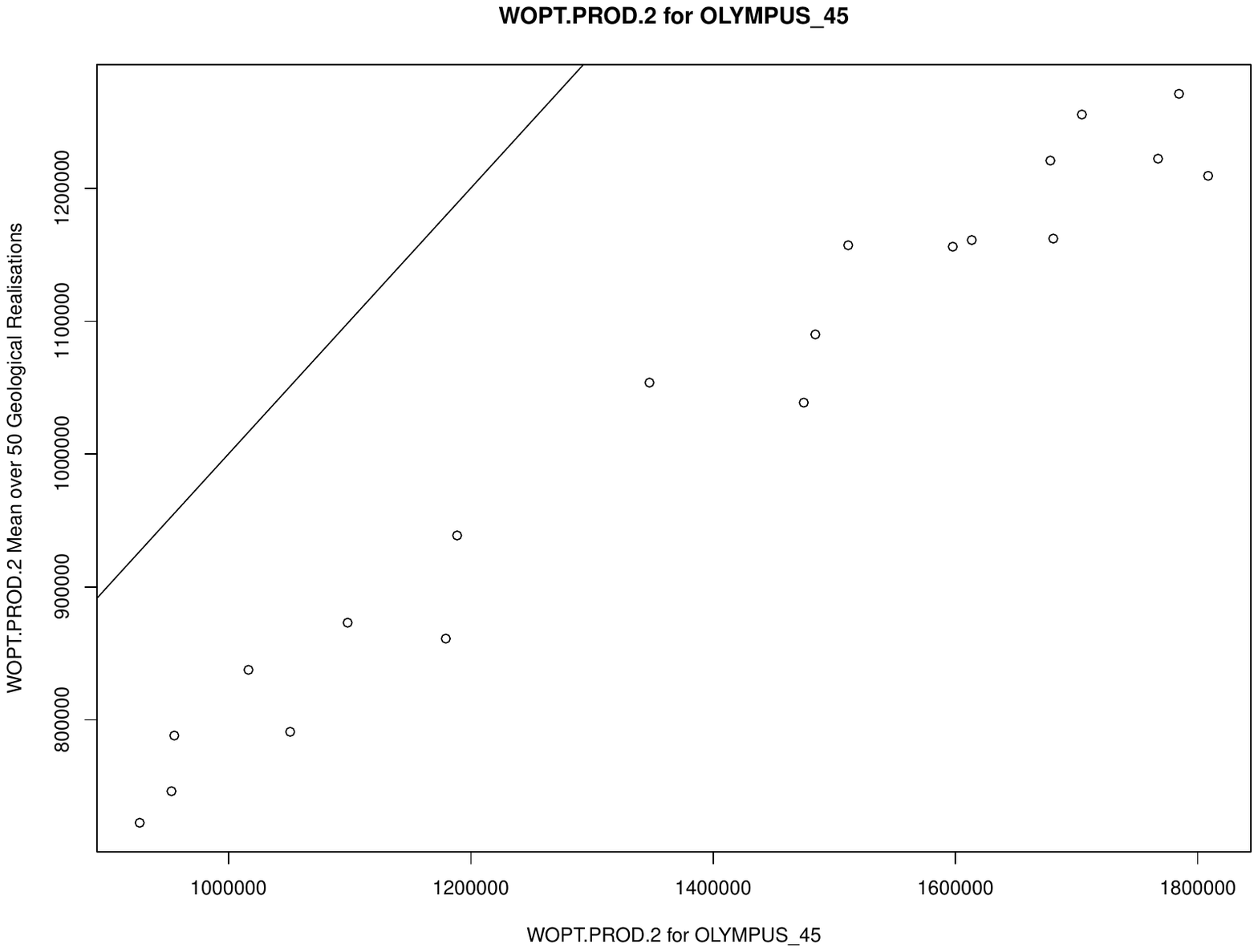}
			\caption{OLYMPUS 45 WOPTPROD2.}
			\label{subfig:OLYMPUS-EGES-WOPTPROD2-ensemble-mean-vs-model-OLYMPUS-45}
		\end{subfigure}
	
	\begin{subfigure}[t]{0.4955\linewidth}
			\centering
			\includegraphics[page=33,width=\linewidth]{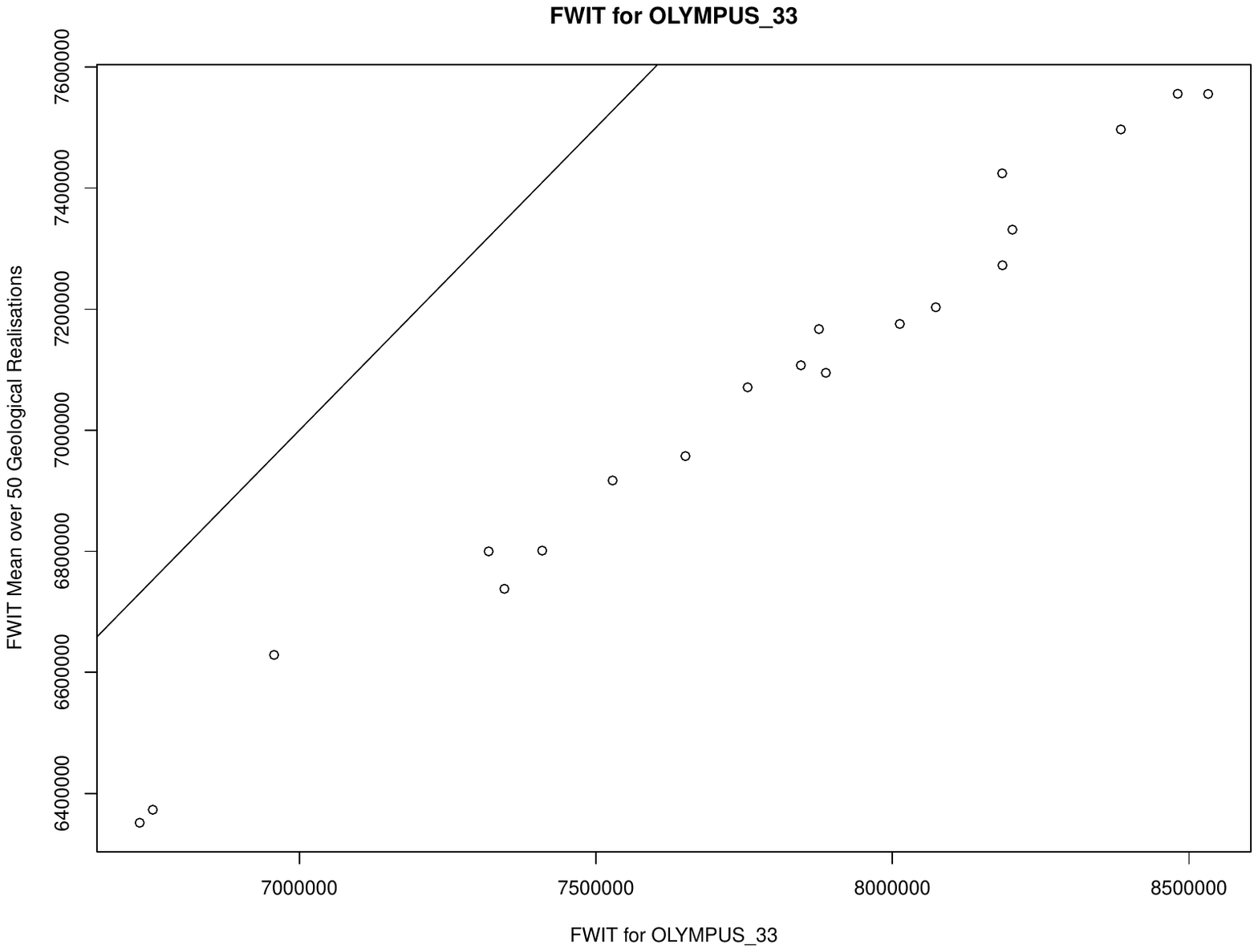}
			\caption{OLYMPUS 33 FWIT.}
			\label{subfig:OLYMPUS-EGES-FWIT-ensemble-mean-vs-model-OLYMPUS-33}
		\end{subfigure}
	\hfill
	\begin{subfigure}[t]{0.4955\linewidth}
			\centering
			\includegraphics[page=50,width=\linewidth]{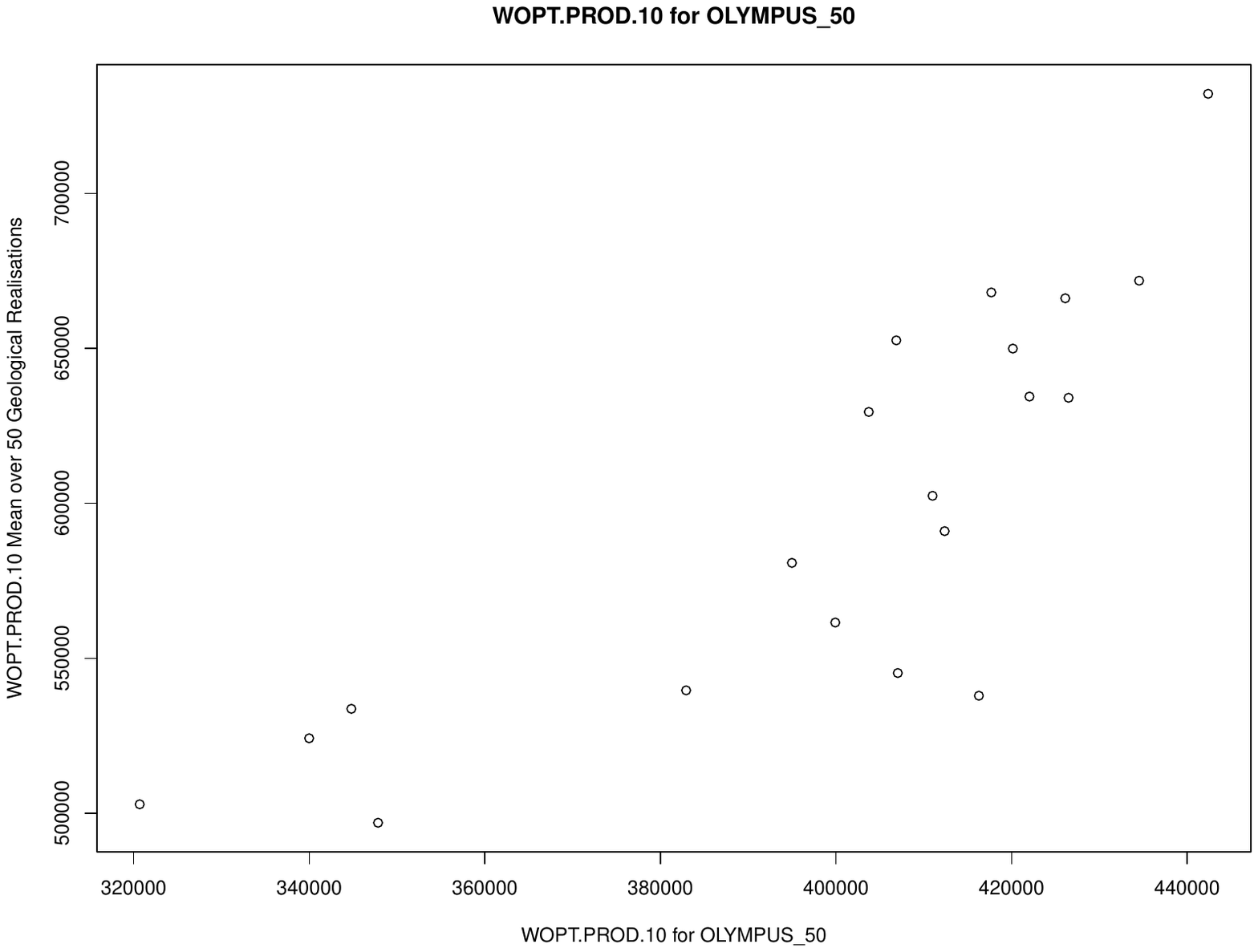}
			\caption{OLYMPUS 50 WOPTPROD10.}
			\label{subfig:OLYMPUS-EGES-WOPTPROD10-ensemble-mean-vs-model-OLYMPUS-50}
	\end{subfigure}
	\caption{Subsampling from the multi-model OLYMPUS ensemble preliminary graphical investigations showing the ensemble mean versus individual model outputs. The black line denotes equality between the ensemble mean and individual model outputs. Note that in \cref{subfig:OLYMPUS-EGES-WOPTPROD10-ensemble-mean-vs-model-OLYMPUS-50} the black line is not shown due to the much smaller values of WOPTPROD10 for OLYMPUS 50 compared with the ensemble mean.}
	\label{fig:OLYMPUS-EGES-ensemble-mean-vs-model-outputs}
\end{figure}

Preliminary graphical investigations utilise plots of the ensemble mean versus the individual model over a range of outputs of interest for the wave 0 simulations. Examples of these plots are shown in \cref{fig:OLYMPUS-EGES-ensemble-mean-vs-model-outputs} where the black line denotes equality between the ensemble mean and individual ensemble member model output. The main outputs of interest stem from the NPV objective function and include: the ensemble mean NPV, oil production, water production and injection totals, both for the field and by well, as well as over the entire field lifetime, and for control intervals. Note that this is a preliminary graphical assessment which is limited to identifying one-dimensional relationships. \Cref{subfig:OLYMPUS-EGES-FOPT-ensemble-mean-vs-model-OLYMPUS-25,subfig:OLYMPUS-EGES-WOPTPROD2-ensemble-mean-vs-model-OLYMPUS-45,subfig:OLYMPUS-EGES-FWIT-ensemble-mean-vs-model-OLYMPUS-33} show strong linear relationships with fairly limited variation providing evidence that even as individual models, OLYMPUS 25, 33 \& 45 are potentially representative for the ensemble mean. An appropriate (linear) transformation may be applied in the cases seen in \cref{subfig:OLYMPUS-EGES-WOPTPROD2-ensemble-mean-vs-model-OLYMPUS-45,subfig:OLYMPUS-EGES-FWIT-ensemble-mean-vs-model-OLYMPUS-33}. In contrast OLYMPUS 50 does not appear to be a good representative model, at least individually, as seen in \cref{subfig:OLYMPUS-EGES-WOPTPROD10-ensemble-mean-vs-model-OLYMPUS-50} where the relationship is more challenging to model. This graphical investigation is also useful as a preliminary screening technique yielding a subset of 9 models to investigate further: OLYMPUS 2, 6, 11, 23, 25, 33, 35, 37, \& 38.

\begin{figure}[!t]
	\centering
	\includegraphics[width=\linewidth]{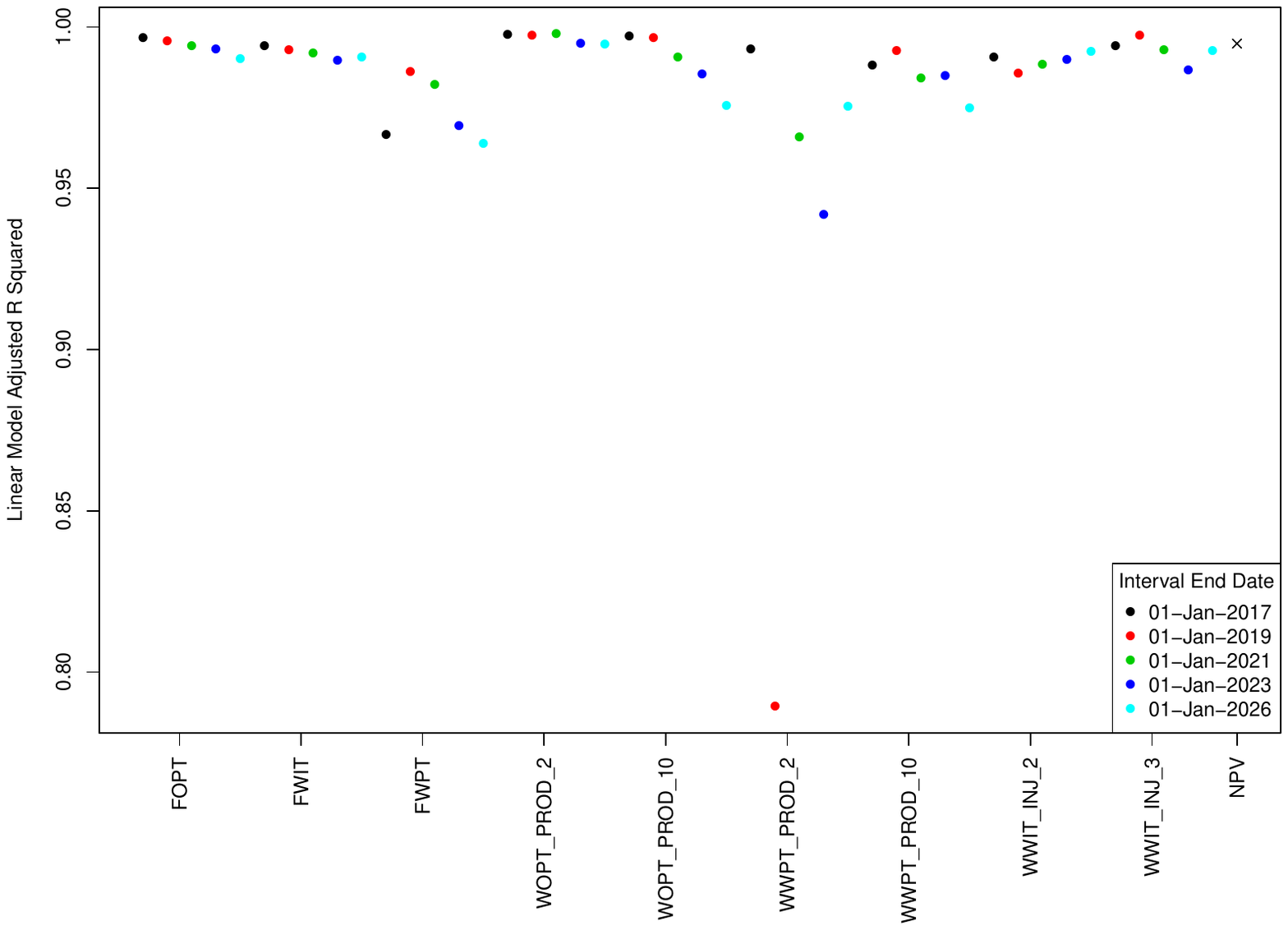}
	\caption{\AdjRsq{} values for the OLYMPUS ensemble subsampling linear models of the form in \cref{eq:EGES-ensemble-mean-resp-linear-model-on-individual-model-resp-N_EGES-models-unknown} for the ensemble mean of various outputs within control intervals using the same subset of $\Ntilde = 3$ OLYMPUS models as predictors.}
	\label{fig:OLYMPUS-EGES-LM-all-resp-and-NPV-adj-R-Sq-3-models}
\end{figure}

The combination of different OLYMPUS models is assessed using the linear model subsampling technique in \ref{eq:EGES-ensemble-mean-resp-linear-model-on-individual-model-resp-N_EGES-models-unknown}. This is first applied to the above proposed subset of 9 OLYMPUS models before considering all models in a both directions stepwise selection with AIC. Only $\Ntilde = 3$ models are necessary for a large number of the investigated outputs, as demonstrated in \cref{fig:OLYMPUS-EGES-LM-all-resp-and-NPV-adj-R-Sq-3-models} showing the linear model \adjRsq{} values for various outputs. All are high with most greater than 0.95 implying that the majority of the ensemble variation can be explained by a small subset. These are OLYMPUS 25, 33, \& 45. It is noted that OLYMPUS 25 \& 33 were identified as part of the proposed subset of models, where as OLYMPUS 45 was not. This is because for certain outputs it was judged that OLYMPUS 45 did not provide a sufficiently good representation of  the ensemble mean, however, in combination with OLYMPUS 25 \& 33 via the linear models, these models collectively provide a good characterisation of the ensemble mean NPV, as well as other outputs.

\subsection{Hierarchical Emulation of the Expected NPV -- Additional Plots} \label{subsec:Extended-Results-Hierarchical-Emulation-of-the-Expected-NPV}

\begin{figure}[!t]
	\centering
	\begin{subfigure}[t]{0.4955\linewidth}
			\centering
			\includegraphics[width=\linewidth]{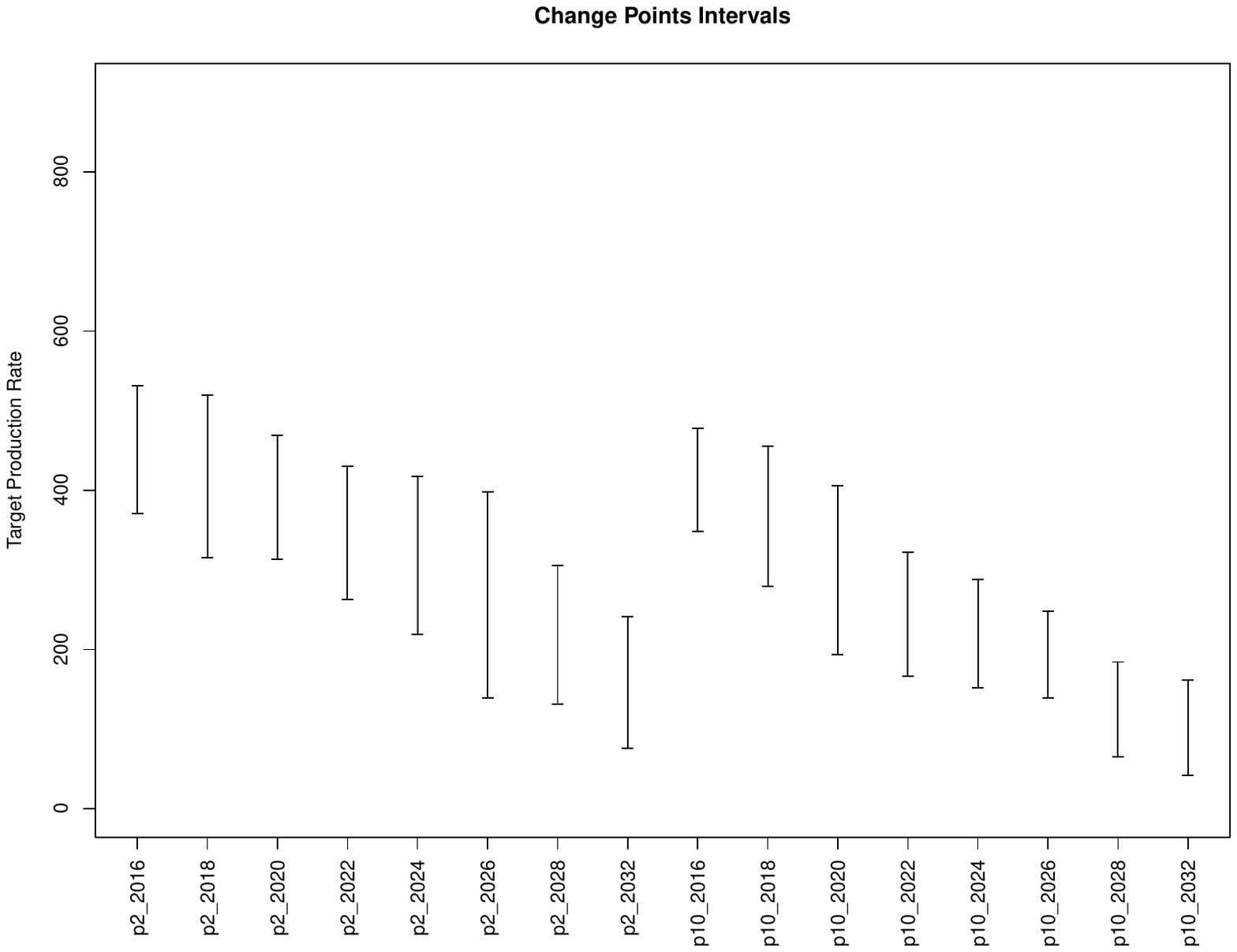}
			\caption{OLYMPUS 25 WOPT}
			\label{subfig:OLYMPUS-25-w1-WOPT-cp-and-extrap-co-intervals}
		\end{subfigure}
	\hfill
	\begin{subfigure}[t]{0.4955\linewidth}
			\centering
			\includegraphics[width=\linewidth]{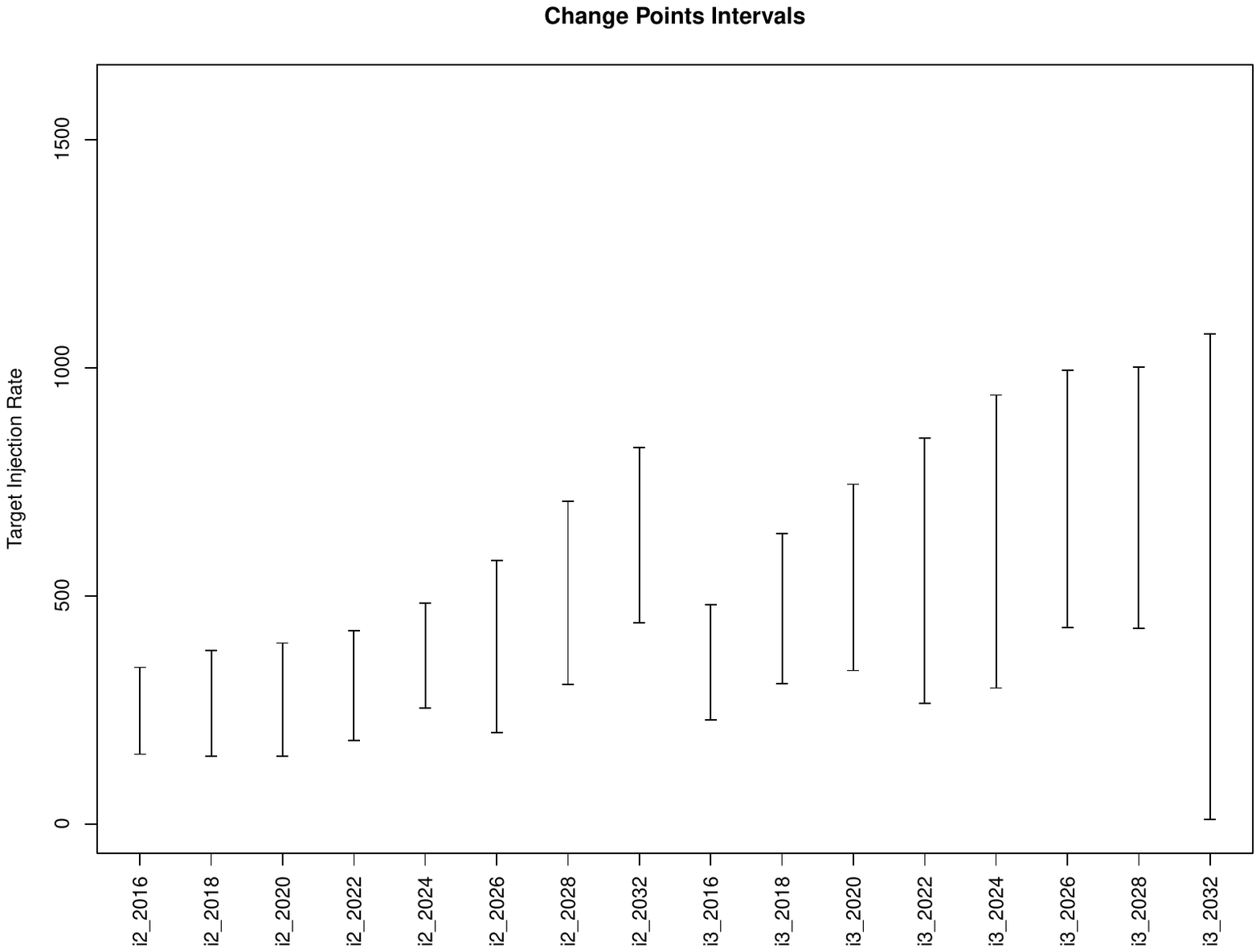}
			\caption{OLYMPUS 25 WWIT}
			\label{subfig:OLYMPUS-25-w1-WWIT-cp-and-extrap-co-intervals}
		\end{subfigure}
	
	\begin{subfigure}[t]{0.4955\linewidth}
			\centering
			\includegraphics[width=\linewidth]{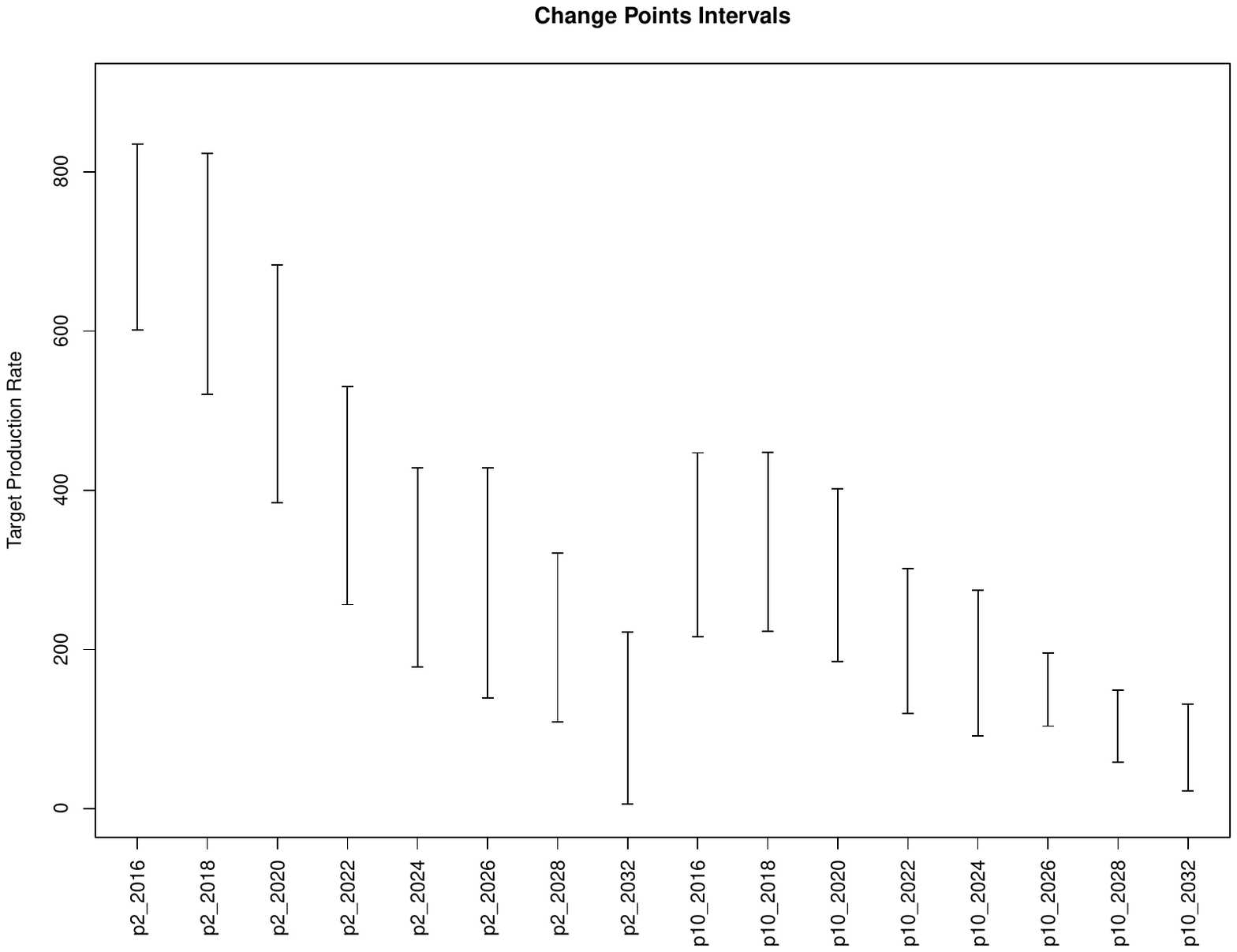}
			\caption{OLYMPUS 33 WOPT}
			\label{subfig:OLYMPUS-33-w1-WOPT-cp-and-extrap-co-intervals}
		\end{subfigure}
	\hfill
	\begin{subfigure}[t]{0.4955\linewidth}
			\centering
			\includegraphics[width=\linewidth]{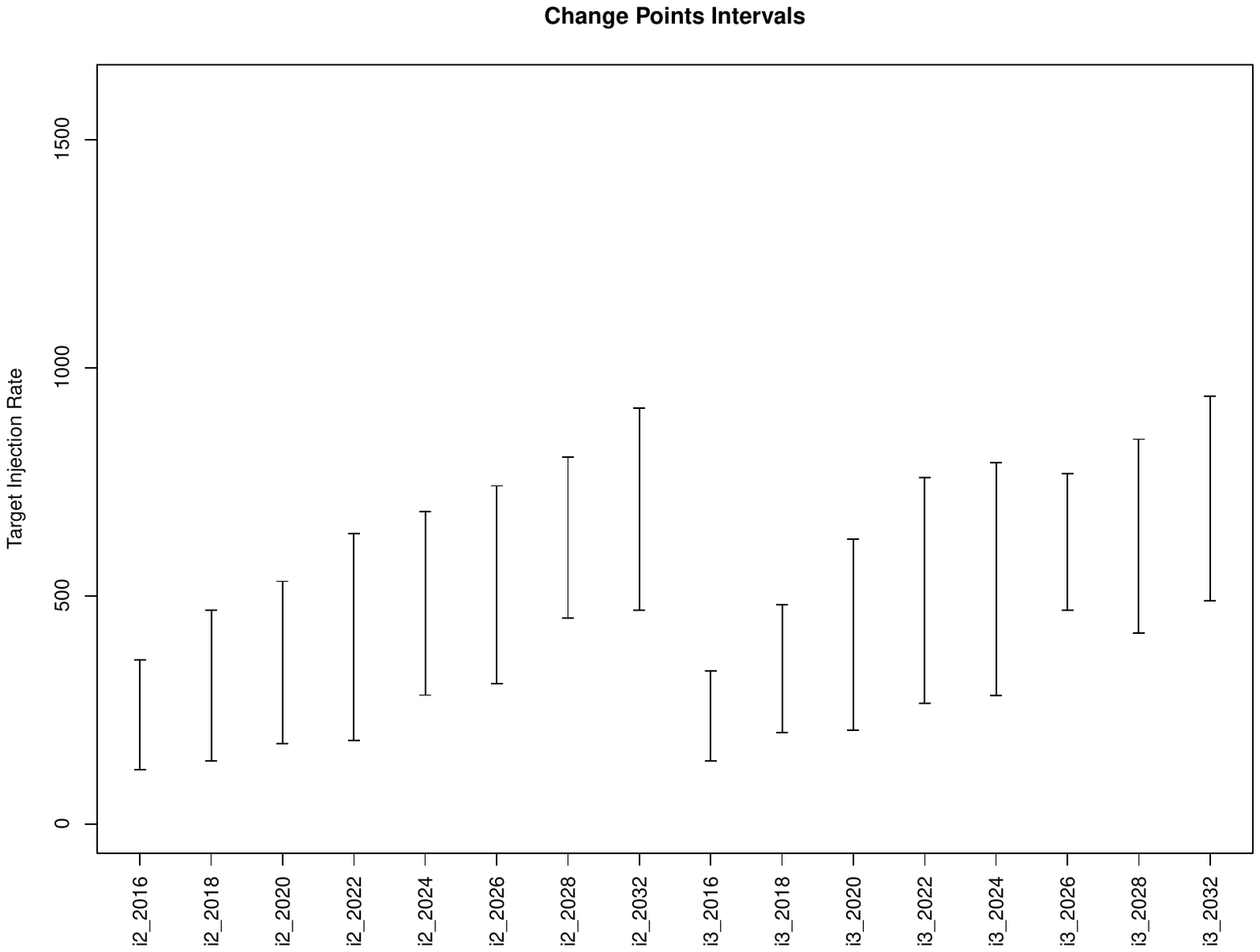}
			\caption{OLYMPUS 33 WWIT}
			\label{subfig:OLYMPUS-33-w1-WWIT-cp-and-extrap-co-intervals}
		\end{subfigure}
	
	\begin{subfigure}[t]{0.4955\linewidth}
			\centering
			\includegraphics[width=\linewidth]{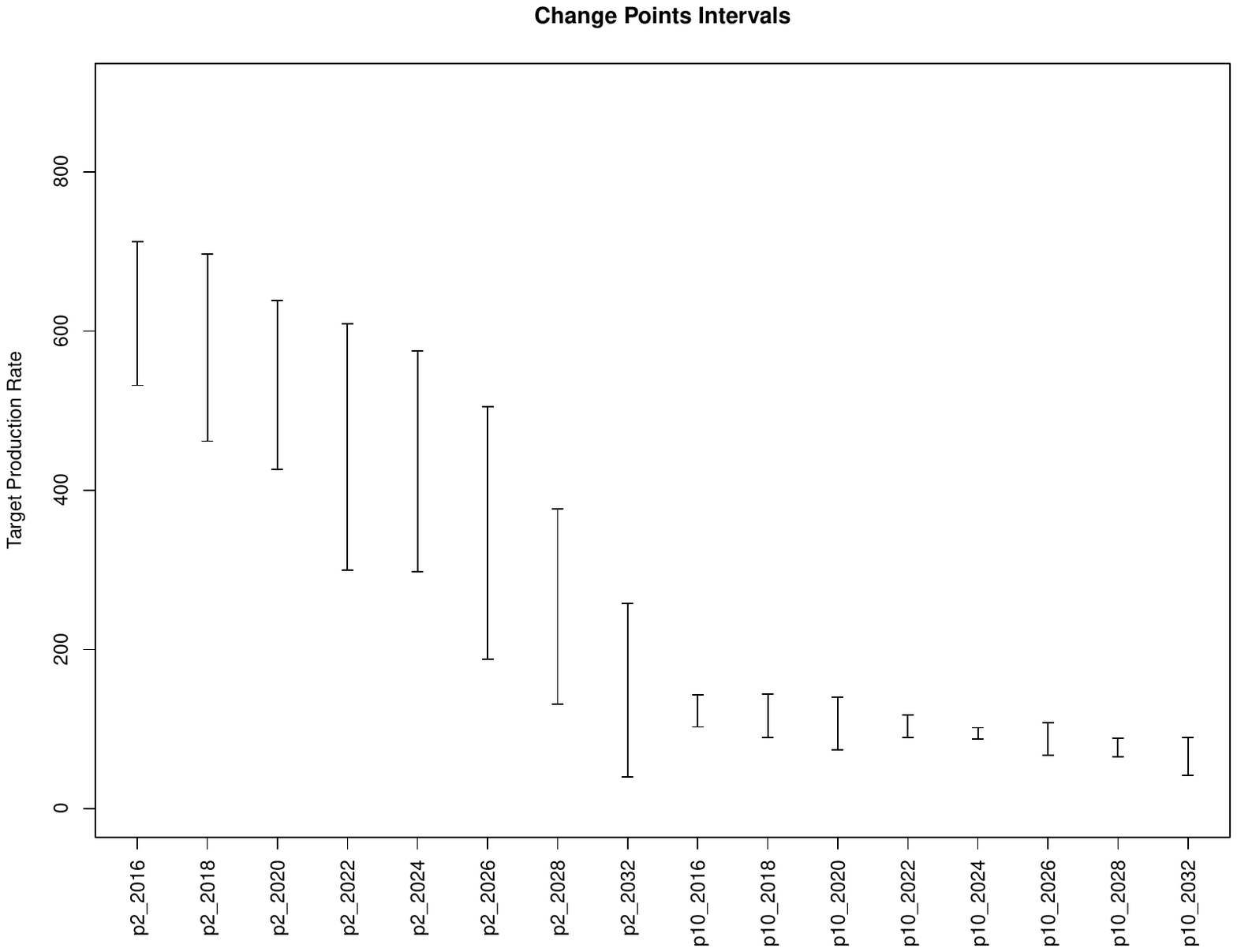}
			\caption{OLYMPUS 45 WOPT}
			\label{subfig:OLYMPUS-45-w1-WOPT-cp-and-extrap-co-intervals}
		\end{subfigure}
	\hfill
	\begin{subfigure}[t]{0.4955\linewidth}
			\centering
			\includegraphics[width=\linewidth]{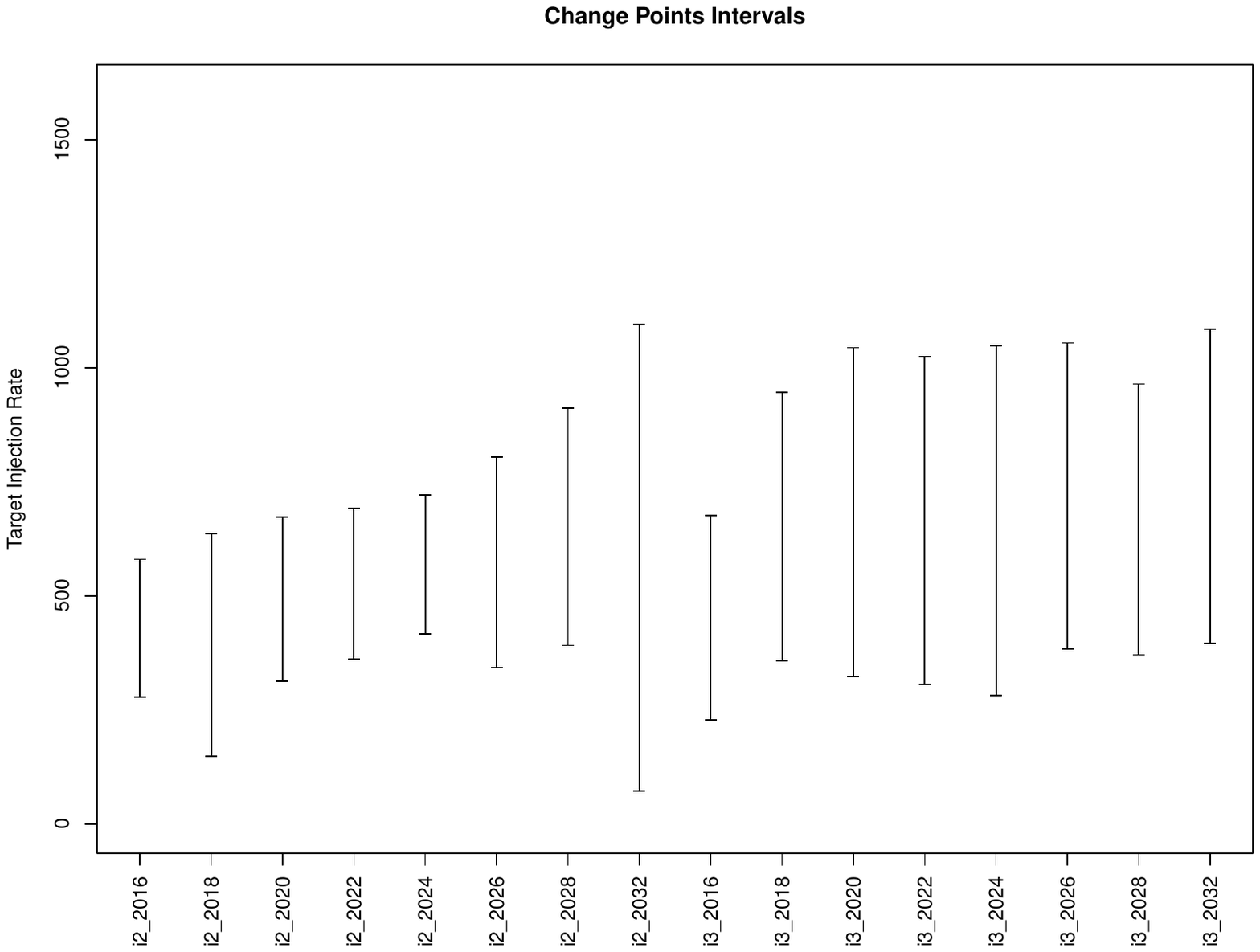}
			\caption{OLYMPUS 45 WWIT}
			\label{subfig:OLYMPUS-45-w1-WWIT-cp-and-extrap-co-intervals}
		\end{subfigure}
	\caption[OLYMPUS wave 1 NPV constituents change point upper bound and extrapolation cut-off intervals]{OLYMPUS wave 1 change point upper bound and extrapolation cut-off intervals for WOPT and WWIT within each control interval with respect to their corresponding decision parameter for each of the three sub-sampled OLYMPUS models.}
	\label{fig:OLYMPUS-w1-WOPT-WWIT-cp-and-extrap-co-intervals}
\end{figure}

The structured emulation technique incorporating known simulator behaviour in \cref{subsec:Structured-Emulators-Exploiting-Known-Simulator-Behaviour-Methodology} is applied separately for each of the OLYMPUS models to the WOPT and WWIT within each control interval for wells in the CWG, with simulator outputs considered as the $\fd$. Firstly, conservative estimates for the change point upper bounds are calculated from the wave 1 simulations using \cref{eq:Change-point-upper-bound}, each time with $\delta_{u} = 10$. This ensures numerical stability and that an upper bound is obtained with all points exceeding this definitely in the plateau region. Next the extrapolation cut-offs are estimated as the change point lower bounds, via \cref{eq:Change-point-lower-bound}, with $\delta_{l} = 10$ to account for numerical precision within the simulations. The change point upper bounds and extrapolation cut-offs are illustrated for all WOPT and WWIT constituents for each wave 1 sub-selected OLYMPUS model in \cref{fig:OLYMPUS-w1-WOPT-WWIT-cp-and-extrap-co-intervals} highlighting the region in which the ``true'' change point is believed to be situated.

\end{document}